\documentclass[a4paper,11pt]{article}

\usepackage[T1]{fontenc}
\usepackage[utf8]{inputenc}
\usepackage{lmodern}

\usepackage[a4paper, hmargin=2.5cm, vmargin=3cm]{geometry}
\usepackage{setspace}
\usepackage{amsmath}
\usepackage{amssymb}

\usepackage{graphicx}
\usepackage{xcolor}
\usepackage{tikz}
\usetikzlibrary{arrows.meta, positioning}

\usepackage{float}
\usepackage{caption}
\usepackage{subcaption}
\usepackage{longtable}
\usepackage{tabularx}
\usepackage{multirow}
\usepackage{makecell}
\usepackage{booktabs}
\usepackage{rotating}
\usepackage{threeparttable}
\usepackage{dcolumn}
\usepackage{pdflscape}

\usepackage{siunitx}
\usepackage[vlined,ruled,linesnumbered]{algorithm2e}

\usepackage{pifont}
\newcommand{\cmark}{\ding{51}}
\newcommand{\xmark}{\ding{55}}

\usepackage{natbib}
\usepackage{comment}
\usepackage{soul}

\usepackage{authblk}

\usepackage[hidelinks]{hyperref}

\usepackage{amsthm}
\theoremstyle{definition}
\newtheorem{definition}{Definition}
\newtheorem{assumption}{Assumption}

\title{Stablecoin Integration and Stress Transmission Through Crypto-Asset Service Providers: Evidence from Austria}

\author[1,2]{Pietro Saggese\thanks{Corresponding author. E-mail: pietro.saggese@imtlucca.it}}
\author[3]{Michael Sigmund}
\author[3]{Burkhard Raunig}
\author[3]{Esther Segalla}
\author[2]{Bernhard Haslhofer}
\author[2,4]{Christos A. Makridis}

\affil[1]{IMT Scuola Alti Studi Lucca}
\affil[2]{Complexity Science Hub}
\affil[3]{Oesterreichische Nationalbank (OeNB)}
\affil[4]{Arizona State University}

\date{\today}

\begin{document}

\maketitle

\begin{abstract}

Stablecoins are increasingly entangled with traditional finance, making their response to stress a financial stability concern.
However, blockchain pseudonymity and the lack of geographic identifiers force existing work to rely on indirect proxies to infer and locate market participants. 
In this paper, we exploit a regulatory registry that directly identifies the on-chain addresses of every crypto-asset service provider (CASP) registered in Austria at the end of 2024, reconstructing their Bitcoin, Ether, USDT, and USDC activity through May 2025. 
Austrian CASPs intermediate roughly \$30~billion in cumulative external transfers, while transfers among them are negligible.
Combining two transaction-level signals, interaction with CASP cold wallets and on-chain transaction frequency, we classify CASP counterparty activity as retail-like or institutionally mediated. 
Around stress events, the two groups exhibit distinct flow responses that are obscured in aggregate data. 
After the FTX bankruptcy, we observe retail-like withdrawals
and institutional deposits across assets. After Silicon Valley Bank (SVB) failure, retail-like USDC deposits and institutionally mediated withdrawals are consistent with USDC’s asymmetric redemption access.
Registry-based, transaction-level measurements thus offer a reproducible basis for cross-jurisdictional monitoring of how cryptoasset markets transmit risk.

\vspace{0.5cm}


\vspace{0.2cm}

\end{abstract}

\clearpage

\section{Introduction}
\label{sec:intro}

When a money-like claim comes under stress, who runs, and through which channels? Variants of this question have organized research on intermediary fragility and runs in banking and shadow banking, from bank deposits to money market funds and asset-backed commercial paper~\citep{covitz2013evolution, schmidt2016runs}.
Stablecoins, now a major asset class in the
financial system with a total market capitalization of roughly \$300~billion in mid-2026,\footnote{See \url{http://defillama.com/stablecoins}, accessed on October 2026.} pose the same questions in a new setting: they are privately issued,
money-like digital claims designed to maintain a fixed value, mostly through cash or safe short-term securities reserve backing, and on-demand convertibility.
The most prominent are fiat-backed, issuer-managed stablecoins like Tether (USDT) or USD Coin (USDC); both are tokenized claims residing on a Distributed Ledger Technology (DLT), but they are often custodied on holders' behalf by crypto-asset service providers (CASPs). 
Like comparable money-like claims, they are thus vulnerable to runs~\citep{gorton2023taming}, determined by the informational quality of the reserve assets and stabilization mechanisms~\citep{ahmed2023public}.
Moreover, while issuers promise par convertibility, the majority of investors can only trade stablecoins on secondary markets~\citep{ma2025stablecoin}, affecting who can redeem the underlying assets, and how, in times of stress.
These two institutional features make this setting a natural laboratory for intermediary fragility, and stablecoins' growing entanglement with the traditional financial system, through reserve composition, custodial intermediaries, and institutional participation, makes their behavior under stress a first-order concern for regulators and central banks.

Yet, measuring the activity of crypto intermediaries like CASPs, and hence understanding their role under stress, faces a binding data constraint: blockchain transactions are pseudonymous and public networks can generate fresh addresses by design, leaving the on-chain footprint of individual holders and their custodians imperfectly observable.
Moreover, blockchain transactions carry no geographic identifier, which opens up a second challenge: CASPs are legal entities, registered and supervised in specific jurisdictions, and the primary channel through which investors interact with cryptoassets off-chain; however, today it is hard to quantify their integration within national financial systems.
Most studies focus on the global cryptocurrency market~\citep{makarov2021blockchain}, while works inferring jurisdiction-level activity rely on indirect proxies like web traffic, IP geolocation, off-chain exchange data, or survey indices~\citep{auer2025crossborder,makridis:cbdcs}, all of which are noisy, partial, and ill-suited to fine-grained behavioral analysis. Basic questions thus remain open: how large CASP-mediated crypto activity is within an economy, how individuals reallocate across assets and custody structures under stress, and whether stablecoins serve the safe-haven function issuers claim during turmoil.

In this paper, we address this gap by measuring at the transaction level how the cryptoasset activity intermediated by Austria's end-2024 cohort of registered CASPs
is structured and how it responds to stress. Our approach exploits a feature of Austrian financial reporting: under current regulation, crypto-asset service providers must report the full set of blockchain addresses they control to the competent authorities~\citep{fma2025aml}. From the resulting pseudonymized address-to-entity registry, we reconstruct the on-chain flows directed to and originating from the 12 CASPs in our sample,
across Bitcoin (BTC), Ether (ETH), and the two largest stablecoins, USDT and USDC, through May 2025. 
Because the registry provides direct attribution rather than inferring ownership through clustering heuristics, we observe these flows with a reliability that proxy-based country-level studies cannot achieve.\footnote{Coverage still depends on the accuracy and completeness of CASP reporting; see Section~\ref{sec:limits} for a discussion.}

We exploit this dataset to characterize the structure of CASP-intermediated activity in Austria, consisting of \num{11980687} asset transfers worth \$50.4~billion in total; next, we classify counterparties into retail-like and institutionally mediated groups based on two structural signals, their interaction with CASP warm/cold wallets and their blockchain-based transaction activity. Then, we use this classification within an event-study setting around three distinct shock types -- an algorithmic stablecoin de-pegging (Terra), a centralized exchange insolvency (FTX), and a banking-sector failure with direct stablecoin reserve exposure (SVB) -- to examine how observed flow responses vary across assets, counterparty groups, and directions, and whether the patterns are consistent with differences in redemption access and intermediary exposure.

We document four main findings. 
First, Austrian CASPs intermediate roughly \$30~billion in on-chain flows with counterparties outside the registry, while direct transfers between registered Austrian CASPs total just \$5.8~million: on-chain inter-CASP links within the jurisdiction are economically negligible. 
A further 41.1\% of total volume reflects within-entity bookkeeping rather than counterparty activity, and would be misclassified as economic activity in aggregate on-chain data.
Second, external activity is highly concentrated: a small set of counterparties transacting through CASP warm and cold wallets accounts for 57.8\% of external volume while representing only 0.77\% of external transfers, whereas about 98\% of addresses on each chain exhibit low-frequency, retail-like behavior.
Third, observed flow responses differ between retail-like and institutionally mediated groups across the three episodes. 
The Terra Luna crash generated broad but short-lived reallocation across assets, concentrated in institutionally mediated flows.
After the FTX collapse, retail-like withdrawals increased persistently, consistent with reduced custodial exposure following the failure of a major exchange,
while institutionally mediated flows are consistent with liquidity management or cross-platform rebalancing rather than a one-way exit. 
The Silicon Valley Bank (SVB) failure produced a narrower response concentrated in USDC, the stablecoin with direct reserve exposure to SVB: consistent with fiat-backed stablecoins' unequal redemption mechanism, retail-like counterparties deposited USDC, while institutional ones operated in the opposite direction; BTC and ETH remained mostly unaffected. 
Fourth, stablecoin flow responses were two-sided and shock-specific, most clearly around FTX and SVB, and do not support the stablecoin-as-safe-haven hypothesis.

Our paper contributes primarily to the empirical literature on runs and fragility in traditional money-like asset pools and its extension to stablecoins and crypto markets.
In mutual funds, money market funds (MMFs), and asset-backed commercial paper (ABCP), strategic complementarities make redemptions self-reinforcing~\citep{chen2010payoff}, and run risk depends on the investor base, the intermediary, and portfolio characteristics: retail and institutional investors in the same fund react differently under stress~\citep{schmidt2016runs}, and vehicles with weaker sponsor support and lower-quality assets are more likely to suffer runs~\citep{covitz2013evolution}.
These dynamics are visible only when flows are observable at a granular level, since aggregate series pool heterogeneous investor behaviors, and inflows can offset outflows~\citep{martin2026deposit}. 
Fiat-backed stablecoins are money-like claims increasingly compared with MMFs~\citep{oefele2024stablecoins,anadu2026framework} and exposed to similar risks~\citep{gorton2023taming,goel2026making}. 
When custodied by CASPs, stablecoins involve a further intermediary layer, and two risks coexist: the issuer's reserve and redemption risk, and the custodian's operational and insolvency risk. CASPs are therefore a relevant channel for stress transmission; yet, evidence on their activity and responses under stress remains incomplete. 
On-chain studies observe stablecoin issuance, transfers, and redemptions~\citep{liu2023anatomy} or cross-asset reallocations~\citep{sankaewtong2026tracing} comprehensively, but across accounts that are unattributed or imperfectly attributed~\citep{griffin2020bitcoin,makarov2021blockchain,lyons2023keeps}; off-chain studies rely instead on prices, aggregate flows, or theoretical models~\citep{briola2023anatomy,galati2024silicon,aldasoro2025stablecoins}.
\citet{ma2025stablecoin}, for instance, identify issuer-eligible arbitrageurs through primary-market interactions on-chain, but observe the secondary market only through exchange prices.

We extend this literature in four ways.
First, we provide evidence on responses to stress in custodial intermediaries at the transaction level and by counterparty type.
We observe each on-chain deposit and withdrawal at the registered CASPs in our sample separately, giving cryptoasset markets the kind of granular data that advanced the literature on market fragility~\citep{covitz2013evolution, martin2026deposit}. Moreover, separating flows into retail-like and institutionally mediated groups, an on-chain analogue of the retail and institutional share classes in \citet{schmidt2016runs}, reveals patterns otherwise concealed in aggregate data.
Second, we document counterparty-specific flow patterns consistent with asymmetric access to stablecoin redemption. For fiat-backed stablecoins like USDT and USDC, direct par redemption requires an eligible issuer account. By separating flows by counterparty, we find that around SVB, retail-like USDC deposits into CASPs coincided with institutional USDC withdrawals. This pattern is consistent with the two-tier redemption mechanism documented by \citet{ma2025stablecoin}, where eligible issuer-account holders can redeem USDC directly at par, whereas other holders rely on intermediaries or secondary markets. 
Third, observed responses differ across shocks with distinct sources of risk. The FTX shock, a custodian failure, reached all four assets, stablecoins included, whereas SVB, a shock to USDC's reserves, impacted primarily USDC and did not produce a systematic response in BTC or ETH; Terra, a design failure in a competing stablecoin, produced primarily short-lived rebalancing across assets.
Fourth, we document responses at intermediaries in a different jurisdiction from the failing institution and not themselves in distress. After FTX, retail-like withdrawals from Austrian CASPs followed the public escalation of a crisis originating abroad, with no visible lag. These withdrawals hit solvent, registered intermediaries other than the failed one, consistent with a sector-wide loss of confidence in custodial intermediation rather than institution-specific information.

Next, we contribute to a related and active macro-finance debate over the role of stablecoins as safe-haven assets~\citep{baur2010gold}.\footnote{
In the broader literature on safe assets and convenience yields, flights-to-safety involve reallocation toward assets providing liquidity, collateral, and protection in bad states, especially government debt and money-like instruments~\citep{baele2020flights,krishnamurthy2012aggregate}. These properties are not intrinsic to dollar denomination alone, but depend on institutional credibility, market depth, liquidity, and state-contingent hedging value \citep{caballero2008equilibrium,he2019model,jiang2021foreign}. Fiat-backed stablecoins are an analogous but distinct case: dollar-denominated private money-like claims whose safe-asset properties depend on reserve composition, redemption rights, issuer credibility, and custodial structure.%
} 
Two strands of evidence reach opposing conclusions.
The flight-to-quality view~\citep{anadu2024runs,oefele2024flight} documents that aggregate stablecoin flows during stress episodes mirror the prime-to-government MMF reallocation patterns familiar from traditional finance; the contrary view~\citep{aldasoro2025stablecoins} finds that stablecoins respond to crypto-specific and monetary-policy shocks differently from MMFs and concludes that they do not consistently serve as safe havens. 
Similarly, \citet{campello2025demand} find evidence that stablecoins are fragile providers of safety and liquidity services under stress in Decentralized Finance (DeFi) lending markets. 
Other works reach mixed findings on whether stablecoins act as safe havens against other cryptoassets or markets~\citep{wang2020stablecoins,kolodziejczyk2023stablecoins,feng2024stablecoins}. 
Both strands rely on aggregate price or flow data, which cannot distinguish distinct responses; using transaction-level data, we document heterogeneous stablecoin transfers across all three shocks rather than a uniform directional flow response in support of the safe-haven hypothesis.\footnote{Because trades within a CASP and final holdings are not observed, these transfers do not by themselves directly establish or reject portfolio-level flights to safety.}

Our paper also contributes to a growing literature on the integration of digital assets into national economies. Empirical work has largely measured cryptoasset activity globally~\citep{ron2013quantitative,makarov2021blockchain} or approximated country-level flows through indirect proxies: IP-address relay data \citep{parino2018analysis}, exchange geolocation and web traffic \citep{feyen2022crypto,auer2025crossborder}, off-chain fiat trading pairs in cryptocurrency exchanges \citep{von2023decrypting}, and survey-based adoption measures \citep{makridis:cbdcs,abramova2023can}. 
Most studies link pseudonymous addresses to real-world actors through clustering heuristics and attribution techniques~\citep{meiklejohn2013fistful}.  
Moreover, direct-observation studies remain rare. \citet{alvarez2023cryptocurrencies} survey Bitcoin adoption in El Salvador after its 2021 legal-tender designation, and \citet{hu2021evading}, \citet{cardozo2024cross}, \citet{saggese2024assessing}, and \citet{cerutti2024primer} use indirect attribution to study Bitcoin flows in China, Brazil, Austria, and globally, respectively. 
None of these studies identifies ownership directly, nor observes the population of licensed CASPs in a national jurisdiction at the transaction level. 
We do so for Austria, covering every CASP registered at the end of 2024 and quantifying the size of the market they intermediate. To our knowledge, this is the first study of this kind.

Finally, our paper contributes to the literature on investor behavior in cryptoasset markets, focusing on retail versus institutional heterogeneity. Existing work on crypto investor characteristics relies primarily on brokerage account data~\citep{hackethal2022characteristics} or survey-based behavioral measures~\citep{almeida2023systematic,fessler2024oenb}, neither of which can be matched to on-chain transactions. We import retail-identification techniques from fixed-income~\citep{schultz2012market,bessembinder2020,oharazhou2021,dehaan2023retail} and equity microstructure~\citep{boehmer2021tracking,barber2024subpenny}. \citet{hund2025rise}, for instance, show that the standard \$100{,}000 trade-size cutoff misclassifies a substantial share of trades; we thus replace the single threshold approach with two structural signals: warm/cold wallet contact and chain-wide transaction activity. The resulting classification is the basis for an event-study framework that builds on crypto-specific precedents~\citep{abramova2021out,marmora2022does,yousaf2023responses,jalan2023systemic} and on the literature on narrative transmission and investor attention, following evidence that investors respond mainly to shocks with simple, asset-relevant narratives~\citep{narrative2017shiller,philippas2019media, jiang2022investor,FeldkircherMakridis:words}. 
We thus provide a novel approach to classify retail and institutional CASP-mediated flows that contributes both to the cryptocurrency literature, where direct jurisdiction-level attribution is limited, and to the traditional financial one, where typical threshold approaches consist of single rather than multiple structural signals. 
We emphasize that the resulting distinction is not a simple descriptive refinement but a relevant contribution because the composition of the investor base is itself a determinant of fragility.

The findings carry implications for several stakeholders. For supervisors, the registry-based approach is a systematic and reproducible alternative to proxy- and heuristic-based measurement, pointing toward collecting address and counterparty data directly from supervised CASPs, and could form the basis for a cross-jurisdictional observatory of cryptoasset flows. 
For policymakers, the heterogeneous responses observed indicate that aggregate measures can mask offsetting differences across custodial channels, and the limited and two-sided stablecoin response during stress bears on whether privately issued stablecoins can replicate the stabilizing role of traditional safe assets and how they will interact with public digital currencies. For CASPs, abnormal flows appear within days of the selected shocks, making the timing and direction of liquidity needs relevant to operational planning. For retail users, episodes like SVB expose a structural disadvantage: when a peg is threatened, par redemption is reserved for institutional accounts, leaving retail holders to sell on secondary markets, often at a loss~\citep{ma2025stablecoin}.

The remainder of the paper proceeds as follows. Section~\ref{sec:background} reviews the related literature and background. Section~\ref{sec:data} describes the dataset and how we select the events. Section~\ref{sec:flows} documents the structure of Austrian CASP activity and develops the retail and institutional classification, and Section~\ref{sec:event_study} reports the event-study results for the Terra Luna, FTX, and SVB shocks. Section~\ref{sec:discussion} discusses the findings and their implications, and Section~\ref{sec:conclusions} concludes.

\section{Background}
\label{sec:background}

Measuring the on-chain financial activity intermediated by CASPs and customer responses to shocks requires connecting three threads: how CASPs custody funds on-chain and why their flows are observable yet hard to attribute (Section~\ref{sub:casp_backg}); the redemption mechanisms of fiat-backed stablecoins and their relation to other money-like claims (Section~\ref{sec:stablecoins_backg});
and how retail activity can be isolated when identity is not observable or  under pseudonymity (Section~\ref{sec:ident_litrev}).

\subsection{CASPs and Their On-Chain Footprint}
\label{sub:casp_backg}

\begin{figure}[t]
	\centering
	\includegraphics[width=0.97\textwidth]{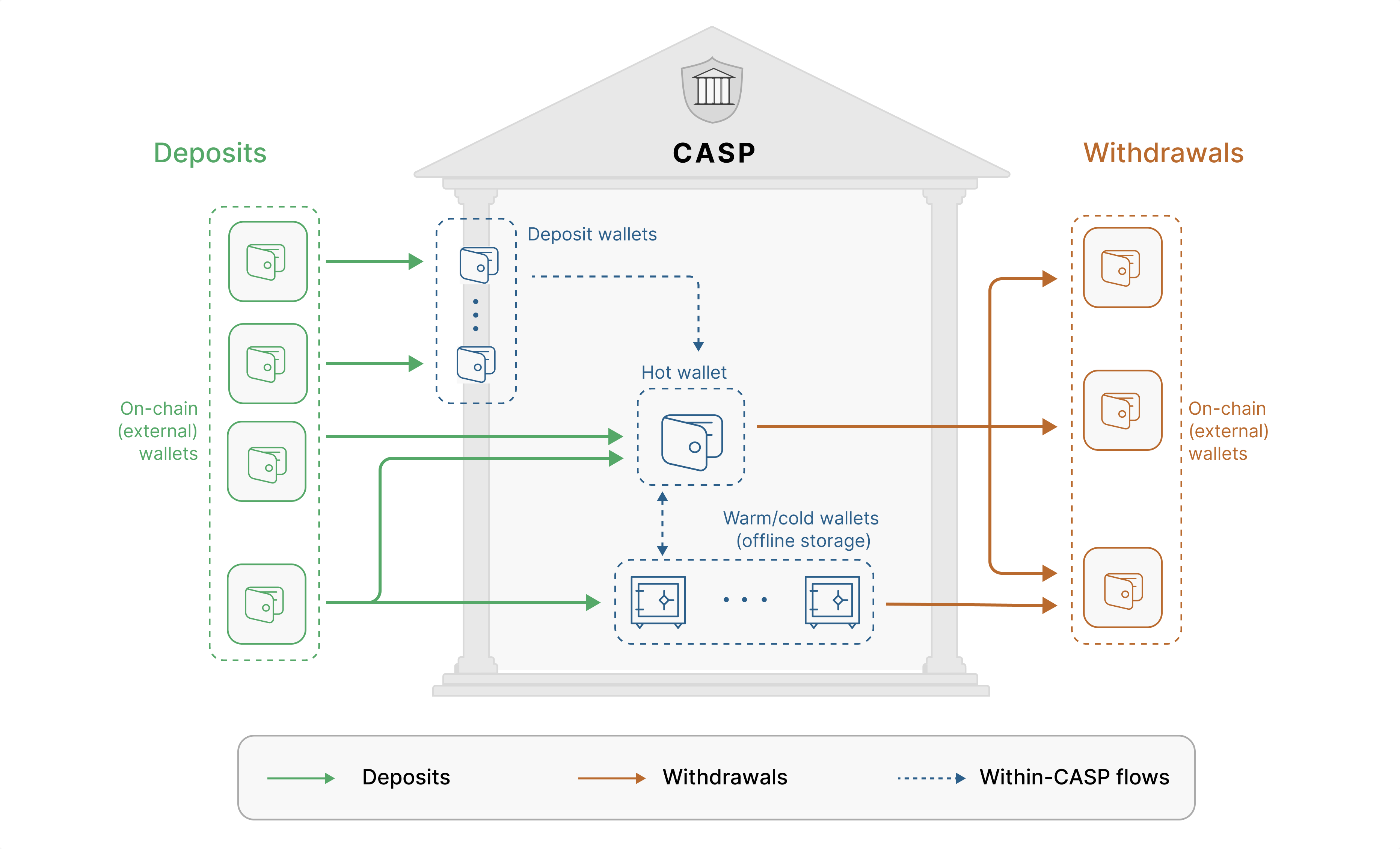}
	\caption{\textbf{CASP on-chain wallet architecture.}
		CASPs use hot wallets for daily deposit and withdrawal operations, and warm and cold wallets for secure storage and large-value transfers. Typical custody patterns are: (i)~users interact with one main (hot) wallet, either directly or through so-called deposit (hot) wallets; this wallet collects funds, manages daily operations, and occasionally interacts with internal secure wallets for rebalancing; (ii)~a few large transactions occur directly with CASP's warm or cold wallets.}	
	\label{fig:within}
\end{figure}

Crypto-Asset Service Providers (CASPs) are defined under the European Union's Markets in Crypto-Assets (MiCA) Regulation as entities that provide one or more cryptoasset services such as custody, order execution, trading, portfolio management, and advisory services, on a professional basis~\citep{eu2023mica}. They operate as centralized intermediaries connecting the cryptoasset and traditional financial systems, and include crypto exchanges, custodial services, and payment providers.

A CASP's wallet architecture refers to the set of on-chain addresses or wallets\footnote{In the following, we use the terms \emph{wallet} and \emph{address} synonymously.} and the key management scheme through which the provider custodies client and operational funds. CASPs operating a custodial model record customer balances in an internal, off-chain ledger, while funds are held on-chain in a smaller set of provider-controlled wallets. They typically control multiple types of wallets serving different functions: \textit{hot wallets} are connected to the internet for client deposits and withdrawals, while \textit{warm} and \textit{cold wallets} follow higher security standards and are employed for offline storage and large-value transactions. Figure~\ref{fig:within} shows a stylized representation of a CASP wallet architecture.

CASPs are linked to institutions in the traditional financial system (e.g., banks) through fiat accounts. This allows customers to transfer funds denominated in a fiat currency (e.g., EUR, USD) from their bank accounts to a CASP and vice versa. Such transactions are not publicly visible and are recorded only in the CASP's internal ledger.

Transfers to other entities in the cryptoasset ecosystem are executed on-chain and are therefore publicly visible asset transfers between addresses. Consequently, and this is the focus of this paper, it is in principle possible to measure on-chain cryptoasset flows \emph{to} and \emph{from} CASPs.

A major challenge lies in linking on-chain addresses to the CASP controlling the underlying private keys. This deanonymization task, commonly known as \emph{attribution}, currently relies mainly on so-called clustering heuristics~\citep{meiklejohn2013fistful}, which, by definition, produce false positives and false negatives~\citep{Froewis:2020a}. Blockchain transactions also carry no geographic information, so localizing on-chain financial activity directly is hardly possible~\citep{auer2025crossborder}. As a consequence, the role of CASPs as intermediaries integrating digital assets into domestic financial systems, and the implications for financial stability, remain understudied.

Since only the CASPs themselves know the addresses they control, this information is best retrieved directly from them. Where a responsible authority within a jurisdiction can compel its disclosure, at least partial geographic bindings become possible. This is the approach we are pursuing in this work.

\subsection{Stablecoins as Private Money: Reserves, Redemption, and Safe-Haven Dynamics}
\label{sec:stablecoins_backg}

Stablecoins play a pivotal role in decentralized finance and crypto markets as an alternative, stable asset class~\citep{auer2024technology}. By anchoring their value through reserves or algorithms, they offer a solution to the wide fluctuations that characterize cryptoassets like Bitcoin and Ethereum~\citep{catalini2022some}. A large body of work examined their design choices, price stability, and impact on traditional and crypto markets, summarized in two recent reviews~\citep{dionysopoulos202410,mahrous2026stablecoins}.

Fiat-backed stablecoins like USDT and USDC
are issued by private firms (e.g., Tether, Circle) and maintain their peg by promising a $1{:}1$ par-value redemption and full collateralization with highly liquid fiat reserves.
This promise does not give every holder direct access to issuer redemption. 
In practice, direct redemption is restricted to only a small set of vetted institutional counterparties, subject to KYC/AML screening, minimum thresholds, fees, and jurisdictional limits. Retail holders can interact only indirectly through secondary markets, where the price is set by supply and demand~\citep{ashwanth2025redemptions,ma2025stablecoin}. This produces a two-tiered structure in which the primary-market peg can hold for institutional actors even as the secondary-market peg can break for other participants, and the primary peg holds only as long as the issuer remains solvent and able to convert reserves into fiat quickly.

Stablecoins thus inherit the fragility of the money-like claims studied in the literature on runs in banking and shadow banking.
The safety of a private money-like claim rests on its reserves and the issuer incentives: \citet{kacperczyk2013safe} show that money market funds (MMFs), long regarded as
nearly as safe as cash, have strong incentives to take on risk and are thus vulnerable to runs, and \citet{covitz2013evolution}
show that the probability and severity of a run is affected by asset-level features like sponsor support and asset quality.
Fragility also depends on a claim's redemption terms. In mutual funds, redemptions impose costs on the investors who remain, so outflows respond more strongly to bad performance when fund assets are illiquid~\citep{chen2010payoff}; in corporate bond funds, the mismatch between redeemable shares and illiquid assets likewise creates a first-mover advantage~\citep{goldstein2017investor}.
Redemption dynamics depend on investor composition, too. ~\cite{schmidt2016runs} show that institutional and more sophisticated investors withdrew from MMFs more heavily than retail investors holding the same portfolio.
Stablecoins add two features to this picture.
The first is unequal redemption access, which ties redemption terms to investor type: institutions can redeem at par with the issuer, whereas retail holders must sell on secondary markets. The second is layered risk: a stablecoin held at a CASP carries the risk of its issuer and that of the custodian holding it, so reserve quality, redemption rights, and custodial arrangements jointly determine how stress is transmitted; the question is not just whether stablecoins are safe, but for whom.

Stablecoins are thus exposed to known risks, such as the classic run risk on private money if confidence in their backing falters; moreover, they represent a channel for the transmission of shocks across the crypto and traditional finance systems~\citep{briola2023anatomy,griffin2020bitcoin}. Their fragility bears directly on a debated policy question: do stablecoins serve as a safe haven for investors during periods of market stress and extreme price movements? Aggregate flows suggest a safe-haven role relative to Bitcoin and occasional flight-to-quality behavior similar to MMFs~\citep{baur2021crypto,anadu2024runs,oefele2024flight}, but more recent evidence reports heterogeneous and inconsistent dynamics across events~\citep{aldasoro2025stablecoins,campello2025demand}, or contrasting evidence~\citep[see e.g.][]{wang2020stablecoins,kolodziejczyk2023stablecoins,feng2024stablecoins}.
Existing research thus remains mixed and inconclusive.

\subsection{Identification of Retail Activity}
\label{sec:ident_litrev}

Because blockchain activity is pseudonymous and a single entity can control multiple addresses, investor identity cannot be directly observed~\citep{moser2013inquiry}. Two strands of literature have developed standard methods for identifying retail trades under similar conditions in traditional finance. In municipal and corporate bond markets, the convention is a trade-size threshold: trades below \$100{,}000 par are treated as retail and trades above \$1 million as institutional~\citep{schultz2012market,oharazhou2021,bessembinder2020,dehaan2023retail}. In U.S. equities, \citet{boehmer2021tracking} flag retail orders by the sub-penny price improvements that wholesalers apply when internalizing retail order flow, though \citet{barber2024subpenny} document non-trivial misclassification in validation tests. Both reduce investor type to a single observable proxy, but this approach is fragile: using a structural model with latent investor type on municipal trades from 2005 to 2023, \citet{hund2025rise} find that the \$100{,}000 cutoff separates informed from uninformed trades poorly, with roughly half of uninformed (retail-like) trades occurring above the threshold and sophisticated investors routinely placing retail-sized orders.

Comparable strategies have been applied to cryptoasset markets. \citet{divakaruni2024uncovering} exploit trade-size thresholds, reflecting the assumption that retailers place smaller trades than professional investors; \citet{baur2019bitcoin} infer investor type from time-dependent trading patterns; and \citet{john2025bitcoin} proxy retail activity via aggregate user counts on trading platforms such as RobinHood. Other work relies on brokerage account data~\citep{hackethal2022characteristics} or survey-based behavioral measures~\citep{almeida2023systematic,fessler2024oenb}, neither of which can be matched to on-chain transactions. Across domains, investor type is recovered through indirect proxies rather than observed directly.

Our setting shares the central constraint of this literature. While the CASP-controlled addresses in our data are attributed through disclosure rather than heuristics, the individual counterparties transacting with them remain pseudonymous and carry no strong identity information. Investor type therefore cannot be observed directly and must be inferred from observable transaction characteristics. We retain the proxy-based approach but combine two signals, full-chain transaction counts and interaction with warm or cold CASP wallets, to distinguish retail-like from institutionally mediated activity.

\section{Data}
\label{sec:data}

\subsection{Address Reporting and CASP Identification}

As motivated in Section~\ref{sub:casp_backg}, the set of addresses a CASP controls is reliably known only to the provider itself and is best obtained directly, wherever an authority can compel its disclosure. Austria's recent reporting requirements establish such an obligation: crypto-asset service providers must report the blockchain addresses under their control to the competent financial authority~\citep[Austrian Financial Market Authority or][]{fma2025aml}. Our analysis builds on the resulting registry of address-to-entity mappings, comprising pseudonymized Bitcoin and Ethereum addresses reported by the 12 Austrian CASPs registered with the FMA at the end of 2024.

The registry is pseudonymized to protect the privacy of both reporting entities and individual address holders: each service is represented solely by the set of addresses it controls, with no disclosure of user identities or other sensitive information. Since these mappings originate from a mandatory regulatory report instead of voluntary disclosures or clustering heuristics, they avoid the false positives and false negatives inherent to heuristic attribution and can be treated as highly trustworthy data for the population of registered Austrian providers.\footnote{We note that reporting obligations bind on the CASP, not on the underlying blockchain, so the registry can omit addresses a provider controls but fails to disclose: the registry is thus only as complete as the providers' own disclosures. In the course of the analysis we identified and reported a small number of addresses likely controlled by Austrian CASPs that were absent from the original mapping (see Section~\ref{sec:limits} for further discussion).} We therefore do not apply any address clustering to expand the registry. To the best of our knowledge, this is the first study to draw on a dataset of this kind.

Further details on the registry, including the share of addresses by CASP and asset type, are provided in Appendix~\ref{app:registry}. The exact number of addresses is withheld for confidentiality reasons, as agreed with the Austrian financial authorities.\footnote{Aggregate data are available on request, subject to the Authority's approval. The code is available on request and will be released upon publication (see the Data availability statement).
}

\subsection{Asset Flow Extraction}

From this registry, we reconstruct the on-chain activity of the reported wallets. We parse the full Bitcoin and Ethereum ledgers and extract every transaction directed to or originating from a registered address, up to a cutoff date of 31 May 2025. We restrict attention to four major cryptoassets: Bitcoin (BTC) and Ether (ETH), the two largest by market capitalization, and Tether (USDT) and USD Coin (USDC), the two largest fiat-backed stablecoins by market capitalization. This yields \num{11980687} asset transfer records, corresponding to \num{3214643} unique transactions. To express values in a common currency, we convert each transfer using daily USD prices retrieved from the \citeauthor{coingecko2026api} API. The total volume routed through Austrian CASP wallets exceeds \$50.4 billion, distributed across assets as 65.53\% BTC, 15.90\% ETH, 15.47\% USDT, and 3.10\% USDC. 

We note that our unit of observation is on-chain transfers to and from CASP-controlled addresses. We do not observe customers' off-chain ledger balances, the trades executed on a CASP's platform, or the ultimate disposition of withdrawn funds. Deposits and withdrawals are therefore best read as changes in on-chain custodial exposure rather than as direct evidence of off-chain activity or final asset custody. Where we describe a flow pattern as consistent with selling, redemption, or self-custody, we mean that the observed on-chain movement matches what such behavior would produce, not that we directly observe the underlying trade, redemption, or custody outcome.

\subsection{Event Selection}
\label{sec:event_sel}

Crypto investors tend to respond to shocks they can readily interpret, often through media intermediaries~\citep{philippas2019media,FeldkircherMakridis:words}.
We thus focus on crypto-related incidents, hacks, failures, and bankruptcies, since such shocks lend themselves to simple narratives that are easy to grasp and communicate, even when the underlying mechanics are complex.
Among these, we select those that attracted wide public attention.

\begin{table}
	\small
	\centering
	\begin{tabularx}{1\textwidth}{@{}lX@{}}
	\toprule
	\multicolumn{2}{@{}l}{\textbf{Terra (UST) and Luna (LUNA) collapse}}  \\
	\midrule
	2022-05-05 & Selling pressure on LUNA aimed at destabilising markets \\
	2022-05-07 & UST loses its peg to the dollar for the first time \\
	2022-05-09 & UST loses its peg to the dollar for the second and final time \\
	2022-05-11 & Final attempt by Terra governance to restore the peg \\

	\midrule
	\multicolumn{2}{@{}l}{\textbf{FTX bankruptcy}}  \\
	\midrule
	2022-11-02 & Media report that Alameda, FTX's sister firm, is heavily exposed to the FTX token FTT  \\
	2022-11-06 & Binance announces sale of its FTT holdings \\
	2022-11-08 & FTT price falls below the critical \$22 level \\
	2022-11-09 & Binance withdraws from tentative deal to acquire FTX \\
	2022-11-10 & Regulators freeze FTX Digital Markets assets \\
	2022-11-11 & FTX and Alameda file for bankruptcy \\

	\midrule
	\multicolumn{2}{@{}l}{\textbf{Silicon Valley Bank failure}}  \\
	\midrule
	2023-03-08 & SVB announces a \$1.8\,bn loss \\
	2023-03-10 & SVB fails; FDIC takes control \\
	2023-03-11 & Circle discloses \$3.3\,bn exposure to SVB \\
	2023-03-12 & Treasury, Federal Reserve, and FDIC announce full depositor protection \\
	\bottomrule
\end{tabularx}
	\caption{\textbf{Timeline of the three selected events:} the Terra Luna collapse (May 2022), the FTX bankruptcy (November 2022), and the Silicon Valley Bank failure (March 2023).}
	\label{tab:sel_events_timeline}
\end{table}

Starting from a published repository of \num{1141} crypto crime events~\citep{carpentier2025mapping}, supplemented with data from \citeauthor{DeFiLlama_hacks} and \citeauthor{coinmarketcap2025events}, we select 19 events that (i) occurred between 2019 and 2025, when all four assets were actively traded; (ii) caused estimated losses exceeding \$100 million; and (iii) affected crypto exchanges (CEXs), DeFi, or centralized finance (CeFi) entities. Using Google Trends search volumes as a proxy for public attention, queried with event-specific keywords across two geographical areas, Austria and worldwide, three events stand out consistently: the May 2022 collapse of the algorithmic stablecoin Terra and its Luna token; the November 2022 solvency crisis of the crypto exchange FTX; and the March 2023 run on Silicon Valley Bank (SVB), a lender closely tied to the crypto sector. Table~\ref{tab:sel_events_timeline} provides a detailed timeline drawn from prior studies~\citep{briola2023anatomy,liu2023anatomy,conlon2023collapse,akyildirim2023understanding,galati2024silicon,metrick2024failure}. Further details on the selection procedure are reported in Appendix~\ref{app:registry}.

\section{Cryptoasset Flow Analysis}
\label{sec:flows}

This section characterizes the on-chain activity intermediated by Austrian CASPs, from aggregate flows down to individual counterparties. We first decompose total volume into three components, separating market-facing activity from inter-CASP and internal flows (Section~\ref{sub:volume_flow_decomposition}). We then examine transaction-level structure, identifying the recurring patterns through which CASPs route customer funds and the wallet tiers they use (Section~\ref{sub:txn_level}). Finally, we classify the external counterparties that interact with Austrian CASPs into retail-like and institutional, using a proxy-based heuristic that combines wallet-tier interaction with overall on-chain activity (Section~\ref{sub:retail_identification}). These steps characterize counterparties' observed activity, transfer volumes, and custody structures through which they interact with the CASPs.

\subsection{Decomposing Flows by Counterparty}
\label{sub:volume_flow_decomposition}

We decompose transaction volumes into three components according to the counterparties involved. \textit{CASP-to-market} activity comprises all transactions between an Austrian CASP and an external (non-Austrian-CASP) actor. \textit{CASP-to-CASP} activity isolates transactions in which the sender address belongs to one Austrian CASP and the recipient to another Austrian CASP. \textit{Within-CASP} activity captures transactions in which sender and recipient are controlled by the same Austrian CASP.

\paragraph{CASP-to-Market}
\label{sec:casp-to-mkt}

Separating CASP-to-Market flows from internal bookkeeping isolates the economic activity that Austrian CASPs conduct with the wider market, which amounts to roughly \$30 billion in total. 

\begin{figure}[t]
	\centering
	\includegraphics[width=\columnwidth]{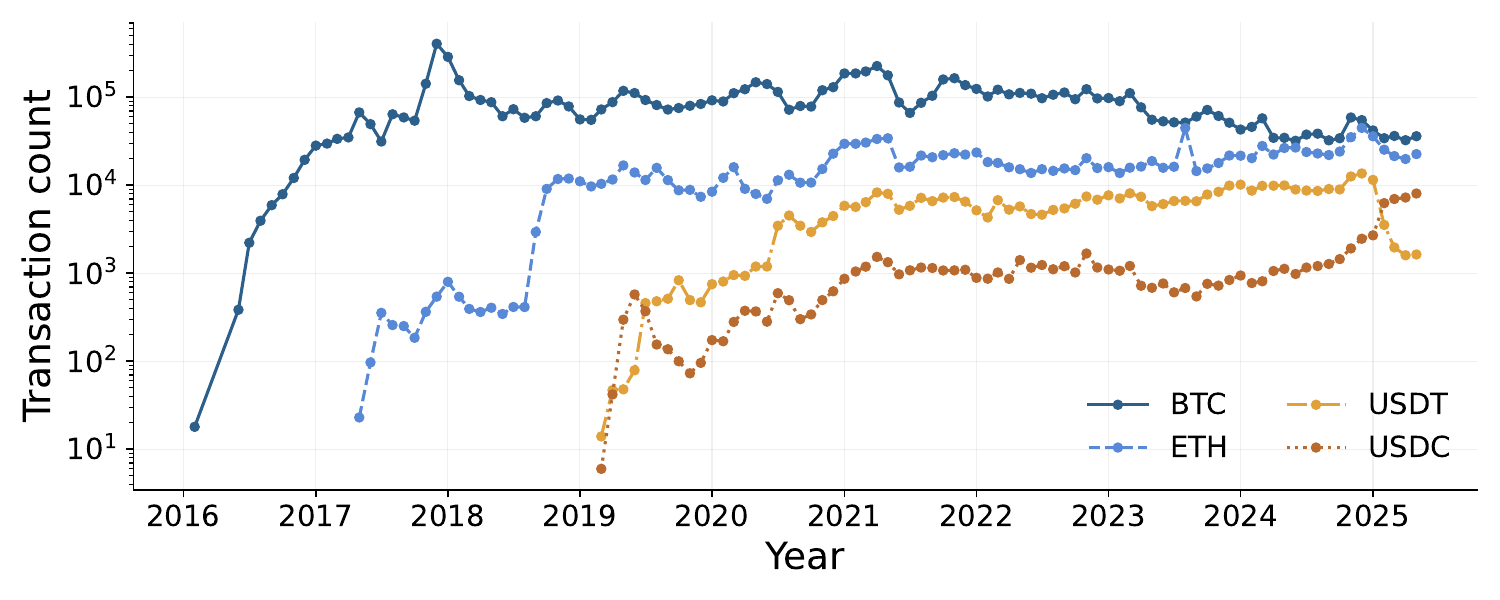}
	\caption{\textbf{Number of on-chain asset transfers by month and asset.} Bitcoin plays a predominant role; stablecoins first appear in CASP activity in 2019. Activity on the Ethereum chain grew steadily from 2017 on.}
	\label{fig:timeline}
\end{figure}

\begin{figure}[h]
	\centering
	\begin{subfigure}[b]{0.49\textwidth}
		\centering
		\includegraphics[width=\columnwidth]{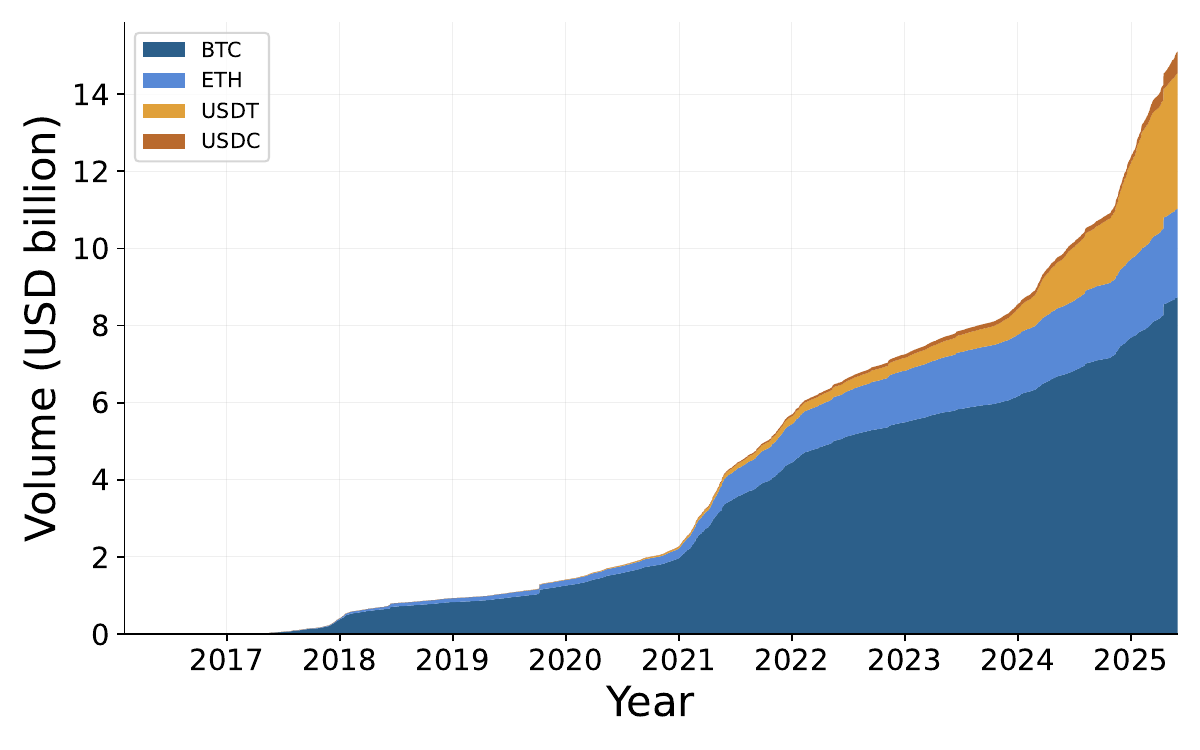}
		\caption{}
		\label{fig:cumsum_dep}
	\end{subfigure}
	\begin{subfigure}[b]{0.49\textwidth}
		\centering
		\includegraphics[width=\columnwidth]{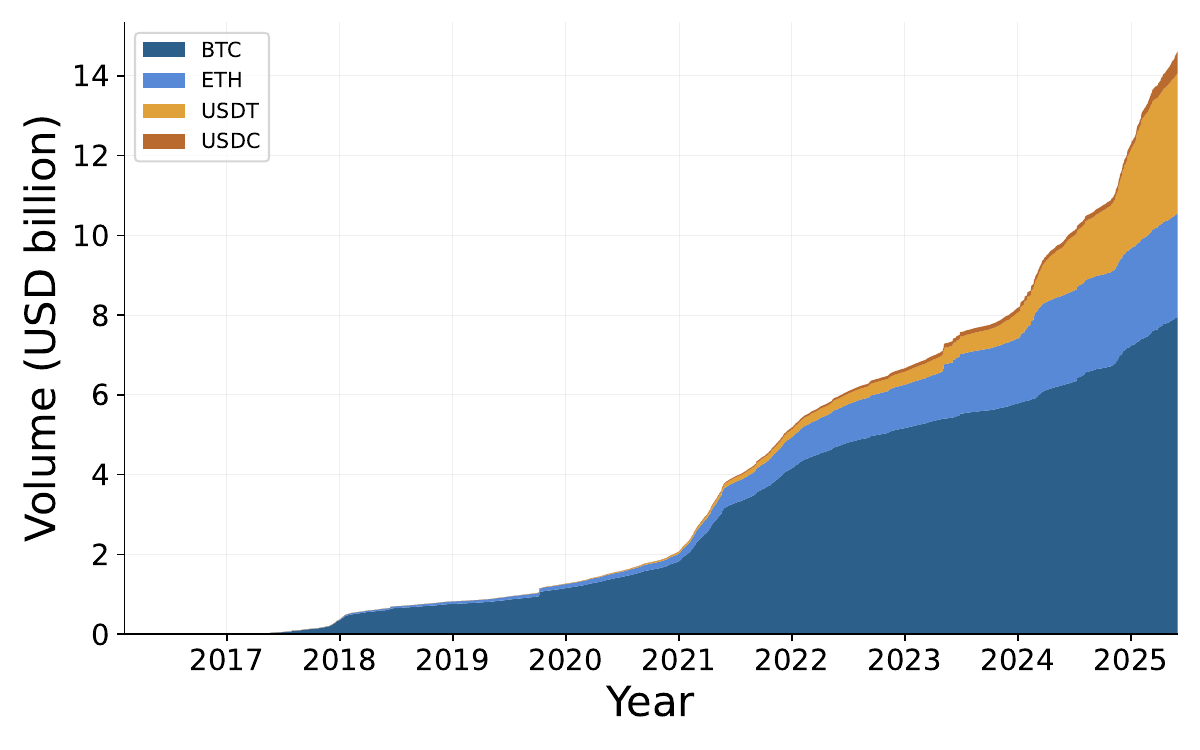}
		\caption{}
		\label{fig:cumsum_wit}
	\end{subfigure}
	\caption{\textbf{Cumulative deposits and withdrawals over time by asset.} Panels (\subref{fig:cumsum_dep}) and (\subref{fig:cumsum_wit}) respectively show the cumulative inflows directed to known CASP addresses and outflows for BTC, ETH, USDT, and USDC. Values are denominated in US dollars at historical prices.} 
	\label{fig:cumsums}
\end{figure}

Austrian CASPs operate primarily in Bitcoin and Ether. 
Figure~\ref{fig:timeline} shows the number of asset transfers over time, broken down by asset and month. Bitcoin has played a major role since 2016, while Ether and stablecoins first appeared in CASP activity in 2017 and 2019, respectively. Activity peaked around 2021 and the Bitcoin share fell relative to other assets thereafter, whereas stablecoin activity grew over the full period.
Figure~\ref{fig:cumsums} presents cumulative inflows and outflows by asset, respectively in the left and right panels, and shows that total inflows amount to \$15.1 billion, while the total outflows to \$14.6 billion.

CASP counterparties are highly heterogeneous: both transaction sizes and activity levels follow heavy-tailed distributions. Transaction values (aggregated by unique transaction and separated by asset) are strongly right-skewed, with median values of only a few hundred dollars (\$206 for ETH to \$544 for BTC) and 95th percentiles from roughly \$5{,}000 (ETH) to \$21{,}000 (BTC). Aggregate volumes are driven by a few very large transactions: the largest Bitcoin transaction (\$287.5M) exceeds the largest transaction in any other asset (\$13.5M--\$20.8M) by an order of magnitude. Activity is similarly concentrated: most addresses transact fewer than ten times, with the corresponding bin holding 95--98\% of addresses across all assets (see Table~\ref{tab:num_tx_addrs}); only a small fraction are highly active, a pattern that holds across assets and both blockchains.

\paragraph{CASP-to-CASP}
\label{sec:casp-to-casp}

CASP-to-CASP activity captures volume flows between distinct Austrian CASPs, identified as transactions whose sender and recipient addresses are associated with two different Austrian CASPs. We find that such inter-CASP activity is limited in both frequency and volume: just \num{3040} transfers, totaling \$5.774 million. Most CASP-to-CASP flows are in Bitcoin, though its share has declined over time; transaction counts peaked in 2021, while transferred value was highest in 2022 and 2023.

These magnitudes are economically negligible relative to the \$30 billion flows with counterparties outside the registry, indicating that financial integration among Austrian CASPs is marginal and that the network is instead integrated primarily with the global market. This is consistent with the relatively small size of the Austrian market. Appendix~\ref{app:casp_network} reports CASP-to-CASP activity broken down by asset and year, together with quarterly CASP-to-CASP interaction networks.

\paragraph{Within-CASP}
\label{sec:within-casp}

Within-CASP activity captures asset flows between a CASP's own hot, warm, and cold wallets. We identify \num{798196} such interactions, which account for only 6.7\% of all asset transfers but $\sim$\$20.7 billion, or 41.1\% of total volume. Internal bookkeeping thus involves comparatively few transfers, each moving large amounts on average. Makarov and Schoar~(\citeyear{makarov2021blockchain}) report that within-entity transfers make up around 90\% of global Bitcoin on-chain volume. The two figures are not directly comparable,\footnote{\citet{makarov2021blockchain} consider a broader set of global entities, including the largest centralized exchanges, focus exclusively on Bitcoin, and aggregate within-entity volumes together with peeling chains and change-of-address transactions.} but both indicate that within-entity activity accounts for a substantial share of total on-chain volume.

\begin{table}[h]
	\small
	\centering
	\begin{tabular*}{\textwidth}{@{\extracolsep{\fill}}lrrrrrr}
	\toprule
	& \multicolumn{6}{c}{Number of transactions} \\
	\cmidrule(lr){2-7}
	Asset & $[1,10)$ & $[10,10^2)$ & $[10^2,10^3)$ & $[10^3,10^4)$ & $[10^4,10^5)$ & $\geq 10^5$ \\
	\midrule
	BTC  & \num{1259786} & \num{57560} & \num{2833} & 398 & 64 & 7 \\
	ETH  & \num{390036}  & \num{13752} & 598        & 40  & 8  & 0 \\
	USDT & \num{134123}  & \num{4135}  & 140        & 24  & 0  & 0 \\
	USDC & \num{33709}   & 951         & 38         & 1   & 0  & 0 \\
	\bottomrule
\end{tabular*}
	\caption{\textbf{Distribution of non-CASP addresses by transaction activity.} For each asset, addresses are grouped into bins by their total number of transactions, from $[1,10)$ up to $\geq 10^5$. Low-activity addresses dominate.}
	\label{tab:num_tx_addrs}
\end{table}

\clearpage

\subsection{Transaction-Level Analysis}
\label{sub:txn_level}

Having examined aggregate flows, we now turn to the transaction level to identify how CASPs internally manage customer activity. Prior work shows that CASPs may use deposit addresses to credit customer funds to the correct account before forwarding them to a main exchange wallet under the same CASP's control~\citep{victor2020address}. Building on this, we develop tracing algorithms, adapted to heterogeneity across CASPs and blockchains, that proceed in two steps: we first construct network-based indicators (degree, betweenness, PageRank, eigenvector centrality) to detect hub wallets within a CASP, then match flows by transaction identifier, asset, and transferred value to trace chained movements across same-CASP addresses that terminate at a hub wallet. We restrict the analysis to CASPs with sufficient within-CASP activity, excluding those controlling few on-chain addresses. This sample covers more than 94\% of the Austrian BTC market, 91\% ETH, 98\% USDC, and 99\% USDT.

We identify two patterns in transaction-level flows. \emph{Pattern~1}, followed by 89\% of analyzed transactions, routes funds either to a deposit hot wallet and immediately on to the main hot wallet, or directly to the main wallet; main wallets occasionally interact with the same CASP's warm or cold wallets for rebalancing. This resembles the ``fan-in'' transactions that merge funds from many wallets into one~\citep{baqer2016stressing}.
\emph{Pattern~2} is a distinct mode in which a restricted set of market-controlled addresses interacts directly with the warm or cold wallets of Austrian CASPs. It represents only 0.75\% of asset records yet accounts for over 55.8\% of volume directed from external addresses to Austrian CASPs.\footnote{\emph{Pattern~1} and \emph{Pattern~2} do not cover the full sample; the remaining $\sim$10\% could not be classified into a specific pattern. We also note that while addresses in the second group can interact with CASPs through the first identified pattern, we do not observe the opposite.}

Table~\ref{tab:wallet_types} reports statistics for the wallet-tier interaction on the full sample of CASP-to-Market interactions: for every asset, the large majority of interactions target hot wallets (99.23\% overall), yet those reaching warm or cold wallets account for more than 57.8\% of total volume. 
Taken together, these findings suggest that interactions are concentrated among a small set of actors and are associated with high-value or operational flows, whereas routine activity is mediated through hub wallets. Appendix~\ref{app:casp_network} describes the tracing algorithms, shows representative within-CASP networks and per-asset breakdowns, and documents minor additional patterns among Bitcoin-operating CASPs.

\begin{table}
	\small
	\centering
	\begin{tabular*}{\textwidth}{ll@{\extracolsep{\fill}}rrr}
	\toprule
	Asset & Wallet type & Transfers & Volume (\$M) & \% of volume \\
	\midrule
	\multirow[t]{2}{*}{BTC}  & hot       & \num{9149248} & \num{8796}  & 52.7 \\
	                         & warm/cold & \num{39844}   & \num{7887}  & 47.3 \\
	\addlinespace
	\multirow[t]{2}{*}{ETH}  & hot       & \num{1464273} & \num{2612}  & 53.3 \\
	                         & warm/cold & \num{31720}   & \num{2284}  & 46.7 \\
	\addlinespace
	\multirow[t]{2}{*}{USDC} & hot       & \num{87857}   & 263         & 23.8 \\
	                         & warm/cold & \num{1866}    & 841         & 76.2 \\
	\addlinespace
	\multirow[t]{2}{*}{USDT} & hot       & \num{395336}  & 872         & 12.4 \\
	                         & warm/cold & \num{12347}   & \num{6157}  & 87.6 \\
	\midrule
	\multirow[t]{2}{*}{Total}& hot       & \num{11096714}& \num{12543} & 42.2 \\
	                         & warm/cold & \num{85777}   & \num{17170} & 57.8 \\
	\bottomrule
\end{tabular*}
	\caption{\textbf{CASP-to-Market interactions by wallet tier and asset.} Most asset transfers involve hot wallets, regardless of chain and asset; the few interactions with warm or cold wallets account for a large share of total volume.}
	\label{tab:wallet_types}
\end{table}

\subsection{Separating Retail-Like from Institutional Activity}
\label{sub:retail_identification}

The preceding analysis showed that activity across external addresses is highly skewed: a few addresses account for most volume and transactions, while the remainder interact with Austrian CASPs only a handful of times, typically for a few hundred to a few thousand dollars. We ask whether the body of this distribution corresponds to retail-like on-chain activity and the head to institutional counterparties such as other CASPs, market makers, custodians, and OTC desks.

Standard methods identify retail trades through a single observable proxy (Section~\ref{sec:ident_litrev}); such a classifier misassigns addresses that resemble retail on that one margin. We therefore combine two signals. The first is wallet-tier interaction: warm/cold flows originate almost exclusively from large counterparties (0.77\% of external transfers, 57.8\% of external volume), so an address interacting with a warm or cold wallet is unlikely to be retail. The second is overall on-chain activity, measured by transaction count rather than dollar volume to remove denomination effects~\citep{makarov2021blockchain,auer2025crossborder}; retail-like addresses are assumed to fall in the body of the count distribution among CASP counterparties, whereas exchanges, market makers, and other CASPs occupy the upper tail.

\subsubsection{Definitions and Assumptions}

Our classification approach is proxy-based and heuristic. We do not observe directly whether a non-Austrian-CASP address belongs to a retail or an institutional entity;
instead, we classify addresses by two observable proxies for institutional behavior. The signals are proxies for an unobserved actor type, and the thresholds that turn them into binary labels are conventional rather than theoretically derived. The definitions below should therefore be read as operational rules over these proxies, not as exact identification of actor types.

Let $A_c$ denote the set of external (non-Austrian-CASP) addresses\footnote{
The CASP addresses in our registry require no clustering; the external counterparties, however, are not in the registry and thus require clustering. Consistent with existing approaches~\citep{makarov2021blockchain,capponi2026price,cong2025centralized}, we apply standard multi-input clustering heuristic to Bitcoin counterparties,
while we treat each Ethereum counterparty address as a persistent entity under the account-based model and do not cluster. So each $i \in A_{\text{BTC}}$ is a clustered set of addresses rather than a single address, and the resulting clusters are used to compute network statistics such as transaction count and degree.
For address-level blockchain information,
we use the same observation window used for the main analysis.
} that interact with at least one Austrian CASP wallet on chain $c \in \{\text{BTC},\text{ETH}\}$. 
For each element $i \in A_c$, we define two binary signals.
The first, $\text{cold}_i \in \{0,1\}$, equals $1$ if $i$ ever transacts with a warm or cold Austrian CASP wallet, and $0$ otherwise. The second, $\text{active}_i \in \{0,1\}$, flags whether $i$ ranks among the most active CASP counterparty addresses on chain $c$. Let $n_i$ be the total number of transactions of $i$ on chain $c$, and define its log-transformed count $\ell_i = \log(1 + n_i)$. Because transaction counts are heavy-tailed, spanning several orders of magnitude, we measure activity on the log scale, where the bulk of the distribution is approximately symmetric and a dispersion-based threshold is meaningful. 
With
\[
\mu_c = \frac{1}{|A_c|}\sum_{j \in A_c} \ell_j,
\qquad
\sigma_c = \sqrt{\frac{1}{|A_c|-1}\sum_{j \in A_c} \left(\ell_j - \mu_c\right)^2},
\]
the mean and standard deviation of log activity across $A_c$, we compute the z-score and set
\[
\text{active}_i =
\begin{cases}
	1 & \text{if } z_i = \dfrac{\ell_i - \mu_c}{\sigma_c} > 3,\\[6pt]
	0 & \text{otherwise,}
\end{cases}
\]
so that $i$ is active when its log transaction count exceeds the mean among CASP counterparty addresses on the same chain by more than three standard deviations. The threshold of three is the conventional outlier cutoff; we assess sensitivity to it in the robustness checks in Appendix~\ref{app:robustness}. We thus define:

\begin{definition}[Retail-like (RL)]
	An external address $i$ is \emph{retail-like} if $\text{cold}_i = 0$ and $\text{active}_i = 0$. Such an address never interacts with a warm or cold wallet and is not among the most active on its chain, the on-chain footprint expected of individual retail activity.
\end{definition}

\begin{definition}[High-confidence institutional (HI)]
	An external address $i$ is \emph{high-confidence institutional} if $\text{cold}_i = 1$ and $\text{active}_i = 1$. Both signals are present: the address transacts at the warm/cold tier and ranks among the most active on its chain, a combination characteristic of exchanges, market makers, or other large intermediaries.
\end{definition}

\begin{definition}[Low-confidence institutional (LI)]
	An external address $i$ is \emph{low-confidence institutional} if $\text{cold}_i + \text{active}_i = 1$. We partition this group into $\text{LI}_{c}$ ($\text{cold}_i = 1$, $\text{active}_i = 0$) and $\text{LI}_{a}$ ($\text{cold}_i = 0$, $\text{active}_i = 1$). Exactly one signal is present: the address shows a single institutional marker, either warm/cold contact ($\text{LI}_c$) or high activity ($\text{LI}_a$), but not both.
\end{definition}

Together, the HI, LI$_c$, and LI$_a$ groups constitute institutionally mediated activity. Below, ``institutional'' is shorthand for this proxy-defined group. The two signals are jointly informative because each recovers a way an institutional address can resemble retail on the other margin. An institutional counterparty with modest aggregate activity, such as an OTC desk or a custodian that transacts on-chain only when settling large flows, is not flagged by the activity signal but is recovered through warm/cold contact (LI$_c$). A highly active intermediary that never reaches the warm/cold tier, such as a hot-wallet-only aggregator or arbitrageur, is not flagged by the wallet-tier signal but is recovered through the activity threshold (LI$_a$). The two subgroups are therefore not residual cases but addresses that a single-signal classifier would have assigned to retail.
The classification rests on two assumptions:

\begin{assumption}[Wallet-tier segmentation]
	The warm/cold wallet tier of Austrian CASPs services high-value, low-frequency flows that originate predominantly from institutional rather than retail counterparties.
\end{assumption}

\begin{assumption}[Activity dispersion]
		Retail on-chain activity does not reach the upper tail of the activity distribution among CASP counterparty addresses on that chain; addresses in the upper tail are therefore not retail.
\end{assumption}

Assumption~1 is supported by the evidence above that few transfers reach warm or cold wallets, yet those transfers carry a disproportionate share of volume (0.77\% of external transfers, 57.8\% of external volume), a profile not compatible with retail use. Assumption~2 follows \citet{makarov2021blockchain}, who identify exchanges and large Bitcoin entities through centrality- and activity-based heuristics.

Three limitations apply and are further discussed in Section~\ref{sec:limits} and in the appendix. First, the rule misclassifies any retail address that interacts with a warm or cold wallet under atypical circumstances; such cases are bounded by the small set of bookkeeping transactions in which CASPs route funds through these wallets (Table~\ref{tab:seed_addresses}, Appendix~\ref{app:registry}), whose volume is economically negligible. Second, a single actor's activity may be fragmented across many addresses, which mechanically lowers per-address activity and biases classification toward retail rather than away from it; this bias operates in the same direction across all subsamples. Third, thresholds are not theoretically derived; as a robustness check we use alternative classifications that change both the activity threshold and the activity measure, obtaining broadly similar group profiles (see Appendix~\ref{app:robustness}).

\subsubsection{Classification Results}

\begin{table}
    \small
	\centering
    \begin{tabular*}{\textwidth}{@{\extracolsep{\fill}}llrrrrrr}
	\toprule
	&  & \multicolumn{1}{c}{} & \multicolumn{2}{c}{Avg.\ transactions} & \multicolumn{1}{c}{Avg.\ degree} & \multicolumn{1}{c}{Avg.\ USD} & \multicolumn{1}{c}{Hot wallet} \\
	\cmidrule(lr){4-5}
	Group & Chain & \multicolumn{1}{c}{Addresses} & \multicolumn{1}{c}{overall} & \multicolumn{1}{c}{in sample} & \multicolumn{1}{c}{overall} & \multicolumn{1}{c}{in sample} & \multicolumn{1}{c}{(fraction)} \\
	\midrule
	\multirow{2}{*}{RL}      & ETH & \num{534062}  & 29        & 2.6   & 11        & \num{1568}   & 1.00 \\
	                         & BTC & \num{1286555} & 10        & 2.9   & 15        & \num{1516}   & 1.00 \\
	\addlinespace
	\multirow{2}{*}{HI}      & ETH & 472           & \num{6114749} & 644   & \num{292149}  & \num{53539}  & 0.16 \\
	                         & BTC & 82            & \num{5044178} & \num{25425} & \num{2636324} & \num{80604}  & 0.53 \\
	\addlinespace
	\multirow{2}{*}{LI$_c$}& ETH & 552           & 88        & 4.8   & 25        & \num{72170}  & 0.03 \\
	                         & BTC & \num{2815}    & 3.7       & 1.8   & 46        & \num{378333} & 0.02 \\
	\addlinespace
	\multirow{2}{*}{LI$_a$}& ETH & \num{6957}    & \num{349525}  & 44    & \num{32886}   & \num{4458}   & 1.00 \\
	                         & BTC & \num{31196}   & \num{17479}   & 108   & \num{10749}   & 975          & 1.00 \\
	\bottomrule
\end{tabular*}
	\caption{\textbf{External addresses by group and chain.} Addresses are grouped into retail-like (RL), high-confidence institutional (HI), and low-confidence institutional (LI$_c$, LI$_a$), separately for each chain (Bitcoin, Ethereum).}
	\label{tab:users_by_group}
\end{table}

Table~\ref{tab:users_by_group} reports the four groups by chain. First, retail-like addresses dominate the count but not the value: they make up 98.6\% (Ethereum) and 97.4\% (Bitcoin) of external addresses, each transacting only a few times with Austrian CASPs, for roughly \$1{,}500 per transaction on average. The cross-chain symmetry is notable: despite the UTXO/account-based distinction between the two blockchains, the retail-like profile is the same.

Second, the HI tier is small and disproportionately large in value: 554 addresses across both chains, with overall transaction counts in the millions, network degree from the hundreds of thousands to the millions, and average per-transaction values between \$53{,}539 (ETH) and \$80{,}604 (BTC). A representative BTC-HI address transacts roughly \num{25000} times with Austrian CASPs but about 5 million times overall, a ratio consistent with exchanges, market makers, or large bridges for which Austrian CASPs are one node in a far larger network.

Third, the two LI subgroups capture distinct profiles. LI$_c$ addresses reach the warm/cold tier but are not highly active; they are large, infrequent counterparties consistent with OTC desks, custodians, or other CASPs operating at modest on-chain frequency, with average per-transaction values one to two orders of magnitude above retail-like addresses. LI$_a$ addresses are highly active but never reach the warm/cold tier; they are consistent with the lightly regulated exchange intermediaries that \citet{auer2022banking} term the ``shadow crypto financial system'', plausibly including aggregators, arbitrageurs, or hot-wallet-only services, with small per-transaction values but large aggregate activity. That the two subgroups are populated by economically distinct entities is consistent with the value of the bivariate scheme: a single-signal rule would collapse them into one residual category.

\begin{figure}
	\begin{subfigure}{0.49\textwidth}
		\includegraphics[width=\textwidth]{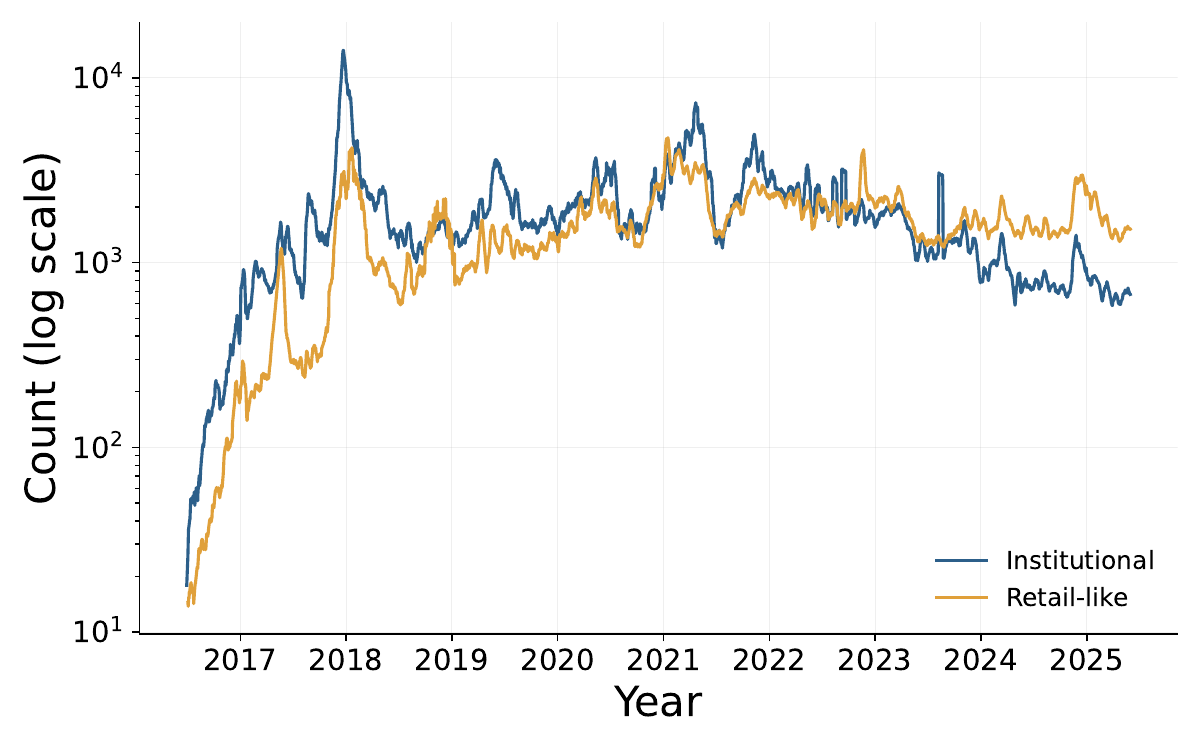}
		\caption{}
		\label{fig:group_count}
	\end{subfigure}
	\hfill
	\begin{subfigure}{0.49\textwidth}
		\includegraphics[width=\textwidth]{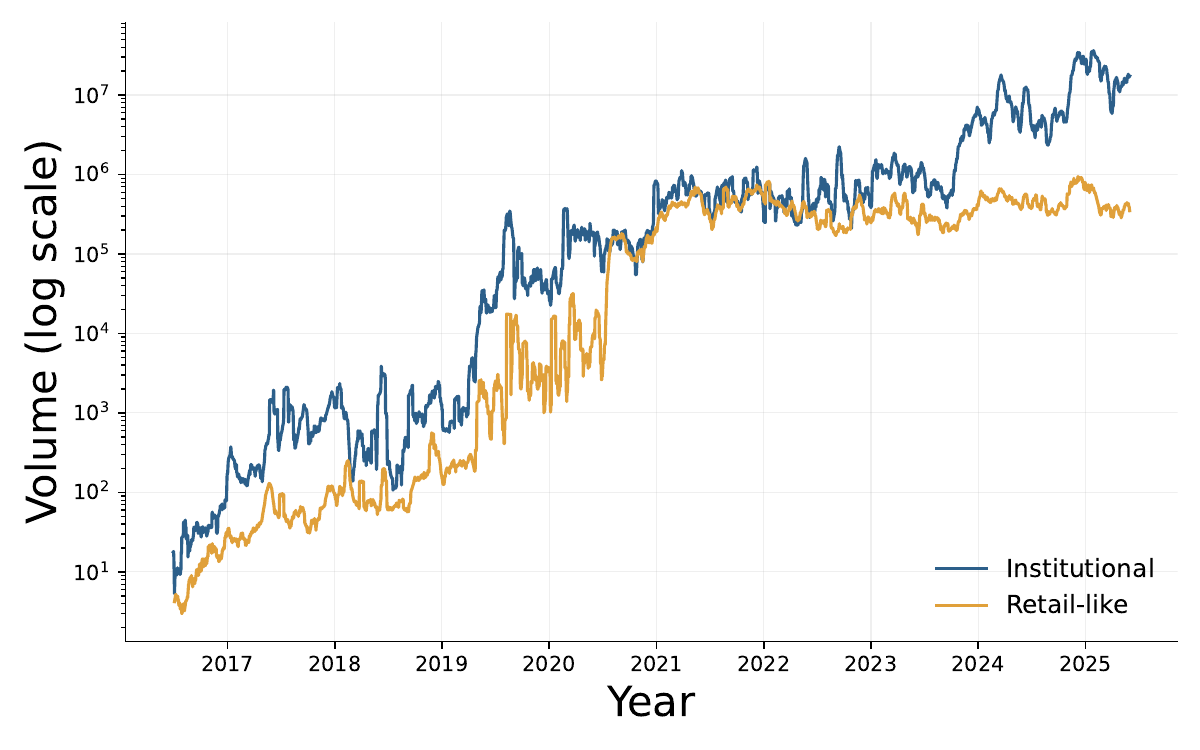}
		\caption{}
		\label{fig:group_vol}
	\end{subfigure}
	\caption{\textbf{Transaction count and volumes by group}. Retail-like activity is shown in orange, institutional is in blue. Panel (\subref{fig:group_count}): Transaction count. Panel (\subref{fig:group_vol}): Volume flows.}
	\label{fig:groups}
\end{figure}

Figure~\ref{fig:groups} aggregates the non-retail groups (HI\,+\,LI) and plots daily counts and volumes on a log scale for retail-like and institutional addresses. In Panel~(\subref{fig:group_count}), institutional activity leads in the early period (2016--2018), exceeding retail-like counts by up to an order of magnitude and peaking near $10^4$ daily transactions at the end of 2017. From roughly 2019 the two series track each other closely, settling near $10^3$ through 2022. After 2023 they separate again, but in the opposite order: retail-like counts hold near $1.5\times10^3$ while institutional counts decline to below $10^3$, so retail-like activity overtakes institutional on the count margin. This comovement over 2019--2022 is consistent with institutionally mediated flow partly intermediating retail demand, though it does not rule out autonomous institutional activity such as cross-venue settlement or rebalancing.

Panel~(\subref{fig:group_vol}) shows a different pattern for volume. The two series move together until 2021 and then diverge sharply from late 2023: institutional daily volume rises by roughly two orders of magnitude, reaching ${\sim}10^7$~USD by 2025, while retail-like volume remains near $10^5$--$10^6$~USD. Two observations follow. First, the marginal dollar of CASP-mediated activity in recent years is institutional rather than retail. Second, this divergence follows in time the shocks we study (Terra Luna, May 2022; FTX, November 2022; SVB, March 2023).
Institutional addresses dominate on volume while retail-like addresses dominate on count in the recent period, and the two margins can move independently.

\subsection{Summary}
\label{sub:flows_summary}

Three findings stand out. First, market-facing activity dominates: of the volume intermediated by Austrian CASPs, roughly \$30~billion is conducted with external actors, while inter-CASP flows are negligible (\$5.8~million) and the remainder is internal bookkeeping that moves few transfers but a large share of value. Second, both the size and the activity of CASP counterparties are heavy-tailed, and a structural divide separates them: warm and cold wallets account for only 0.77\% of external transfers yet 57.8\% of external volume, indicating that a small set of large counterparties handles most of the value. Third, classifying external addresses on these two margins (wallet-tier and activity) lets us examine flows separately and yields a population that is overwhelmingly retail-like by count
but institutional by value.

Retail activity is informationally and behaviorally distinct from the institutional~\citep{boehmer2021tracking,barber2024subpenny}; aggregate data, pooling the two, obscures any group-specific pattern.
The stablecoin-redemption channel (Section~\ref{sec:stablecoins_backg}) is the cleanest case: \citet{ma2025stablecoin} document that direct par redemption of USDC is restricted to verified institutional accounts, so under stress retail users can only sell USDC at secondary-market prices while institutions arbitrage par redemption. 
Separating the two groups therefore lets us examine flow patterns consistent with such a mechanism.
This and the patterns documented above motivate the next question: how retail and institutional flows respond differentially to market-wide shocks.

\section{Event Study: Responses to Major Crypto Shocks}
\label{sec:event_study}

This section presents the paper's main behavioral evidence. We estimate whether major crypto-market shocks produce abnormal changes in CASP flows, and whether retail-like and institutionally mediated activity respond differently across assets and flow directions. We also examine whether stablecoin transfers display a common directional response, relating to the safe-haven question considered in aggregate-flow studies~\citep{anadu2024runs,oefele2024stablecoins}. Since deposits and withdrawals do not reveal trades executed off-chain within the CASP or final holdings, however, we do not test directly portfolio-level flight-to-safety patterns.
We study three episodes identified in Section~\ref{sec:event_sel}, in chronological order: the Terra Luna crash (Section~\ref{sec:TerraLunaResults}), the FTX collapse (Section~\ref{sec:FTXCollapseResults}), and the Silicon Valley Bank failure (Section~\ref{sec:SVBFailureResults}).

\subsection{Event Study Design}
\label{sec:CASPLevelEventStudy}

We estimate abnormal CASP flows using daily time series of on-chain activity in BTC, ETH, USDC, and USDT. For each asset and group, we construct three outcome measures: transferred volumes in native asset units, transaction counts, and USD-denominated values. We focus on the first since they measure the on-chain quantity response directly. Additional results on other outcome variables are reported in Appendix~\ref{app:robustness}.

Following~\cite{kothari2007eventstudies}, abnormal activity is the difference between the observed outcome $x_t$ and the normal outcome expected absent the event:

\begin{equation*}
	x^{*}_t = x_t - E[x_t \mid X_t],
\end{equation*}

where $X_t$ denotes conditioning information at time $t$.

We model normal activity using a return-adjusted specification. We use a 45-day baseline pre-event estimation window $L_1$ and an event window $L_2$ covering the event day and ten subsequent days. As a robustness check, we also consider pre-event windows ranging from 30 to 60 days. We test the null hypothesis that abnormal activity equals zero with OLS regressions over the combined sample $L_1 + L_2$. Since the figures report daily post-event coefficients, we estimate specifications with separate dummy variables for each post-event day:

\begin{equation}\label{equ:CASPEventStudy}
	x_{t} = \alpha + \beta r_t + \sum_{k=0}^{10} \gamma_k D^{k}_{t} + \varepsilon_{t},
\end{equation}

where $x_t$ denotes the respective outcome variable for a given cryptoasset, flow direction, and group; $\alpha$ is a constant; $r_t$ is an asset-specific return control (daily BTC returns for BTC, USDC, and USDT, and daily ETH returns for Ether, expressed in percentage points), accounting for normal variation associated with broader crypto-market movements; and $D^{k}_{t}$ is an indicator equal to one on day $k$ after the event and zero otherwise. The coefficients $\gamma_k$ estimate abnormal activity on event days $k=0,\ldots,10$.\footnote{Coefficients should be interpreted, for each shock, as reactions to an unfolding crisis rather than to a single isolated announcement. This is by design: the daily-coefficient specification, Eq.~(\ref{equ:CASPEventStudy}), traces the dynamic response day by day and does not presume a one-time jump.
Because $\gamma_k$ are conditional on contemporaneous returns that can themselves respond to the crisis, these coefficients need not equal the total flow response to the event.
} The horizontal line in the event-study figures corresponds to $\gamma_k=0$. Inference is based on a bootstrap procedure with 1{,}000 resamples of the regression residuals~\citep{kothari2007eventstudies}.\footnote{We assess serial dependence using Durbin--Watson and Breusch--Godfrey tests at lag orders 1, 5, and 10. The diagnostics generally provide no evidence of statistically significant residual autocorrelation. We therefore use independent residual resampling as our baseline bootstrap procedure.}

We estimate Eq.~(\ref{equ:CASPEventStudy}) separately for each asset, decomposing deposit and withdrawal flows as well as retail-like and 
institutionally mediated activity.
Rather than reporting every possible combination of asset, group, and flow measure, we focus on the margins that best summarize the economic response to each event. 
The remaining combinations of asset, group, and flow direction are reported in Appendix~\ref{app:robustness}, together with robustness checks that control for the dependence on the length of the pre-event estimation window (30 to 60 days), on the outcome measure (transaction counts instead of asset units), and on spurious responses generated by the specification itself (placebo events unrelated to each shock). The results are broadly consistent with those reported below.

\subsection{Terra Luna Crash}
\label{sec:TerraLunaResults}

The Terra Luna crash in May 2022 was a crypto-native confidence shock: the algorithmic stablecoin broke its dollar peg after markets started exploiting its design vulnerabilities on May~5, entering a destabilizing spiral that culminated in a failed attempt to restore the peg on May~11. The event undermined the credibility of algorithmic stablecoins as a class and triggered a broader shock within crypto markets (see Table~\ref{tab:sel_events_timeline}).

\begin{figure}
	\centering
	\includegraphics[width=\textwidth]{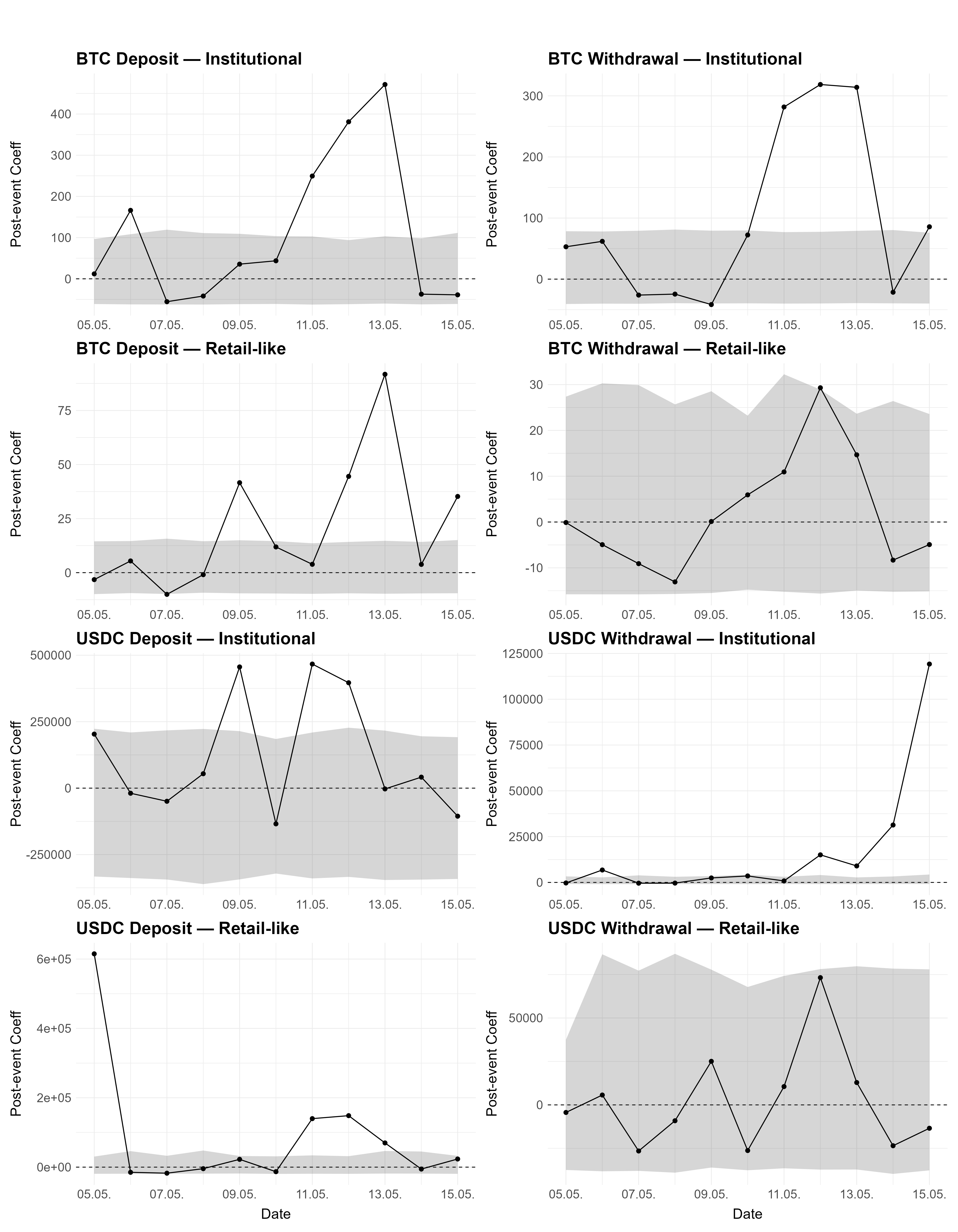}
	\caption{\textbf{Terra Luna crash: selected asset-flow responses}. {\small \textit{Notes:} Selected post-event coefficients from the Terra Luna crash event-study regressions, in asset units. Each panel shows coefficients for days 0 to 10 after the event, using a 45-day pre-event estimation window. The dashed horizontal line marks zero. Shaded areas show bootstrap bands based on 1,000 replications. Rows correspond to BTC and USDC, each split into institutionally mediated and retail-like flows; the left column reports deposits and the right column withdrawals.}}
	\label{figure:EventStudyTerraLunaCrash}
\end{figure}

Figure~\ref{figure:EventStudyTerraLunaCrash} shows the estimated responses, which are pronounced but heterogeneous across assets, groups, and flow directions. The clearest response is in BTC among institutionally mediated flows: both deposits and withdrawals rise well above the displayed bootstrap bands after May~11, after the final attempt to restore Terra's peg. Both directions move together; this points to active portfolio rebalancing rather than a one-sided movement into or out of BTC. Retail-like BTC flows show a weaker version of the same pattern, with deposits rising above the displayed band around May 9, when Terra lost its peg for the last time, and withdrawals staying largely within their bands. The BTC response is therefore driven mainly by institutional actors, consistent with high-volume entities repositioning during the shock, and its lagged timing matches Terra's progressive depeg.

Stablecoin flows are more mixed. Institutional USDC deposits are volatile, with positive spikes around May~9, when UST lost its peg for the second and final time, and again in the following days, coinciding with the last governance attempt to restore the peg, but show no persistent inflow. Institutional withdrawals remain limited for most of the window but rise strongly at the end, suggesting delayed liquidity adjustment. Retail-like USDC deposits spike on the event day and show smaller positive movements around May 11, while withdrawals fluctuate within the bootstrap band, with a temporary positive peak around May 12 rather than a sustained increase. The USDC results are thus consistent with short-run liquidity adjustment around key moments in the unfolding Terra crisis rather than a persistent safe-haven movement.

In sum, the Terra Luna results indicate broad reallocation and increased volatility after a crypto-native confidence shock. The most systematic response appears in BTC flows among institutional actors, where deposits and withdrawals move together, pointing to intensified trading or rebalancing. Stablecoin flows are more volatile and less directional, especially for USDC. The Terra Luna episode, thus, did not generate a uniform custodial-flow response consistent with a flight into stablecoins, but rather heterogeneous stablecoin transfers across groups and flow directions.

\subsection{FTX Collapse}
\label{sec:FTXCollapseResults}

The FTX collapse in November 2022 was a major custodial exchange shock, caused by the failure of one of the largest centralized exchanges active at the time, raising broad concerns about the safety of assets held on custodial platforms. The crisis unfolded on November 2, as doubts about FTX solvency and ties to its affiliated trading firm Alameda triggered a rapid loss of confidence and large withdrawal requests. The post-event window captures the escalation of the crisis, unfolding with a sharp fall in FTT price on November 8, Binance's withdrawal from the tentative acquisition of FTX on November 9, and the bankruptcy filing of FTX and Alameda on November 11.

\begin{figure}
	\centering
	\includegraphics[width=\textwidth]{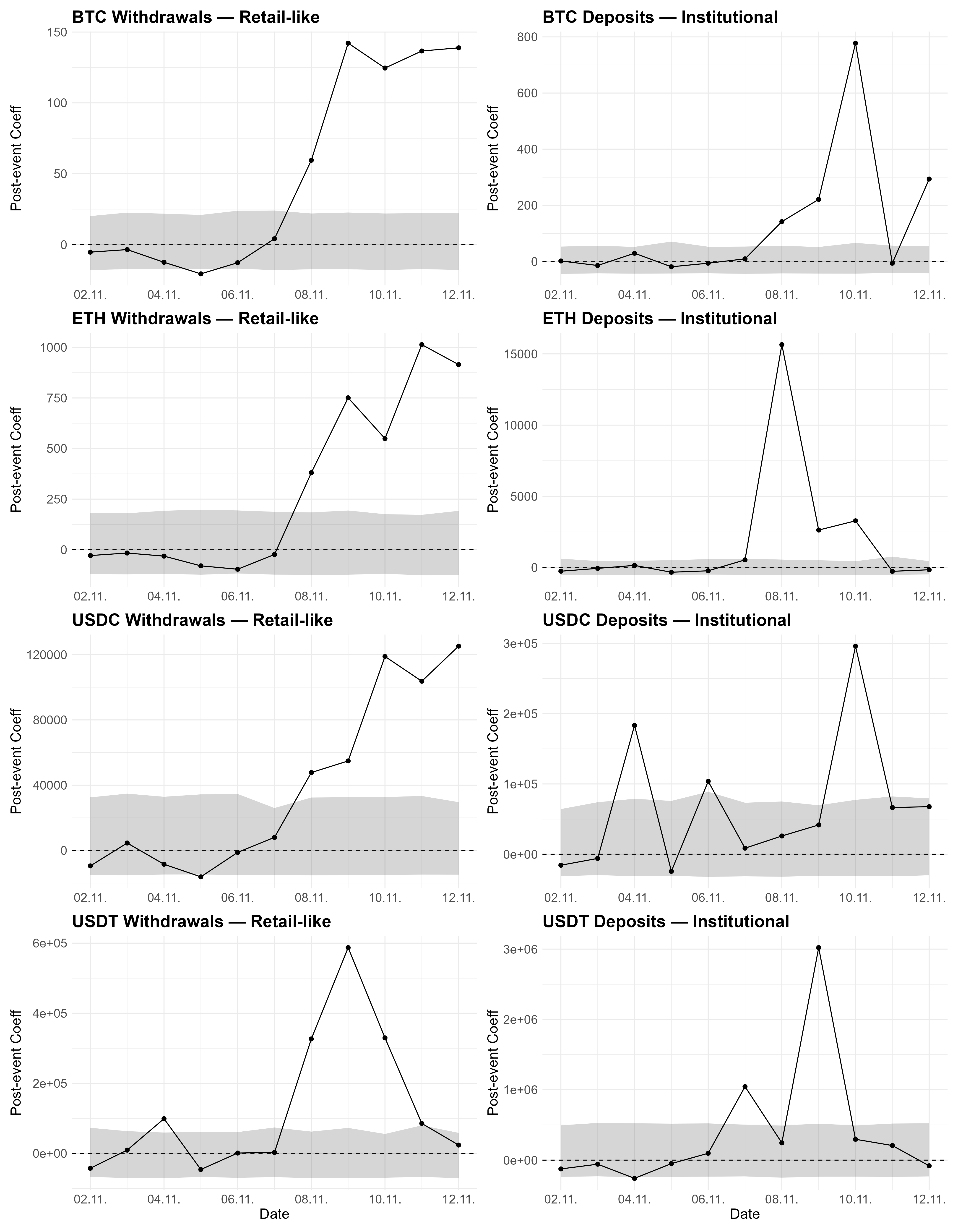}
	\caption{\textbf{FTX collapse: selected asset-flow responses}. {\small \textit{Notes:} Selected post-event coefficients from the FTX collapse event-study regressions, in asset units. Each panel shows coefficients for days 0 to 10 after the event, using a 45-day pre-event estimation window. The dashed horizontal line marks zero. Shaded areas show bootstrap bands based on 1,000 replications. The left column reports retail-like withdrawals. The right column reports institutional deposits. Rows correspond to BTC, ETH, USDC, and USDT, respectively.}}
	\label{figure:EventStudyFTXCollapse}
\end{figure}

Figure~\ref{figure:EventStudyFTXCollapse} reports the results. The left column shows withdrawals by retail-like entities, while the right column shows deposits by institutional actors. The figure reveals a clear heterogeneous response to the FTX collapse. Retail-like withdrawals increase after the event, especially for BTC and ETH. The response becomes most visible around Nov 8-12, coinciding with the intensification of the FTX crisis. This timing is consistent with a reduction in custodial exposure as the severity of the crisis became fully visible.

Institutional activity displays a different pattern. The response is concentrated in large deposit spikes occurring around days Nov 8-10. 
This suggests that institutional actors or large ecosystem participants engaged in substantial reallocation after the shock. One plausible mechanism is that these flows reflect CASP liquidity or reserve-management operations: customer trades and withdrawal requests may be settled internally first, while CASPs adjust their on-chain inventories only periodically. During stress episodes such as the FTX collapse, such rebalancing may become more pronounced.

Overall, the selected event-study results suggest that the FTX collapse increased cryptoasset mobility at Austrian CASPs, but through different adjustment margins. Retail-like activity was conducted mainly through withdrawals, particularly after November 8, consistent with reduced exposure to the sampled custodians and reallocation to self-custody wallets. Institutional actors generated large deposit flows, consistent with high-volume reallocation or platform-level liquidity management. The evidence therefore points to both a broad withdrawal response among retail-like actors and large reallocations by institutional actors following the collapse of FTX.

\subsection{Silicon Valley Bank Failure}
\label{sec:SVBFailureResults}

The Silicon Valley Bank (SVB) failure was a banking-sector shock that reached crypto markets primarily through the stablecoin channel. It began on 8 March 2023, when SVB announced a \$1.8~billion loss, escalated when the bank failed on 10 March, and hit the crypto ecosystem in particular on 11 March, when Circle disclosed \$3.3~billion of exposure to SVB; tensions eased after uninsured depositors were guaranteed on 12 March. Since Circle issues USDC, this episode is especially relevant for stablecoins, with a weaker response expected for BTC and ETH.

\begin{figure}
	\centering
	\includegraphics[width=\textwidth]{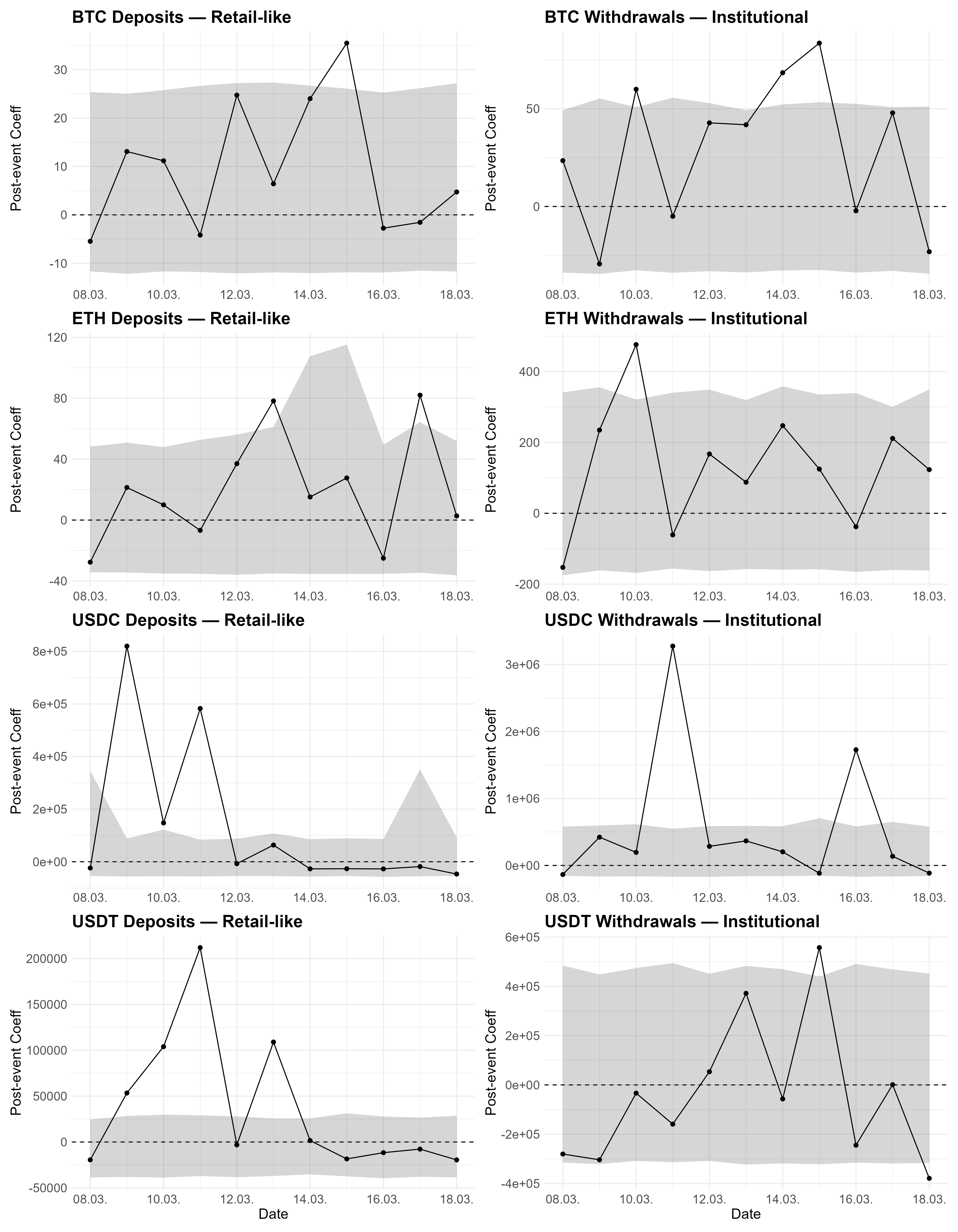}
	\caption{\textbf{SVB failure: selected asset-flow responses}. {\small \textit{Notes:} Selected post-event coefficients from the SVB failure event-study regressions, in asset units. Each panel shows coefficients for days 0 to 10 after the event, using a 45-day pre-event estimation window. The dashed horizontal line marks zero. Shaded areas show bootstrap bands based on 1,000 replications. The left column reports deposits by retail-like entities, while the right column reports withdrawals by institutional actors. Rows correspond to BTC, ETH, USDC, and USDT, respectively.}}
	\label{figure:EventStudySVBFailure}
\end{figure}

Figure~\ref{figure:EventStudySVBFailure} shows the estimated responses, which are consistent with a shock transmitted mainly through stablecoins.
BTC retail-like deposits stay within their bootstrap bands throughout, with no sustained directional response. BTC institutional withdrawals are noisier, with several points above the displayed band, but show no sustained direction. ETH is similar: retail-like deposits and institutional withdrawals fluctuate and occasionally exceed the bootstrap bands on single days, but no response persists and the bands remain wide.

The clearest reactions appear in USDC, and align with the timing of the sub-events. 
Retail-like USDC deposits show large spikes above the displayed band on March 9 and 11 before falling back. The March 9 movement follows SVB's announcement; the March 11 movement coincides with Circle's disclosure date. Institutional USDC withdrawals exhibit instead a large spike on March 11, against a tight band, with a further above-band spike later in the window. 
This pattern matches the redemption asymmetry documented by \citet{ma2025stablecoin}: retail-like USDC deposits are consistent with preparing sales within a CASP, while institutionally mediated withdrawals are consistent with pulling USDC off-venue to redeem at par.
Further evidence in Appendix~\ref{app:robustness} shows that the event also coincided with an abnormally high number of retail-like deposits, pointing to a run-like exit from USDC by many small holders rather than a few large transfers.
Although we do not observe issuer-account eligibility nor the subsequent trades and redemptions, the timing, direction, and breadth of the flows jointly support this interpretation.

USDT behaves differently. Retail-like USDT deposits rise above the bootstrap band over the first post-event days, consistent with the pattern observed with USDC. Institutional USDT withdrawals, instead, are volatile, with a single above-band spike, but the pattern is less clear than for USDC, where both margins move sharply against tight bands.

Overall, the SVB failure operated through a narrower stablecoin channel. As expected, BTC and ETH responses are weak, noisy, and do not show persistent directional movements. The strongest reactions appear in USDC, and to a lesser extent USDT, around the key SVB and Circle-related disclosures. USDC retail-like deposits and institutional withdrawals both increase, consistent with Circle's restriction of USDC direct redemption to verified institutional accounts.

\subsection{Summary}
\label{sub:events_summary}

The three event studies document distinct responses around the selected crises.
The Terra Luna crash, a crypto-native confidence shock, triggered broad reallocation and elevated volatility, most clearly in the comovement of institutional BTC deposits and withdrawals, rather than uniform flights into stablecoins. These patterns are consistent with active portfolio rebalancing. 
After the FTX collapse, a custodial exchange shock, retail-like actors reduced custodial exposure through sustained withdrawals once the crisis escalated, while institutional actors generated large episodic deposits consistent with high-volume reallocation or CASP inventory management. 
The SVB failure operated through the stablecoin channel, with less systematic responses in BTC and ETH and the strongest reaction in USDC, where retail-like deposits and institutional withdrawals are consistent with the restriction of par redemption to institutional accounts~\citep{ma2025stablecoin}. 
Across all three shocks, cryptoasset transfers were two-sided and shock-specific, and do not support consistent reallocation toward stablecoins as a within-crypto safe haven.
This is consistent with evidence from DeFi lending markets~\citep{campello2025demand} and the dynamics documented by \citet{aldasoro2025stablecoins}.
Taken together, the episodes differ in their observed response margins, and retail-like and institutionally mediated flow differences would largely cancel in aggregate data, making separation by group and transfer direction informative. The robustness checks reported in Appendix~\ref{app:robustness} on alternative pre-event windows and outcome variables, as well as placebo tests, are broadly consistent with the results discussed above.

\section{Discussion}
\label{sec:discussion}

\subsection{Key Findings}

This paper studies how cryptoasset activity intermediated by Austria's registered CASPs integrates into the national financial system, using a regulatory dataset covering all crypto-asset service providers registered in Austria at the end of 2024. Under Austrian anti-money-laundering law, CASPs must report the full set of blockchain addresses they control~\citep{fma2025aml}. From the resulting pseudonymized address-to-entity registry, we reconstruct the on-chain activity of the 12 registered CASPs across Bitcoin (BTC), Ether (ETH), and the two largest stablecoins (USDT and USDC) through May 2025: 11.98~million asset transfers and \$50.4~billion in cumulative volume. The registry directly establishes ownership rather than inferring it through clustering heuristics, which by construction produce false positives and negatives~\citep{meiklejohn2013fistful,Froewis:2020a}. We observe these flows with a reliability that
country-level studies relying on indirect proxies have not achieved.

We document four main findings. First, Austrian CASPs intermediate substantial economic activity, roughly \$30~billion in on-chain flows with external counterparties, but the domestic network is only weakly interconnected: flows between Austrian CASPs total just \$5.8~million. 
The Austrian ecosystem is integrated globally rather than domestically, consistent with the cross-border orientation of crypto flows documented at the aggregate level~\citep{auer2025crossborder}.
A further \$20.7~billion, or 41.1\% of total volume, reflects within-entity bookkeeping rather than counterparty activity. This echoes, at a jurisdiction-specific scale, the much larger share of network-wide Bitcoin volume that \citet{makarov2021blockchain} attribute to within-entity and protocol-related transfers, and it confirms that internal wallet management may be misclassified as economic activity in aggregate on-chain data.

Second, external activity is highly concentrated. A small set of transactions conducted directly with the warm and cold wallets of Austrian CASPs accounts for 57.8\% of external volume while representing only 0.77\% of external transfers, whereas the overwhelming majority of addresses, about 98\% on each chain, exhibit low-frequency, retail-like behavior, transacting two to three times for a few hundred to a few thousand dollars. Exploiting this structural divide, we classify external counterparties using two signals, warm/cold-wallet contact and chain-wide transaction activity, rather than the single trade-size threshold standard in the microstructure literature~\citep{schultz2012market,boehmer2021tracking}. This responds to the documented fragility of one-dimensional proxies: trade-size cutoffs misclassify a substantial share of trades, as informed and uninformed counterparties increasingly transact at similar sizes~\citep{hund2025rise}.

Third, in event studies around three broadly covered shocks, namely a run on an algorithmic stablecoin (Terra Luna), the failure of a centralized exchange (FTX), and the failure of a bank intertwined with the crypto sector (SVB), retail-like and institutional actors respond through different margins. After the FTX collapse, retail-like withdrawals increased persistently and across assets, consistent with reduced custodial exposure following a major exchange failure, while institutional flows are consistent with liquidity management and cross-platform rebalancing. The SVB episode produced a narrower response concentrated in USDC, consistent with Circle's two-tiered redemption mechanism, and split sharply by group, with retail-like depositing and institutional withdrawing; instead, BTC and ETH remained mostly unaffected. The Terra Luna crash generated broader but less persistent reallocation across assets, leading to heightened volatility. 
This group-level heterogeneity is the behavioral counterpart to the distinction between retail and institutional order flow established in equity markets~\citep{boehmer2021tracking,barber2024subpenny} and parallels the evidence from money market funds, showing that institutional and retail investors holding the same portfolio responded differently to the same shock~\citep{schmidt2016runs}; importantly, since the two groups can move in opposite directions, their responses partly offset each other and would be largely masked in aggregate flows~\citep{martin2026deposit}.

Fourth, stablecoin flow responses were two-sided and shock-specific, rather than a uniform inflow; our findings do not support the view that stablecoins function as a systematic within-crypto safe haven during stress.
These findings complement work on stablecoin safe-haven properties~\citep{baur2021crypto,anadu2024runs,oefele2024stablecoins} based on aggregate data
and align with the more recent evidence of heterogeneous, inconsistent safe-haven dynamics across events~\citep{aldasoro2025stablecoins}.

\subsection{Impact}

Our findings speak to several groups of stakeholders.

\vspace{-0.25cm}
\paragraph{Financial Supervisors and Regulators}

The registry-based approach offers a systematic and repeatable way to measure CASP-mediated cryptoasset activity and attribute it to a specific jurisdiction. 
By deriving flows directly from addresses reported by supervised CASPs, rather than from third-party attribution or clustering heuristics, it yields reliable and reproducible measurements once the reporting pipeline is in place.
This suggests a concrete supervisory practice: collecting address-registry data directly from supervised entities, and updating it as their wallet infrastructure changes.
Such data could be used, for instance, to automate audit processes
and improve CASP solvency assessment~\citep{saggese2024assessing}. 
Supervisors could further require CASPs to report available attribution on their counterparties, since reporting entities can often distinguish retail customers from institutional ones through their own KYC and account records.
Such information would sharpen the retail-versus-institutional distinction and let supervisors observe heterogeneous responses directly rather than through proxies. 
The same design could, in principle, be extended across jurisdictions to support a broader observatory of cryptoasset flows, European or even global in scope.

\vspace{-0.25cm}
\paragraph{Policymakers}

The results bear on debates over financial stability, monetary transmission, and public digital money (e.g., central bank digital currencies). 
The heterogeneous responses to shocks show that aggregate transfer measures can obscure differences across counterparties, as well as channels through which risk may transmit; 
the two-sided stablecoin response during stress, rather than a clean flight to quality, adds to the question of how far privately issued stablecoins can replicate the stabilizing role of traditional safe assets, and how they will interact with future public digital currencies. 
At the same time, the pronounced withdrawal response following the custodial failure of FTX is consistent with how quickly confidence in custodial intermediaries can shift, another channel relevant to any framework that treats CASPs as points of systemic exposure.

\vspace{-0.25cm}
\paragraph{Crypto-Asset Service Providers}

For CASPs, the timing of the responses is relevant: abnormal flows appear within days of each shock, with no visible lag.
Stress propagated through the on-chain system rapidly, leaving little time for operational adjustment once an event unfolds.
This argues for ex-ante safeguards rather than reactive ones, in particular liquidity and redemption arrangements robust to sudden two-sided stablecoin activity of the kind observed around the FTX failure and Circle's SVB disclosures.

\vspace{-0.25cm}
\paragraph{Retail Users}

The evidence indicates that retail-like and institutional flows operate on different margins, with retail dominating by transaction count and institutions by volume, and that the two can move in opposite directions during stress. For retail users specifically, the SVB episode illustrates a structural disadvantage: when a stablecoin's peg is threatened, par redemption is available only to a restricted set of institutional accounts, so retail holders are left to sell on secondary markets, often at a loss~\citep{ma2025stablecoin}. This asymmetry is a consumer-protection consideration that aggregate stability metrics do not capture.

\subsection{Limitations and Future Work}
\label{sec:limits}

Several limitations qualify our results, each pointing to a direction for future work. First, treating the dataset as a trustworthy source depends on the accuracy and completeness of CASP reporting. In the course of the analysis we identified
a few addresses likely controlled by Austrian CASPs that were absent from the original mapping, which indicates that reporting is not fully exhaustive. Future work could pair the registry with independent attribution or clustering as a cross-check, quantifying reporting gaps rather than assuming completeness.
Relatedly, because the registry reflects only CASPs registered at the end of 2024, our analysis excludes providers that exited the market earlier. While this survivorship concern is real, it is mitigated by the fact that entry and exit concentrate among small CASPs, while the large providers driving most volumes are stable over the sample period; furthermore, the Austrian CASP market itself was negligible before 2017.
A related concern is that address control may change over time. This issue is mitigated by the fact that the registry is retroactive, that is, it includes both addresses controlled at the collection date and at earlier dates. 

Second, the retail-institutional classification is proxy-based: it cannot fully account for a single actor's activity being fragmented across many addresses, and the activity threshold ($z>3$) is conventional rather than theoretically derived. 
More generally, without reliable attribution of counterparty addresses, finer analyses (e.g., at the level of individual customers) are out of reach; likewise, with only 12 CASPs and activity concentrated in a few, we cannot reliably exploit heterogeneity in shock responses across them.
As we argue above, the most direct remedy lies with supervisors, who could attach verified counterparty attribution from CASP KYC records; absent that, future work could validate and refine the classification against partial ground-truth labels and assess sensitivity to alternative thresholds and address-linkage methods.

Third, the analysis covers a single jurisdiction and may not generalize to larger or structurally different markets. Because the approach depends only on registry data that other supervisors could in principle collect, the natural extension is cross-jurisdictional replication, which would also enable the comparative observatory discussed above.

Fourth, we cover four major cryptoassets, omitting the long tail of smaller assets, and we observe USDT and USDC only on Ethereum although both are deployed on several chains. Extending coverage to additional assets and to the other chains on which stablecoins circulate would capture cross-chain substitution that our single-chain view of stablecoins may miss, an aspect directly relevant to the safe-haven question.

Fifth, we observe only Austrian CASPs. We therefore do not see Austrian holders who use non-Austrian providers or who transact solely through self-custodial wallets, so our flows are a partial view of the national cryptoasset activity. Linking the registry to cross-border or self-custody flows, where data-sharing arrangements permit, would help place the Austrian segment, or that of another jurisdiction, within the wider ecosystem.

Beyond addressing these limitations, this study offers a tool for assessing which contagion channels are most active during stress. Applying it systematically across events and jurisdictions could help supervisors assess where cryptoasset markets transmit risk into the wider financial system, which is central to evaluating financial-stability risk.

\section{Conclusions}
\label{sec:conclusions}

Using a regulatory registry that identifies the on-chain addresses of all crypto-asset service providers (CASPs) registered in Austria at the end of 2024, we provide a transaction-level view of how CASP-intermediated cryptoasset activity integrates into a national financial system and how it reacts under stress. The registry lets us measure these flows directly, without the clustering heuristics or indirect proxies on which prior country-level work has relied. We also propose a simple approach to separate retail-like from institutionally mediated activity using on-chain behavior rather than a single trade-size threshold. 
Two pictures emerge. Structurally, Austrian CASPs intermediate tens of billions of dollars but are integrated globally rather than with one another, and their external activity is dominated in value by a small set of institutional counterparties and in number by a large base of retail-like ones. Dynamically, the two groups exhibit different flow patterns around the Terra Luna, FTX, and SVB shocks, and stablecoins do not act as a uniform within-crypto safe haven. 
The clearest stablecoin response, around SVB, was two-sided: retail-like counterparties deposited USDC while institutionally mediated ones withdrew it, consistent with the asymmetric redemption access of fiat-backed stablecoins. 
After FTX, retail-like outflows involved all assets, including stablecoins, pointing to a sector-wide loss of confidence in custodial intermediation. 
These patterns are visible only once we separate flows by counterparty and direction. Overall, registry-based, transaction-level measurement is a promising foundation for monitoring how the cryptoasset activity intermediated by a national jurisdiction's registered CASPs integrates into the financial system.

\newpage

\paragraph*{Acknowledgments}
We thank the Austrian Financial Market Authority (FMA) for providing access to the address registry. We are grateful to  Raphael Auer, Nicola Dimitri, Lorenzo Schoenleber, Massimiliano Sala, Michela Iezzi, and to participants at the Crypto-Asset Monitoring Expert Group (CAMEG) 2025, the European Securities and Markets Authority (ESMA) 2026 Research Conference, and Financial Cryptography in Rome 2026 at the Bank of Italy, for helpful comments and suggestions.

\paragraph*{Data availability}
The address-to-entity registry was provided by the Austrian Financial Market Authority under a confidentiality agreement and pseudonymized, thus cannot be made public. Code and aggregate data are available from the corresponding author upon request, the latter subject to approval by the Austrian Financial Market Authority; the code will be released publicly upon publication.

\paragraph*{Declaration of competing interests}
The authors declare that they have no known competing financial interests or personal relationships that could have appeared to influence the work reported in this paper.

\paragraph*{Disclaimer}
The views expressed in this paper are those of the authors and do not necessarily reflect those of their affiliated institutions.

\bibliographystyle{apalike}
\bibliography{references}

\newpage

\appendix

\section{Data -- Supplemental Material}
\label{app:registry}
\label{app:google_trends}

This section provides additional information on the address-to-entity mapping and the analysis on Google Trends data conducted to select important crypto-related events. 

\subsection{Address Registry}

Table~\ref{tab:seed_addresses} provides a breakdown of the addresses by CASP and type (hot, warm, and cold). 
The vast majority are hot wallets and are used to facilitate deposit/withdrawal interactions with customers.
Addresses are neither equally distributed among CASPs nor by wallet type, with a limited number of CASPs controlling the largest number of addresses.

\begin{table}[htbp]
	\centering
	\footnotesize
	\begin{tabular*}{\linewidth}{@{\extracolsep{\fill}}lrrrr}
		\toprule
		CASP & Hot & Warm & Cold & Total \\
		\midrule
		1 & 0 & 0 & 0.001 & 0.001 \\
		2 & 81.846 & 0.033 & 0.252 & 82.132 \\
		3 & 0.148 & 0 & <0.001 & 0.149 \\
		4 & 0.515 & 0 & 0 & 0.515 \\
		5 & 7.751 & 0 & <0.001 & 7.752 \\
		6 & 8.826 & 0 & <0.001 & 8.827 \\
		7 & <0.001 & 0 & 0 & <0.001 \\
		8 & 0.130 & 0 & 0.006 & 0.135 \\
		9 & 0 & <0.001 & <0.001 & 0.002 \\
		10 & 0.483 & 0 & 0.002 & 0.485 \\
		11 & <0.001 & 0 & 0 & <0.001 \\
		12 & 0 & 0.002 & 0 & 0.002 \\
		\midrule 
		Total & 99.702 & 0.036 & 0.263 & 100.000 \\
		\bottomrule
	\end{tabular*}
	\caption{\textbf{Relative amount of wallets controlled by Austrian CASPs}. Wallets are divided into hot, warm, and cold; values are in percentage as agreed with Austrian financial authorities.} 
\label{tab:seed_addresses}
\end{table}

Table~\ref{tab:transaction_by_asset} reports the evolution in time of the number of asset transfers recorded, broken down by cryptoasset and year. Bitcoin, as expected, plays a major role, while Ether and stablecoins entered the market in 2017 and 2019, respectively. The CASP activity peaked in terms of number of transactions in 2021 and decreased thereafter, with the Bitcoin share decreasing compared to other assets after 2022.

\begin{table}[ht]
	\centering
	\scriptsize
	\begin{tabular*}{\linewidth}{@{\extracolsep{\fill}}lrrrrr}
		\toprule
		Year & BTC & ETH & USDC & USDT & Total \\
		\midrule
		2016 & \num{52040} & \num{0} & \num{0} & \num{0} & \num{52040} \\
		2017 & \num{1087398} & \num{2080} & \num{0} & \num{0} & \num{1089478} \\
		2018 & \num{1310338} & \num{39405} & \num{0} & \num{0} & \num{1349743} \\
		2019 & \num{1038242} & \num{137319} & \num{1851} & \num{3441} & \num{1180853} \\
		2020 & \num{1378311} & \num{165967} & \num{5516} & \num{33743} & \num{1583537} \\
		2021 & \num{1864144} & \num{377384} & \num{19094} & \num{100183} & \num{2360805} \\
		2022 & \num{1366961} & \num{246493} & \num{19211} & \num{86160} & \num{1718825} \\
		2023 & \num{870018} & \num{272073} & \num{14071} & \num{112035} & \num{1268197} \\
		2024 & \num{531972} & \num{335149} & \num{16943} & \num{125635} & \num{1009699} \\
		2025 & \num{189021} & \num{126034} & \num{31855} & \num{20600} & \num{367510} \\
		\midrule
		Total & \num{9688445} & \num{1701904} & \num{108541} & \num{481797} & \num{11980687} \\
		\bottomrule
	\end{tabular*}
	\caption{\textbf{Number of asset transfers recorded on-chain by year and asset.} 2025 covers January to May only.}
	\label{tab:transaction_by_asset}
\end{table}

\subsection{Event Selection}

\begin{table}[!ht]
	\footnotesize
	\centering
	\begin{tabular}{ll}
		\toprule
		Event & Keywords searched in Google Trends \\
		\midrule
		
		\multicolumn{2}{l}{\hspace{0.5 cm} CeFi events} \\
		\midrule
		Bitfinex & Bitfinex \\
		Kucoin & Kucoin \\
		Bitmart & Bitmart \\
		FTX & \textit{FTX}, Sam Bankman-Fried, FTX bankruptcy \\
		Poloniex & Poloniex \\
		Silicon Valley Bank Collapse & \textit{SVB}, Silicon Valley Bank \\
		DMM Bitcoin & DMM Bitcoin \\
		WazirX & \textit{WazirX}, WazirX India \\
		Bybit & Bybit \\
		
		\midrule
		\multicolumn{2}{l}{\hspace{0.5 cm} DeFi events} \\
		\midrule
		Poly Network & Poly Network \\
		Compound V2 & \textit{Compound V2}, Compound protocol, Compound DeFi \\
		Badger DAO &  \textit{Badger DAO}, Badger, BadgerDAO \\
		Portal & \textit{Portal protocol}, Portal, Portal DeFi \\
		Ronin Bridge &  \textit{Ronin Bridge}, Ronin, Ronin DeFi\\
		Terra depeg & \textit{Terra Luna}, UST Terra, Terra stablecoin, Terra crash, UST \\
		Binance Bridge & \makecell[l]{\textit{Binance chain}, BNBChain, BSC, BSC Token Hub,\\ Binance Bridge, Binance hack} \\
		Euler V1 & \textit{Euler protocol}, Euler, Euler DeFi \\
		Multichain & \textit{Multichain}, Multichain DeFi, Multichain protocol \\
		Cetus CLMM & Cetus CLMM \\
		\bottomrule
	\end{tabular}
	\caption{\textbf{List of keywords analyzed with Google Trends}. When two or more were used, the one selected is in italic.}
	\label{tab:keywords}
\end{table}

Table~\ref{tab:keywords} reports the keywords used to measure via Google Trends the media coverage for each of the 19 major events detected as discussed in Section~\ref{sec:data}.
Queries were conducted in June 2026, covering dates from January 2019 to May 2025. To make query batches comparable, we normalized search interest within each geographic area, using the most-searched term as the anchor (Terra Luna for Austria, FTX worldwide).
We note that, since the names of DeFi protocols often correspond with terms widely used in spoken English (an obvious example is the lending protocol Compound), we searched for multiple keywords when ambiguities could arise. Instead, the choice for CeFi actors was more straightforward and led to lesser ambiguity.

\begin{figure}[h]
	\begin{subfigure}{0.49\textwidth}
		\includegraphics[width=\textwidth]{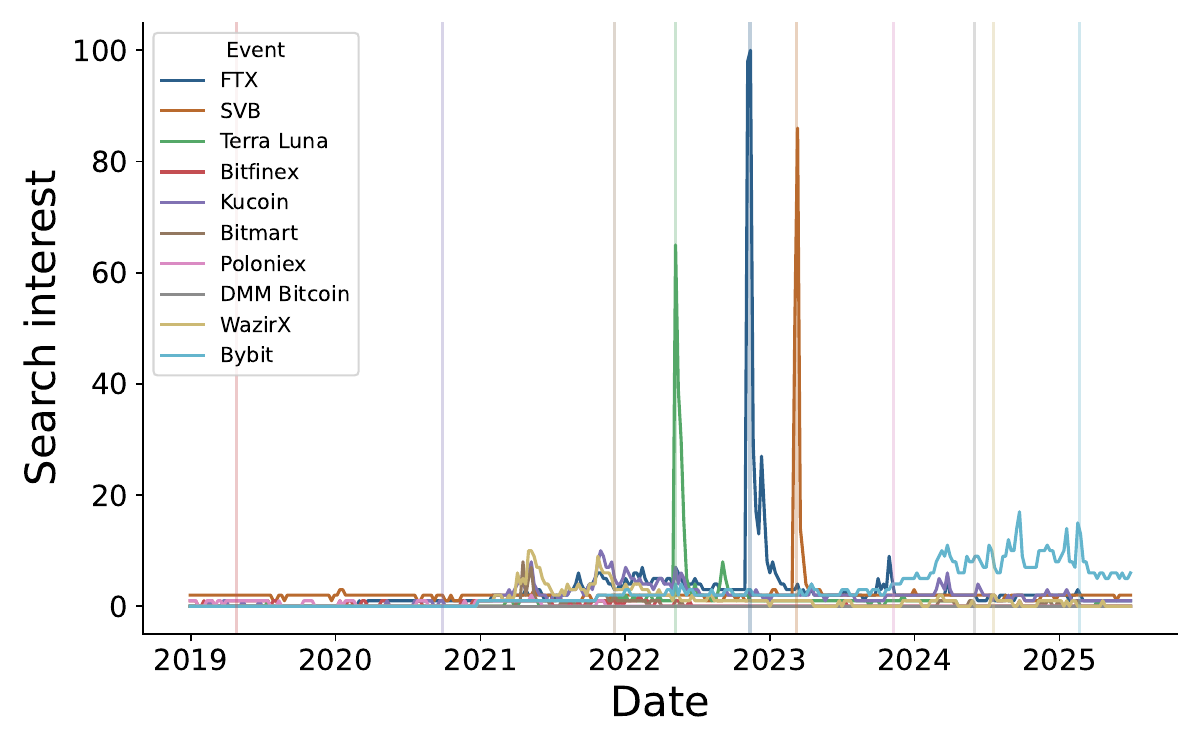}
		\caption{CeFi, World}
		\label{fig:cefi_w}
	\end{subfigure}
	\hfill
	\begin{subfigure}{0.49\textwidth}
		\includegraphics[width=\textwidth]{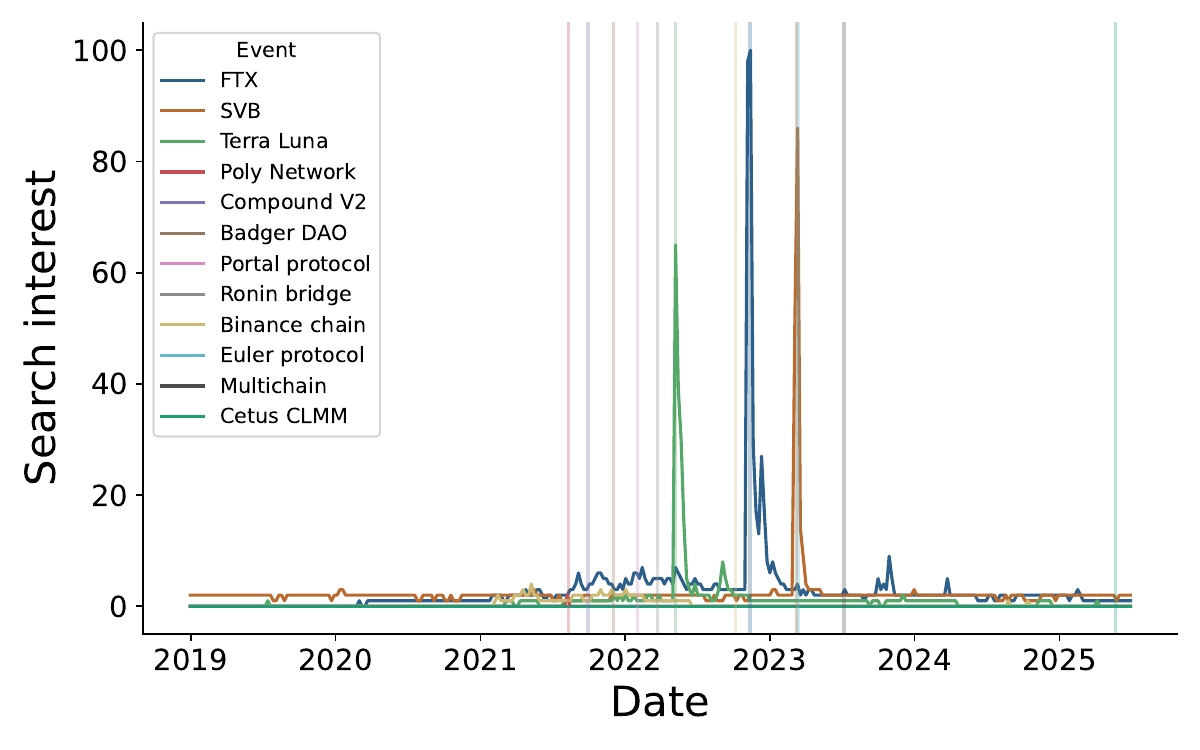}
		\caption{DeFi, World}
		\label{fig:defi_w}
	\end{subfigure}
	
	\medskip
	\begin{subfigure}{0.49\textwidth}
		\includegraphics[width=\textwidth]{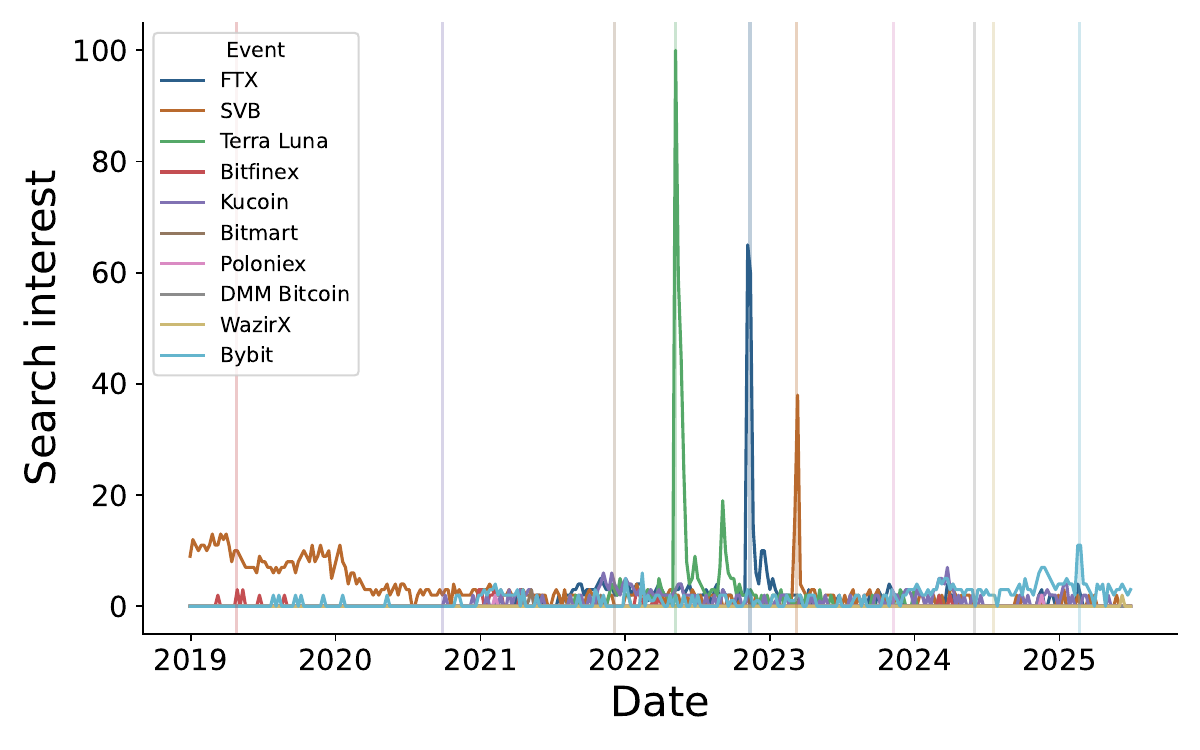}
		\caption{CeFi, Austria}
		\label{fig:cefi_at}
	\end{subfigure}
	\hfill
	\begin{subfigure}{0.49\textwidth}
		\includegraphics[width=\textwidth]{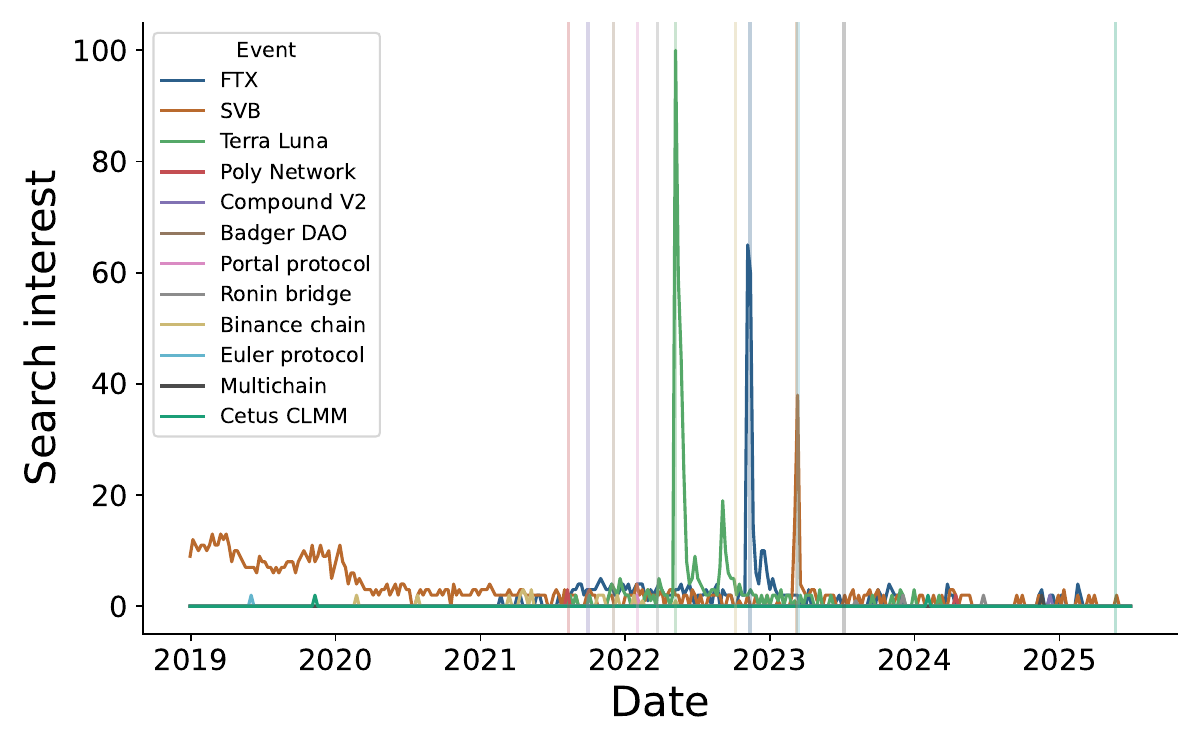}
		\caption{DeFi, Austria}
		\label{fig:defi_at}
	\end{subfigure}	
	\caption{\textbf{Google Trends results for CeFi (left) and DeFi events (right).} Top panels report results on web searches conducted worldwide, bottom panels refer to searches in Austria.} 
	\label{fig:gt}
	
\end{figure}

Figure~\ref{fig:gt} reports the results of the Google Trends analysis. 
For each event, we show the evolution in time of the search volume and a vertical line at each event date (colors of the time series and the vertical line correspond). The plotted series show that Terra Luna, FTX, and SVB received pronounced search-interest peaks around the shock dates compared to all other events.

Next, we describe the series of sub-events that led to each of the three major shocks selected: the Terra Luna crash, the FTX bankruptcy, and the SVB collapse.
The first one refers to the de-pegging from the US dollar of Terra Luna, the largest algorithmic stablecoin at the time (May 2022). 
TerraUSD (UST) maintained its peg through an arbitrage mechanism with its associated token LUNA, rather than through collateral backing, making the system inherently fragile under stress.
In early May 2022, significant selling pressure on both Bitcoin and LUNA contributed to destabilizing market conditions. On May 7, UST lost its peg to the US dollar for the first time, reportedly following large withdrawals from liquidity pools in the Terra ecosystem, triggering a wave of redemptions. Despite temporary recovery attempts, UST de-pegged again on May 9, leading to a self-reinforcing `death spiral' in which the minting of LUNA to defend the peg led to hyperinflation and a collapse in its price. By May 11, efforts by the Terra founders to restore the peg had failed, marking the effective breakdown of the ecosystem.

The second large shock we investigate is the failure of FTX, one of the most prominent cryptocurrency exchanges at the time of its failure in November 2022. Its founder also owned Alameda Research, a trading firm linked with FTX, to which FTX had lent billions of its customers' assets and which held significant amount of FTT, an exchange token issued by FTX.
On November 2 2022, balance-sheet details of Alameda Research were published, showing heavy investments in FTT. On November 6, Binance, a competitor exchange, announced it would sell its remaining FTT holdings; a few days later, on November 8, Binance communicated its intent to acquire FTX, only to back out of the rescue deal one day later. On November 10, the Bahamas Securities Commission froze FTX Digital Markets' assets, and finally on November 11, FTX, FTX US, and Alameda filed for bankruptcy.

The third large shock we investigate is the failure of Silicon Valley Bank. SVB was the sixteenth-largest bank in the United States at the time of its failure in March 2023~\citep{metrick2024failure}. Focusing on the tech and startup sector, SVB had several interlinkages with crypto-related entities, among which Circle, the company that manages USDC.
We identify four sub-events taking place during the failure of Silicon Valley Bank. On March 8 2023, SVB announced a USD 1.8 billion loss on its securities; within the next day, depositors had withdrawn around $25\%$ of total deposits. On March 10, the Federal Deposit Insurance Corporation (FDIC) took SVB into receivership, i.e. the FDIC took control of SVB, promising immediate access only to insured deposits (\$250,000 per depositor). The following day, Circle revealed that it had $\$3.3$ billion in deposits at SVB, causing USDC to depeg for several hours, reaching a value as low as \$0.87. On March 12, the Treasury, Federal Reserve, and FDIC announced that all SVB depositors would be fully protected, including uninsured depositors, with access to their funds from March 13.\footnote{\url{https://www.federalreserve.gov/newsevents/pressreleases/monetary20230312b.htm}}

Finally, in addition to the events described in Section~\ref{sec:data}, we construct a dataset of other events that occurred within the time window of interest and that arguably may have impacted the cryptoasset markets. We extract data on sector-specific events and broader large-scale shocks to the global economy from \cite{koutrouli2025crypto} and Coinmarketcap.
We also gather data from official sources (US government, Federal Reserve, ECB, EU, WHO). 
For each event, we record the date, a brief description, additional information on the type of event, the reference, and an indicator of whether it is considered a major or minor event for the crypto sector.
The dataset (Table~\ref{tab:events_timeline}) covers economic shocks and monetary policy decisions, such as interest rate changes and policy interventions; geopolitical events like conflicts or elections; announcements and milestones, e.g., Bitcoin halvings or price thresholds; and regulatory events, such as the implementation of the MiCA regulation.
The three shocks we analyze do not overlap with any major event, but can overlap with other market-relevant announcements, and the estimates describe abnormal flow patterns around these episodes rather than effects uniquely attributable to a single announcement. In particular, the FTX window includes the November 2 FOMC decision to raise the target range for the federal funds rate;\footnote{\url{https://www.federalreserve.gov/newsevents/pressreleases/monetary20221102a.htm}} yet, Figure~\ref{figure:EventStudyFTXCollapse} shows that the main response coincides with the escalation of the FTX crisis, rather than with the FOMC announcement.

\begin{table}[h!]
	\centering
	{\scriptsize
\begin{tabular}{llcll}
	\toprule
	&   &  &  &   \\
	Date & Identified Event & Major & Type &  Reference \\
	\midrule 
	\multicolumn{2}{l}{\hspace{0.5 cm} Crypto-related announcements, events} & &  & \\
	\midrule
	2020-05-11 & Bitcoin halving  & \xmark  & Technical &  Blockpit \\
	2020-06-30 & USDT reaches 10 bln\$ mkt. cap  & \xmark & Media &   Coinmarketcap  \\ 
	2020-10-21 & Paypal enables BTC payments & \xmark & Media &  ECB et al. \\
	2021-02-07 & 50,000USD/BTC breached  & \xmark & Media &   Coinmarketcap  \\ 
	2021-02-15 & Tesla, Mastercard support BTC  & \xmark & Media &  ECB et al.  \\
	2021-10-15 & SEC approves BTC futures ETFs & \xmark & Media &  Coinmarketcap \\ 
	2021-11-11 & BTC falls 65,000USD/BTC & \xmark & Media &   Coinmarketcap  \\ 
	2023-06-23 & BlackRock files for Spot BTC ETF  &  \xmark & Media & ECB et al.  \\
	2024-01-10 & 11 BTC ETF approved by SEC  &  \xmark & Media &   ECB et al.  \\
	2024-04-20 & Bitcoin halving  & \xmark & Technical &  Blockpit \\
	2024-07-03 & USDT reaches 100 bln\$ mkt.cap  & \xmark & Media &   Coinmarketcap  \\
	2024-12-13 & 100,000USD/BTC breached  & \xmark & Media &  Coinmarketcap  \\
	2025-01-23 & US EO supporting cryptoassets  & \xmark & Regulation &  US govt.
	\\
	
	\midrule
	\multicolumn{2}{l}{\hspace{0.5 cm} Econ. shocks, stimuli, monetary policy} & &   &    \\
	\midrule
	2020-03-15
	& Major FOMC Rate Cut (COVID-19) & \cmark & Monetary & Fed.Res.
	\\
	2020-03-27 & CARES Act (US) approved  & \cmark & Policy &  US govt.
	\\
	2020-07-21 & EU agrees 750 bln recovery package & \cmark & Policy &  EU\\
	2021-03-11 & American Rescue Plan Act approved  & \xmark & Policy &  US govt.
	\\
	2021-07-13 & First EU recovery disbursements & \xmark & Policy &  EU\\
	\makecell[l]{2022-03-16 - \\  2023-07-26}
	& FOMC rate increases & \xmark & Monetary & Fed.Res.\\
	\makecell[l]{2022-07-27 - \\  2023-09-20}
	& ECB interest rate increases & \xmark & Monetary &  ECB \\
	\makecell[l]{2024-09-18 - \\  2024-12-18}
	& FOMC rate cuts & \xmark & Monetary & Fed.Res. \\
	\makecell[l]{2024-06-11 - \\  2025-05-31}
	& ECB interest rate cuts & \xmark & Monetary &  ECB	\\
	2025-03-02 & US announces `Strategic crypto reserve'  & \xmark & Monetary &  US govt.
	\\
	\midrule
	\multicolumn{2}{l}{\hspace{0.5 cm} Geopolitical events} & &   &   \\
	\midrule
	2020-03-02 & \makecell[l]{COVID-19: EC fully activates IPCR} & \xmark & Covid-19 & EU
	\\
	2020-03-11 & COVID-19 as pandemic (WHO) & \cmark & Covid-19 &   WHO
	\\
	2020-11-09& COVID-19 vaccines are ~90\% effective & \cmark & Covid-19 &  Media\\
	2021-09-07 & BTC legal tender in El Salvador & \cmark & Regulation &  Alvarez et al. \\
	2021-09-24 & China crypto ban & \cmark & Regulation &  Coinmarketcap  \\
	2022-02-24 & Russian invasion of Ukraine  & \cmark & War & Media
	\\
	2022-04-27 & BTC legal tender in CAR & \cmark & Regulation &   Coinmarketcap \\ 
	2023-10-07 & Gaza terrorist attack to Israel  & \cmark & War & Media
	\\
	2023-10-27 & Israel invades Gaza  & \xmark & War &  Media
	\\
	2024-11-05 & Trump elected US President  & \cmark & Regulation &  Media
	\\ 
	2025-01-29 & El Salvador amends BTC law & \xmark & Regulation &  Media \\
	\midrule
	\multicolumn{2}{l}{\hspace{0.5 cm} Regulatory events} & &  &    \\
	\midrule
	2020-09-01 & Zug announces tax payments in BTC & \xmark & Regulation &  Media \\
	2021-07-01 & Germany approves Fund Locations Act  & \xmark & Regulation & ECB et al. \\
    2023-04-20 & MiCa EU Parliament approval & \cmark & Regulation &  EU
    \\ 
    2024-06-30 & MiCa applies for stablecoins & \cmark & Regulation &  EU
    \\ 
    2024-12-30 & MiCa applies for all cryptoassets & \cmark & Regulation &   EU
    \\ 
	\bottomrule
\end{tabular}
}

	\caption{\textbf{Timeline of key crypto-related events in the cryptocurrency sector, 2020-2025.} Economic shocks, stimuli, monetary policy; geopolitical events; regulatory events.}
	\label{tab:events_timeline}
\end{table}

\newpage
\clearpage

\section{CASP Volume Flows}
\label{app:casp_network}

This section provides additional information on the cryptoasset flow analysis in Section~\ref{sec:flows}.

\subsection{CASP-to-Market Activity}

\begin{figure}[htbp]
	\centering
	\begin{subfigure}[b]{0.4\textwidth}
		\centering
		\includegraphics[width=\columnwidth]{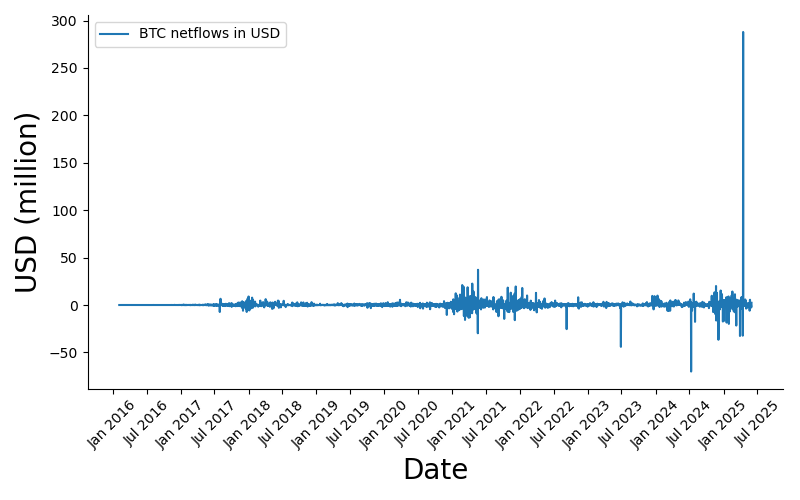}
		\caption{\label{fig:netflow_btc}Bitcoin}
	\end{subfigure}
	\begin{subfigure}[b]{0.4\textwidth}
		\centering
		\includegraphics[width=\columnwidth]{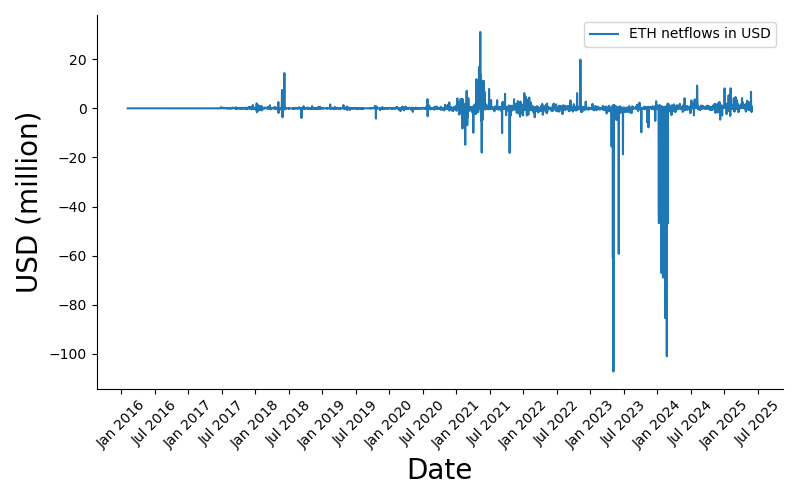}
		\caption{\label{fig:netflow_eth}Ethereum}
	\end{subfigure}
	\begin{subfigure}[b]{0.4\textwidth}
		\centering
		\includegraphics[width=\columnwidth]{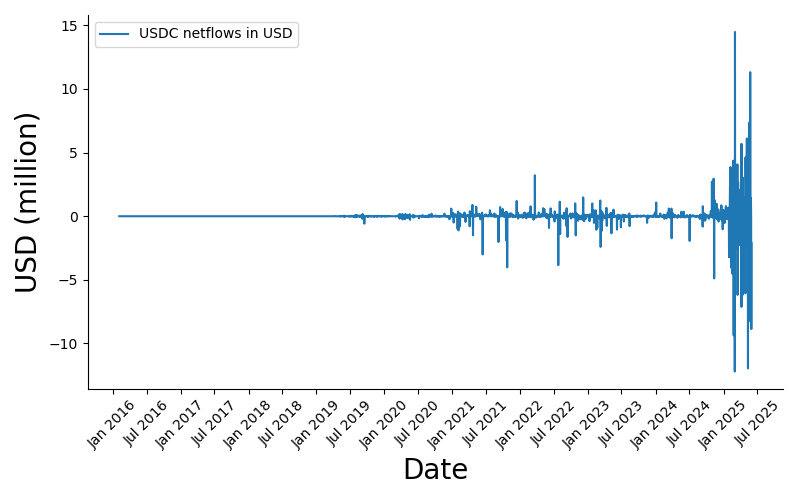}
		\caption{\label{fig:netflow_usdc}USDC}
	\end{subfigure}
	\begin{subfigure}[b]{0.4\textwidth}
		\centering
		\includegraphics[width=\columnwidth]{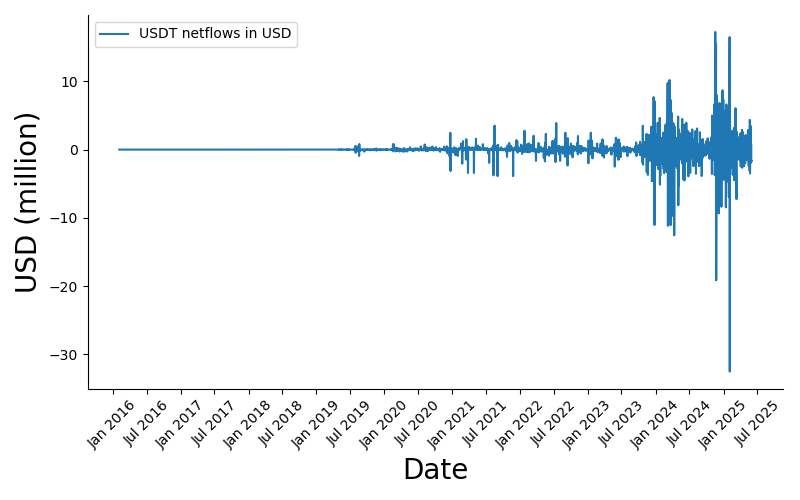}
		\caption{\label{fig:netflow_usdt}USDT}
	\end{subfigure}
	\caption{\textbf{Daily net flows of BTC, ETH, USDC, USDT in Austrian CASPs.} Each plot reports the net flows for one cryptoasset. The y-axis is in millions of US dollars.} 
	\label{fig:netflows}
\end{figure}

For each asset (BTC, ETH, USDT, and USDC), we construct hourly and daily time series of inflows, outflows, net flows, and number of transactions. 
Figure~\ref{fig:netflows} shows net flows by cryptoasset. Three patterns emerge. First, Bitcoin accounts for the largest observed trading volume magnitudes among the four assets. Second, stablecoins played a marginal role until 2023, after which their daily variation increased substantially. Third, daily net flows fluctuate without a uniform directional trend, which serves as a quality check on the underlying data.

Table~\ref{tab:size_distrib} reports descriptive statistics for values aggregated by unique transaction identifier and separated by asset. The distributions are highly right-skewed: most transactions range from hundreds to thousands of dollars, while aggregate volumes are driven by a limited number of very large transactions. The pattern is particularly pronounced for Bitcoin, with the maximum transaction value exceeding by one order of magnitude that of the other assets.

\begin{table}[htbp]
	\scriptsize
	\centering
	\begin{tabular}{lrrrrrrrrr}
	\toprule
	Curr. & Count & Mean & St. Dev. & 10\% & 25\% & 50\% & 75\% & 90\% & Max \\
	\midrule
	BTC  & \num{1351986} & \num{12339.75} & \num{333377.74} & \num{33.58} & \num{111.34} & \num{544.03} & \num{2616.32} & \num{20749.47} & \num{287525662.21} \\
	ETH  & \num{1383287} & \num{3539.92}  & \num{84402.38}  & \num{22.00} & \num{65.03}   & \num{206.44} & \num{712.44}  & \num{5292.05}  & \num{20843506.00} \\
	USDC & \num{85255}   & \num{12951.17} & \num{175071.75} & \num{21.58} & \num{99.99}   & \num{443.76} & \num{1530.84} & \num{12305.15} & \num{14899712.11} \\
	USDT & \num{394115}  & \num{17834.19} & \num{163148.29} & \num{42.13} & \num{111.58}  & \num{449.89} & \num{1384.37} & \num{13582.27} & \num{13509941.63} \\
	\bottomrule
\end{tabular}
	\caption{\textbf{Transaction values aggregated by transaction identifier}. Descriptive statistics. Values are in US dollars.}
\label{tab:size_distrib}
\end{table}

\subsection{CASP-to-CASP}

\begin{figure}[htbp]
	\centering
	\begin{subfigure}{0.4\textwidth}
		\includegraphics[width=\columnwidth]{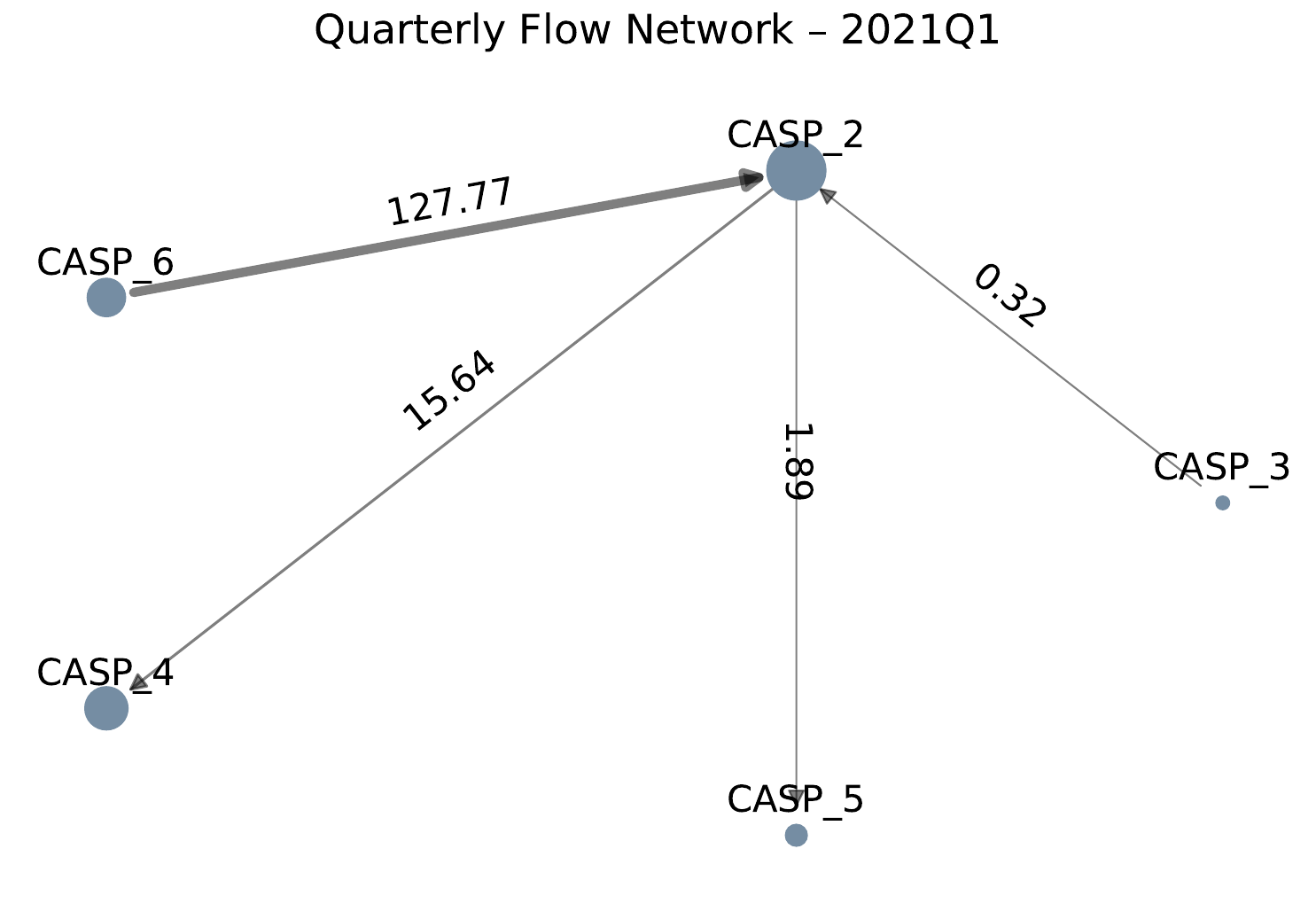}
		\subcaption{}
	\end{subfigure}
	\begin{subfigure}{0.4\textwidth}
		\includegraphics[width=\columnwidth]{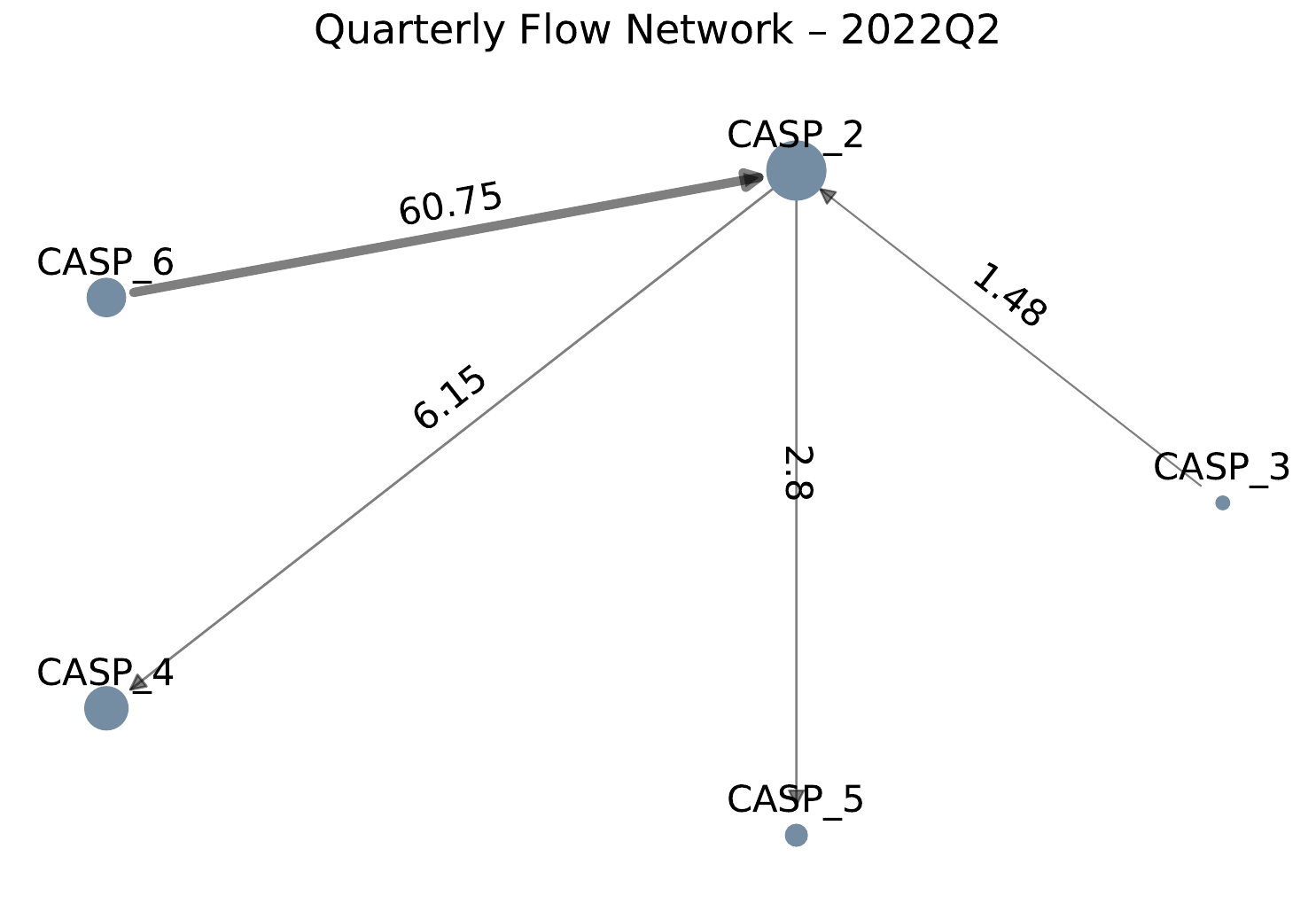}
		\subcaption{}
	\end{subfigure}
	\begin{subfigure}{0.4\textwidth}
		\includegraphics[width=\columnwidth]{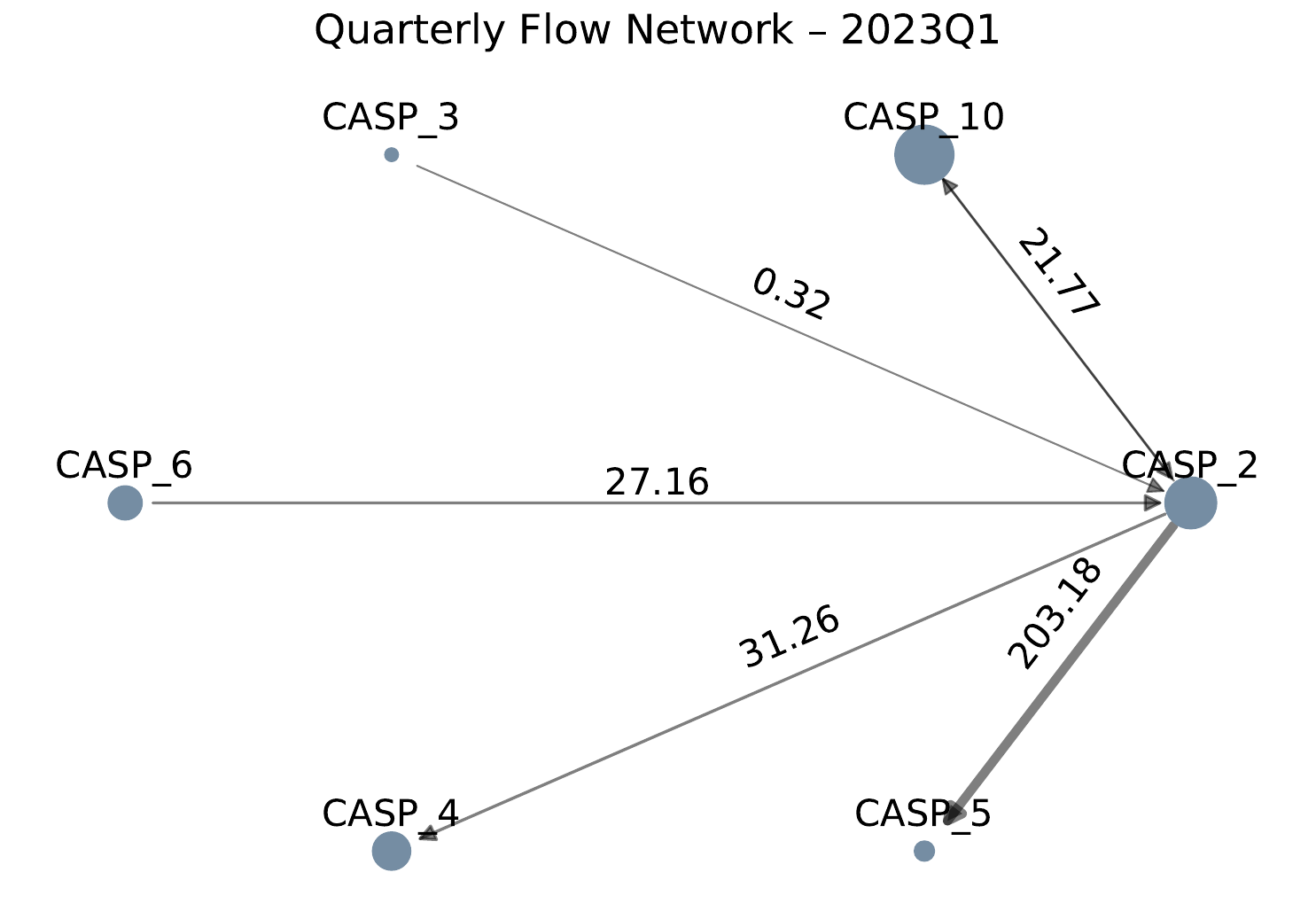}
		\subcaption{}
	\end{subfigure}
	\begin{subfigure}{0.4\textwidth}
		\includegraphics[width=\columnwidth]{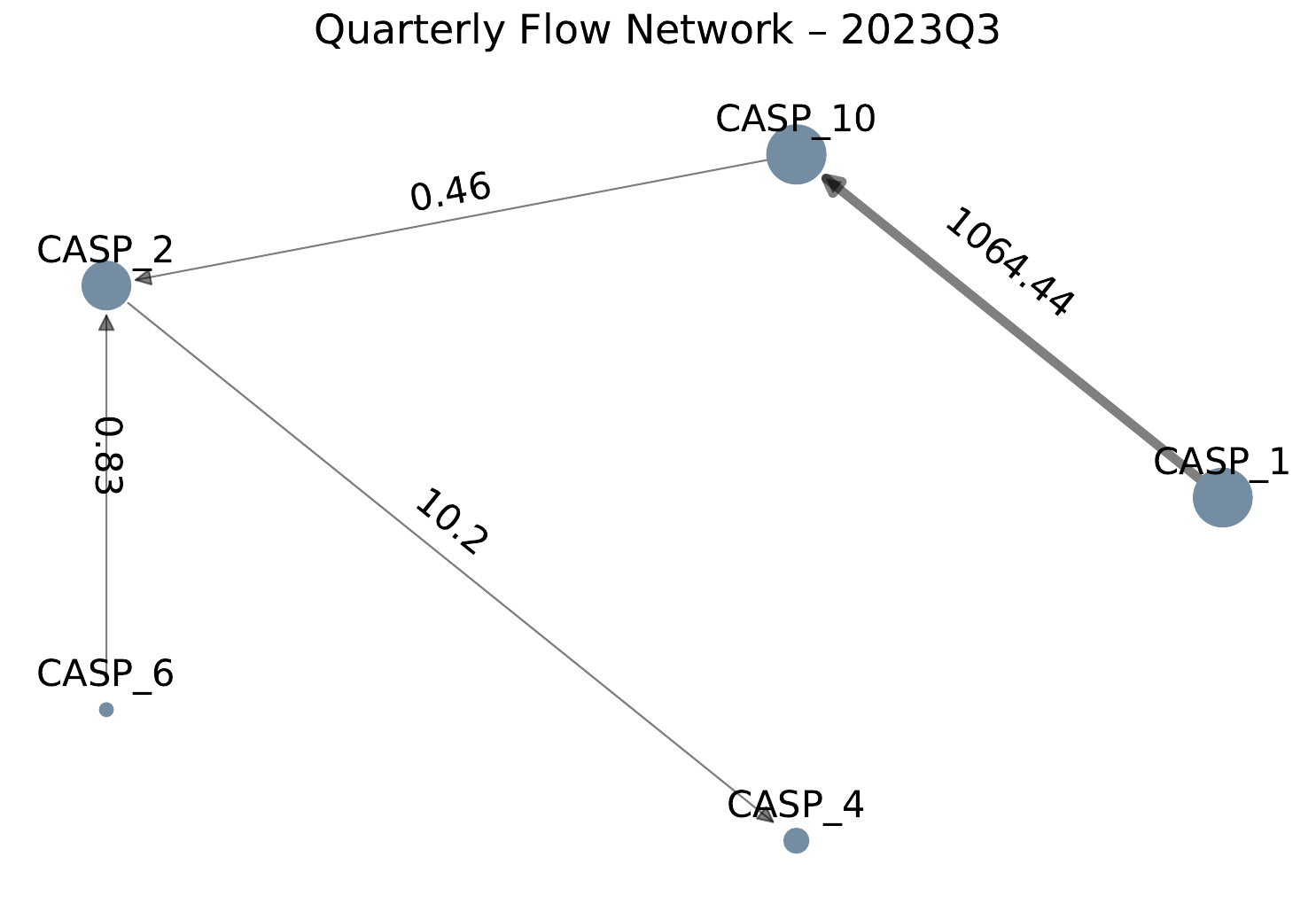}
		\subcaption{}
	\end{subfigure}
	\caption{\textbf{CASP-to-CASP interaction network.} Illustrative examples of interactions across Austrian CASPs aggregated at the quarterly level. Top left: 2021, Q1; top right: 2022, Q2; bottom left: 2023, Q1; bottom right: 2023, Q3. Nodes represent CASPs, and their size is proportional to their relevance in the network. Edges represent value flows; the magnitude is in thousands of USD, and the arrow dimension is proportional to the magnitude of the flow. Arrow heads indicate flow direction and can be bidirectional (in this case, the magnitude is the sum of the two directions).}
	\label{fig:ctc}
\end{figure}

Table~\ref{tab:ctc} reports the Austrian CASP-to-CASP interactions broken down by year and the relative Bitcoin dominance (where 1 indicates that all flows are in Bitcoin and 0 that no flows are in Bitcoin).
Tables~\ref{tab:ctc1} and~\ref{tab:ctc2} respectively report the number of interactions and the volumes moved, broken down by asset and year. 
To produce a visual representation, we construct a network where the nodes are the CASPs, and the edges are the cryptoasset transfers, aggregated at the quarterly level. Figure~\ref{fig:ctc} shows illustrative examples for four different quarters: 2021Q1, 2022Q2, 2023Q1, and 2023Q3.

\begin{table}[h]
	\small
	\centering
	\begin{tabular*}{\textwidth}{@{\extracolsep{\fill}}lccrc}
		\toprule
		Year & Number of interactions & BTC dominance & USD Value & BTC dominance \\
		\midrule
		2019 & 70  & (1.00) & \num{22420}   & (1.00) \\
		2020 & 717 & (0.98) & \num{373578}  & (0.97) \\
		2021 & 851 & (0.96) & \num{439860}  & (0.94) \\
		2022 & 712 & (0.97) & \num{1719969} & (0.99) \\
		2023 & 334 & (0.84) & \num{1558845} & (0.80) \\
		2024 & 277 & (0.68) & \num{1181264} & (0.76) \\
		2025 & 79  & (0.53) & \num{477751}  & (0.81) \\
		\bottomrule
	\end{tabular*}
	\caption{\textbf{Austrian CASP-to-CASP interactions broken down by year}. Numbers in parentheses indicate the fraction that can be attributed to Bitcoin flows. 2025 covers January to May only.}
	\label{tab:ctc}
\end{table}

\begin{table}[h]
	\footnotesize
	\centering
	\begin{tabular*}{\textwidth}{@{\extracolsep{\fill}}lccccc}
		\toprule
		Year & BTC & ETH & USDC & USDT & Total \\
		\midrule
		2019 & 70 & 0 & 0 & 0 & 70 \\
		2020 & 708 & 8 & 0 & 1 & 717 \\
		2021 & 824 & 27 & 0 & 0 & 851 \\
		2022 & 694 & 12 & 4 & 2 & 712 \\
		2023 & 281 & 28 & 10 & 15 & 334 \\
		2024 & 190 & 67 & 5 & 15 & 277 \\
		2025 & 42 & 26 & 11 & 0 & 79 \\
		Total & 2809 & 168 & 30 & 33 & 3040 \\
		\bottomrule
	\end{tabular*}
	\caption{\textbf{Number of CASP-to-CASP interactions broken down by cryptoasset and year.} 2025 covers January to May only.}
	\label{tab:ctc1}
\end{table}

\begin{table}[htbp]
	\footnotesize
	\centering
	\begin{tabular*}{\textwidth}{@{\extracolsep{\fill}}lccccc}
		\toprule
		Year & BTC & ETH & USDC & USDT & Total \\
		\midrule
		2019 & \num{22420} & \num{0} & \num{0} & \num{0} & \num{22420} \\
		2020 & \num{362795} & \num{9551} & \num{0} & \num{1232} & \num{373578} \\
		2021 & \num{416329} & \num{23531} & \num{0} & \num{0} & \num{439860} \\
		2022 & \num{1703211} & \num{8972} & \num{7443} & \num{342} & \num{1719969} \\
		2023 & \num{1253409} & \num{268179} & \num{13918} & \num{23339} & \num{1558845} \\
		2024 & \num{908742} & \num{209927} & \num{17665} & \num{44930} & \num{1181264} \\
		2025 & \num{391626} & \num{62387} & \num{23739} & \num{0} & \num{477751} \\
		Total & \num{5058532} & \num{582546} & \num{62765} & \num{69842} & \num{5773686} \\
		\bottomrule
	\end{tabular*}
	\caption{\textbf{CASP-to-CASP volumes broken down by cryptoasset and year.} 2025 covers January to May only.}
	\label{tab:ctc2}
\end{table}

\newpage

\subsection{Within-CASP}

This section includes additional details on the within-CASP analysis.
We begin by describing the algorithms used to identify the patterns described in Section~\ref{sec:flows}, focusing on the identification of chained movements across addresses controlled by the same CASP. 
These algorithms follow a two-step
process. In a first step, we construct network-based indicators to identify the presence of hub
wallets within a CASP. Algorithm~\ref{alg:hub_wallets} describes the procedure. We first construct for each CASP and asset a directed graph $G(E,V,W,C)$ where $V$ is the set of wallet addresses, and $E$ is the set of edges; a link exists if two addresses are recipient and sender in the same asset transfer. $W$ and $C$ represent the strength of the link and respectively measure the overall volume moved among address pairs and the number of interactions. We then compute a number of network statistics on G and identify, after manual inspection, the list of candidate hub wallets. 

In a second phase, we detect chained movements across addresses controlled by the same CASP, where funds are ultimately routed to a hub wallet. Algorithm~\ref{alg:schema_detection} describes the procedure for deposit addresses. For each CASP and asset, we first separate asset transfers directed to hub wallets (set $T_2$) from other asset transfers (set $T_1$); then, we match rows in the two sets such that the source of a row in $T_2$ equals the destination of a row in $T_1$, under the condition that the two legs have the same transaction identifier and volume. We then flag with a binary $[0,1]$ indicator the rows identified through this approach.
We also develop a variant of the main algorithm to account for small time differences and minimal price changes when the chained flows are not executed within the same transaction. 
Finally, we note that Bitcoin and Ethereum chains work differently (the former is UTXO-based, while the latter is account-based). Since one Bitcoin transaction typically consists of multiple input and output UTXOs, we first preprocess the asset transfer list by aggregating deposits with the same transaction identifier and originating from the same address, as well as withdrawals with the same transaction identifier and directed to the same address.

With one exception, we did not observe changes in address usage from hot to warm or cold over time. One hub wallet shifted from hot-wallet use to cold-wallet storage for large transactions. However, the change in its transaction activity is abrupt, and therefore it was possible to separate its activity before and after the change. This pattern is documented in the codebase and related documentation. 
We also note that the algorithms described above enabled us to identify a pattern in Bitcoin transactions whereby funds are withdrawn immediately after being deposited by the same actors, for a total of \$300 million. We interpret this behavior as the result of the use of change addresses.

\begin{figure}[htbp]
	\centering
	\begin{subfigure}[b]{0.45\textwidth}
		\centering
		\includegraphics[width=\columnwidth]{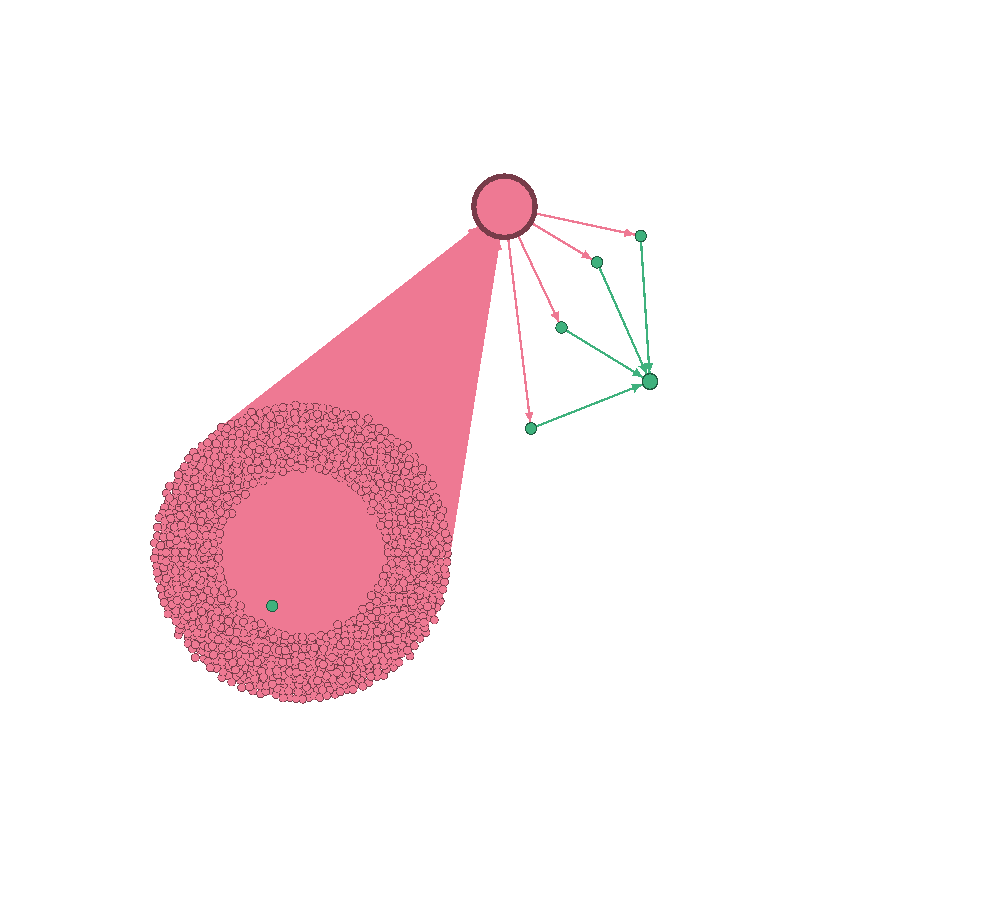}
		\caption{\label{fig:8f2068_btc_within}Bitcoin}
	\end{subfigure}
	\begin{subfigure}[b]{0.45\textwidth}
		\centering
		\includegraphics[width=\columnwidth]{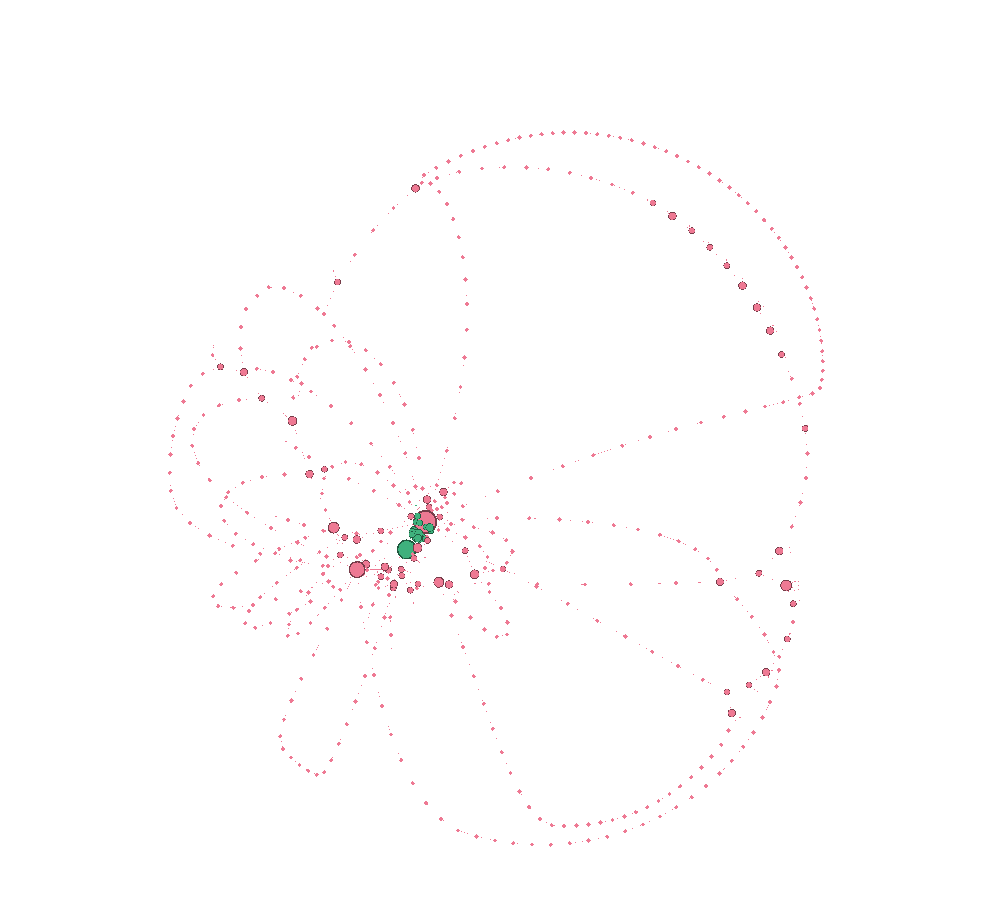}
		\caption{\label{fig:90ea19_btc_within}Bitcoin}
	\end{subfigure}
	\begin{subfigure}[b]{0.45\textwidth}
		\centering
		\includegraphics[width=\columnwidth]{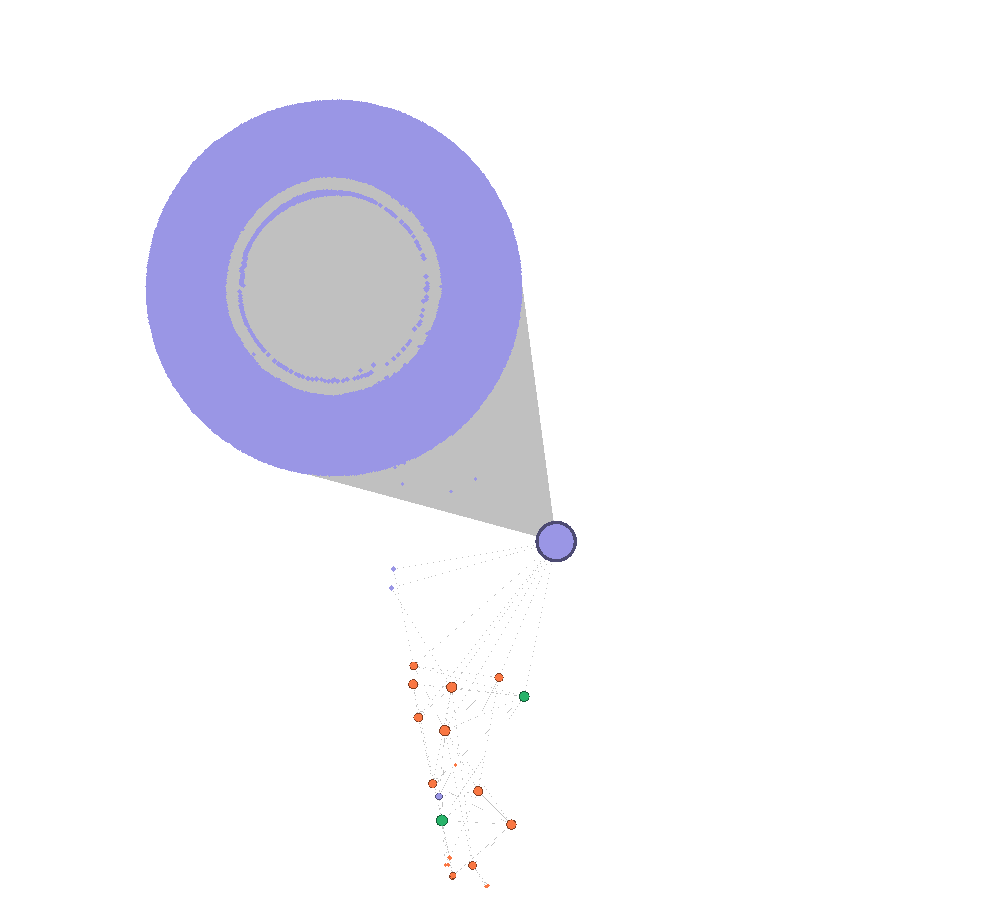}
		\caption{\label{fig:dc5f8e_eth_within}Ethereum}
	\end{subfigure}
	\begin{subfigure}[b]{0.45\textwidth}
		\centering
		\includegraphics[width=\columnwidth]{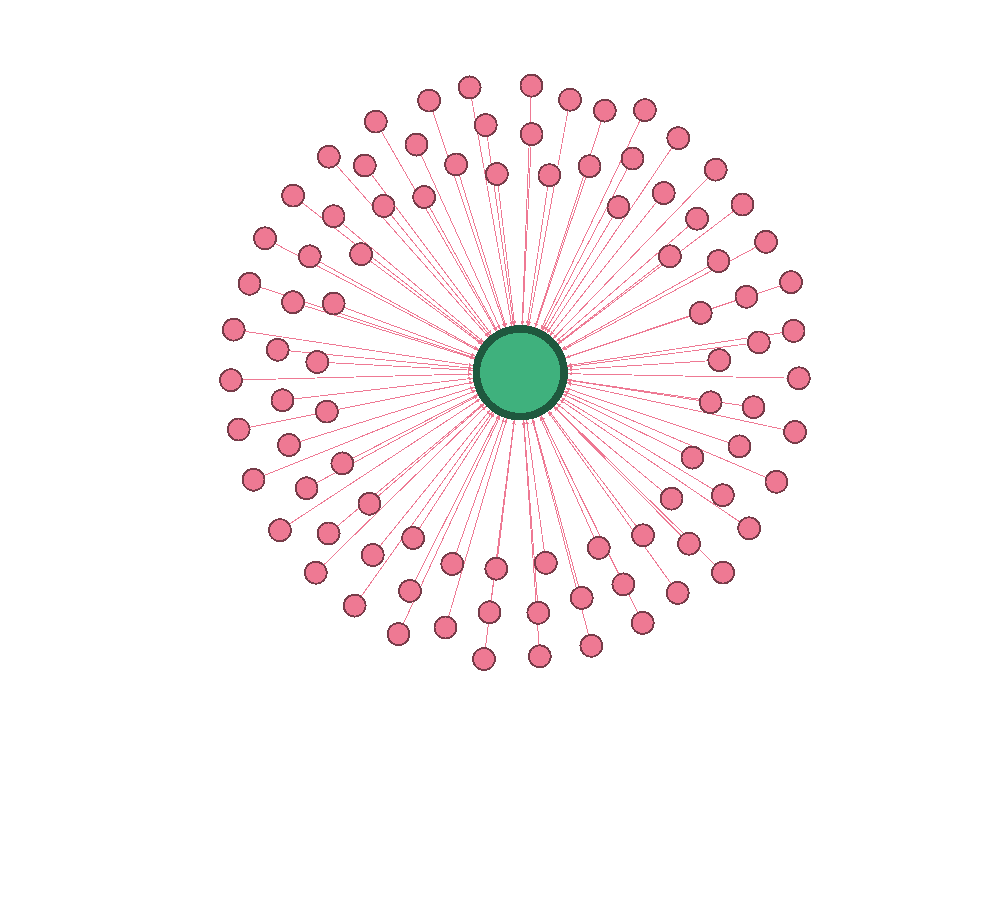}
		\caption{\label{fig:474f53_eth_within}Ethereum}
	\end{subfigure}
	\begin{subfigure}[b]{0.45\textwidth}
		\centering
		\includegraphics[width=\columnwidth]{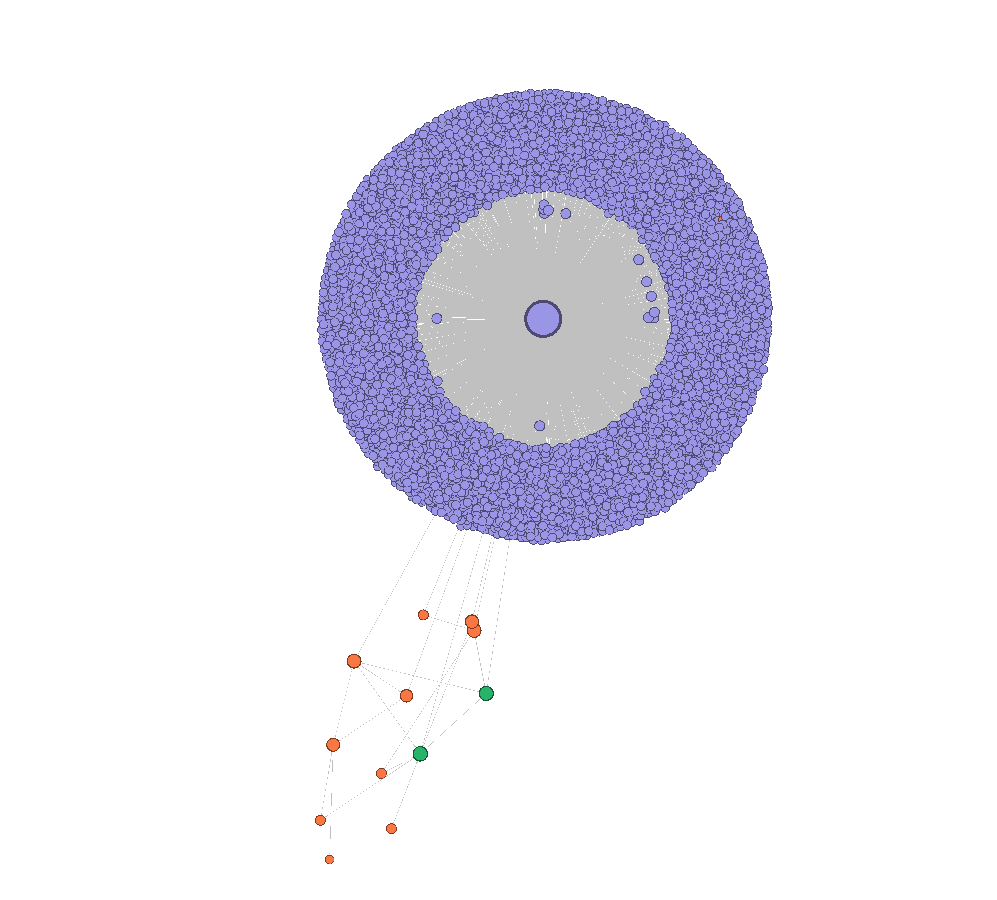}
		\caption{\label{fig:dc5f8e_usdc_within}USDC}
	\end{subfigure}
	\begin{subfigure}[b]{0.45\textwidth}
		\centering
		\includegraphics[width=\columnwidth]{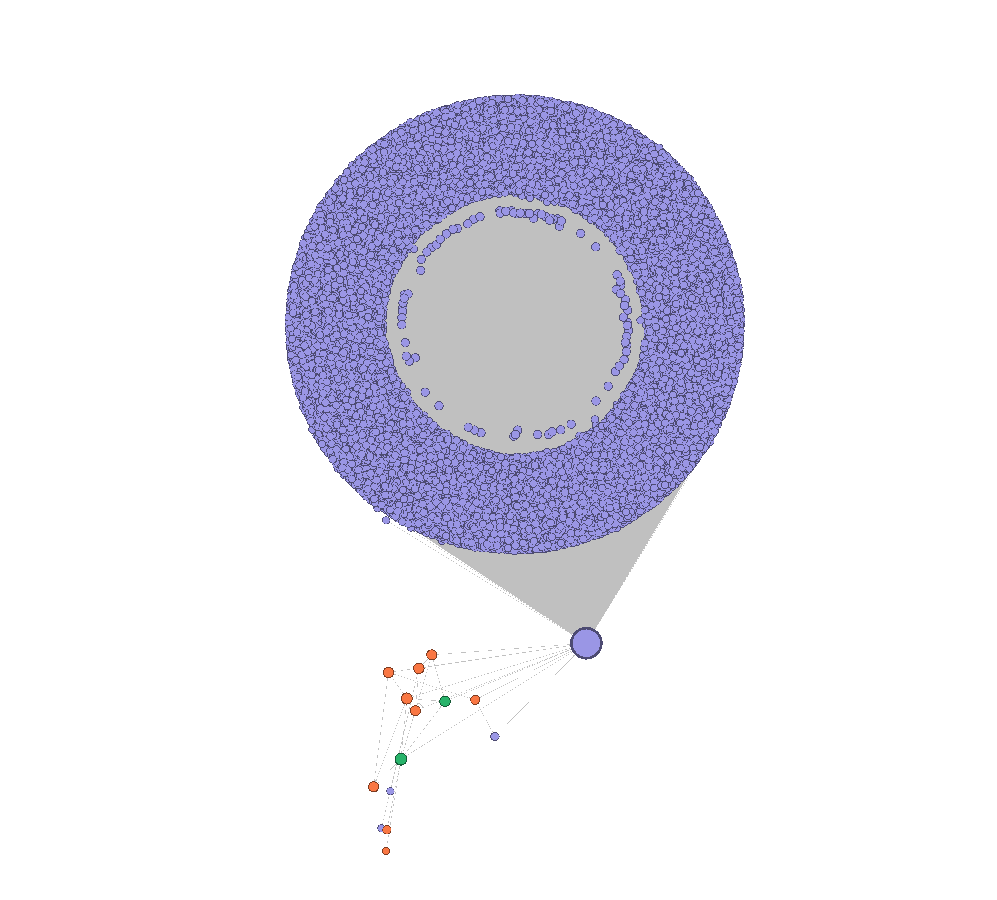}
		\caption{\label{fig:dc5f8e_usdt_within}USDT}
	\end{subfigure}
	\caption{\textbf{Network visualization of within-CASP flows.} Violet and pink indicate hot wallets; orange and green respectively indicate warm and cold wallets.} 
	\label{fig:within_casp_flows}
\end{figure}

Finally, Figure~\ref{fig:within_casp_flows} reports illustrative examples of the within-CASP networks of different CASPs and across different assets. In most of the reported examples, it is possible to identify one hot wallet acting as a hub with many other intermediary hot wallets, and a network of cold or warm wallets connected to the intermediary wallets through the hub wallet (violet and pink indicate hot wallets; orange and green respectively indicate warm and cold wallets).%
We also observed that in a few cases, Bitcoin assets were managed internally following peeling-chain-like patterns (see Panel~\ref{fig:90ea19_btc_within}).

\begin{algorithm}[H]
	\caption{Hub Wallet Identification}
	\label{alg:hub_wallets}
	
	\KwData{(1) Asset transfer list $T = \{(s_i, d_i, v_i)\}$\\
		\hspace{1.25cm} where $s_i = $ sender address,  $d_i = $~destination address, $v_i = $ volume flow\;}
	
	\KwResult{Set of hub wallets $H$}
	
	$P \leftarrow Aggregate(s_i,d_i)$%
	\tcp*{Aggregate transactions by ordered pairs}
	
	\ForEach{$P(s,d)$}{
	$W \leftarrow $ Sum($v_i$)\;
		
	$C \leftarrow $ Count($v_i$)\;
	}
		
	$G(E,V,W,C)$ \tcp*{Construct directed graph G}%
	
	\ForEach{$v \in V$}{
	$N_{ind} \leftarrow $ in-degree centrality\;
	
	$N_{outd} \leftarrow $ out-degree centrality\;
	
	$N_{pr} \leftarrow $ PageRank centrality\;
	
	$N_{ev} \leftarrow $ eigenvector centrality\;
	
	$N_{bw} \leftarrow $ betweenness centrality\;
		
	$N \leftarrow [N_{ind},N_{outd},N_{pr},N_{ev},N_{bw}] $ }

	$H \leftarrow$ filter($V$, by=N) \tcp*{Identify hub wallets based on high values in $N$}
	
	\KwRet{$H$}
	
\end{algorithm}

\begin{algorithm}[H]
	\caption{Matching Transaction Identification: `pattern 1'}
	\label{alg:schema_detection}
	
	\KwData{(1) Asset transfer list 
		$T = \{(tx_i,s_i,d_i,v_i,t_i)\}$,
		where $tx_i = $ transaction identifier,
		$s_i = $~sender address, $d_i = $ destination address,
		$v_i = $ volume flow, $t_i =$~time\; \\ \hspace{1.2cm}(2) $H$ set of hub wallets}
	\KwResult{Label set $L = \{l_i\}, \quad l_i \in \{0, 1\}$}
	\vspace{0.3cm}
	$L \leftarrow \{0\}$ \tcp*{Init. label for all rows}
	
	$Tx \leftarrow \varnothing$ \tcp*{Init. set of matches empty}
	\ForEach{$h \in H$}{
		
		$T_2 \leftarrow$ filter($T$, by=recipient, $d_i = h$) \tcp*{Select rows: recipient is hub wallet}
		
		$T_1 \leftarrow$ filter($T$, by=recipient, $d_i \neq h$) 
		\tcp*{Select other rows}
		
		$M \leftarrow merge(T_1,T_2, on = [tx_i,v_i])$ \tcp*{Match pairs}
		
		$M' \leftarrow$ filter($M$, by=recipient-sender, $d_{i1} = s_{i2}$) \tcp*{Filter chained rows}	
		
		$Tx \leftarrow Tx \cup \{tx_{i} : (i1,i2)\in M'\}$ \tcp*{Retain matches from every hub}	
	}
	
	\ForEach{$tx_i \in T$}{
		\eIf{$tx_i \in Tx$}{$l_i \leftarrow 1$\;}{$l_i \leftarrow 0$\;}
	}
	
	\Return{$L$}\;
	
\end{algorithm}

\newpage

\section{Robustness Checks}
\label{app:robustness}

\begin{table}[h]
    \footnotesize
	\centering
	\begin{tabular*}{\textwidth}{@{\extracolsep{\fill}}llrrrrrr}
	\toprule
	&  & \multicolumn{1}{c}{} & \multicolumn{2}{c}{Avg.\ transactions} & \multicolumn{1}{c}{Avg.\ degree} & \multicolumn{1}{c}{Avg.\ USD} & \multicolumn{1}{c}{Hot wallet} \\
	\cmidrule(lr){4-5}
	Group & Chain & \multicolumn{1}{c}{Addresses} & \multicolumn{1}{c}{overall} & \multicolumn{1}{c}{in sample} & \multicolumn{1}{c}{overall} & \multicolumn{1}{c}{in sample} & \multicolumn{1}{c}{(fraction)} \\
	\midrule
	\multirow{2}{*}{RL}      & ETH & \num{528400}  & 29        & 2.6   & 9         & \num{1566}   & 1.00 \\
	& BTC & \num{1274001} & 10        & 2.7   & 10        & \num{1532}   & 1.00 \\
	\addlinespace
	\multirow{2}{*}{HI}      & ETH & 477           & \num{6049674} & 632   & \num{289096}  & \num{47339}  & 0.16 \\
	& BTC & 149           & \num{2775969} & \num{13991} & \num{1451378} & \num{38903}  & 0.34 \\
	\addlinespace
	\multirow{2}{*}{LI$_c$}& ETH & 547           & 942       & 10.1  & 16        & \num{77747}  & 0.02 \\
	& BTC & \num{2748}   & 5.0       & 1.9   & 19        & \num{387853} & 0.02 \\
	\addlinespace
	\multirow{2}{*}{LI$_a$}& ETH & \num{12619}   & \num{192699} & 25    & \num{18208}   & \num{3216}   & 1.00 \\
	& BTC & \num{43750}  & \num{12467}  & 83    & \num{7814}    & 680          & 1.00 \\
	\bottomrule
\end{tabular*}
	\caption{\textbf{Classification robustness to threshold and activity measure}. Addresses are grouped into retail-like (RL), high-confidence institutional (HI), and low-confidence institutional (LI$_c$, LI$_a$), separately for each chain (Bitcoin, Ethereum). We use an alternative specification with a looser activity threshold (z-score > 2.5) and network degree centrality instead of transaction count. The reported group profiles are broadly comparable to those in Table~\ref{tab:users_by_group}; both the threshold and the activity measure change in this check.}
	\label{tab:users_by_group_alt}
\end{table}

This appendix reports robustness checks for the classification in Section~\ref{sec:flows} and the event study in Section~\ref{sec:event_study}. 
First, we report the classification results under an alternative specification, using a looser activity threshold (z-score > 2.5 instead of > 3) and network degree centrality in place of transaction count as the activity signal. Table~\ref{tab:users_by_group_alt} shows that the resulting groups are comparable to the baseline classification in Table~\ref{tab:users_by_group}: retail-like (RL) addresses still dominate by count, and the low-confidence institutional subgroups (LI$_c$, LI$_a$) retain distinct profiles. The looser threshold and degree-based measure identify more addresses as active;
the qualitative conclusions in Section~\ref{sec:flows} are unaffected.

\begin{figure}
	\centering
	\includegraphics[width=\textwidth]{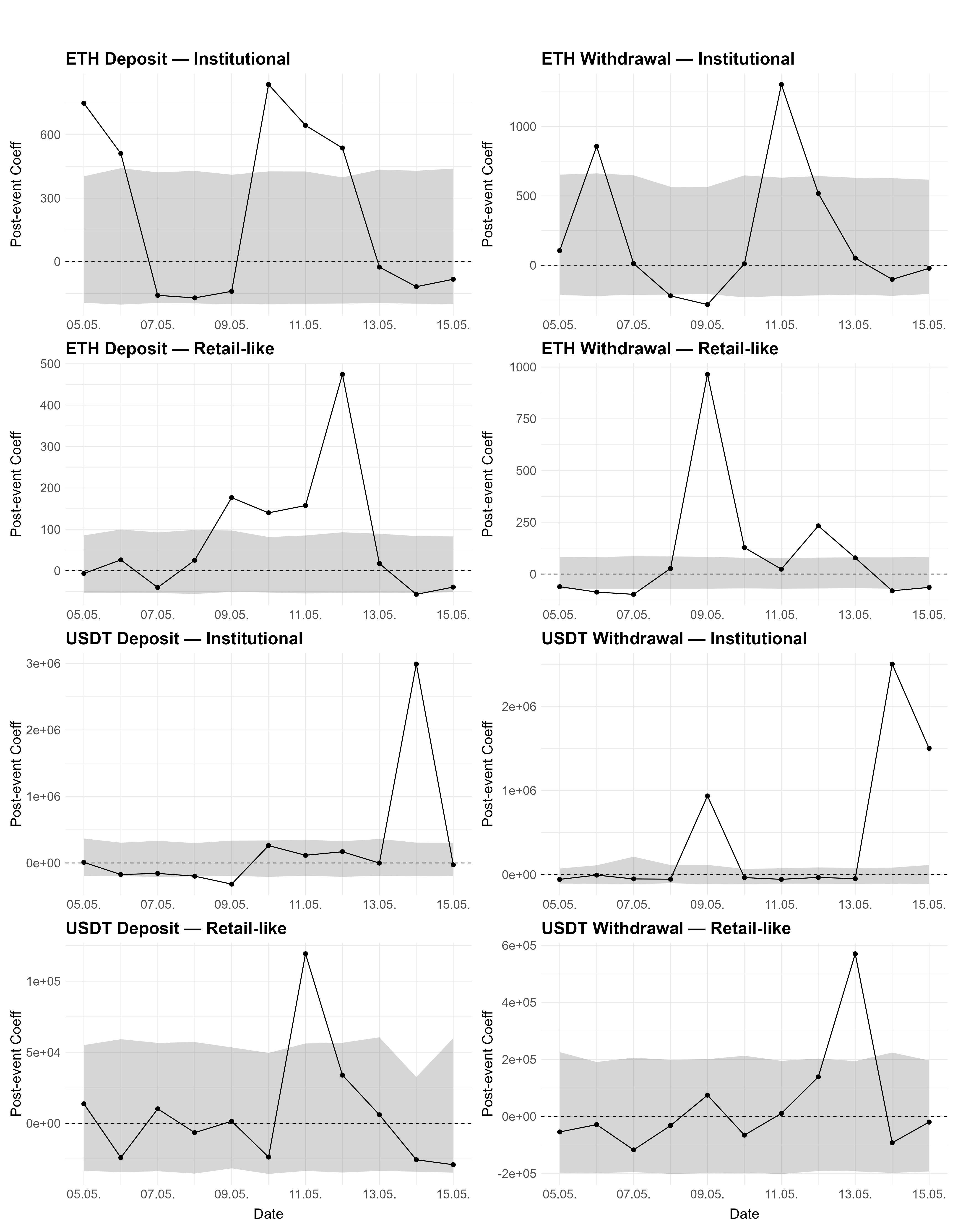}
	\caption{\textbf{Terra Luna crash: alternative asset-flow responses}. {\small \textit{Notes:} post-event coefficients from the Terra Luna event-study regressions, in asset units. Each panel shows coefficients for days 0 to 10 after the event, using a 45-day pre-event estimation window. The dashed horizontal line marks zero. Shaded areas show bootstrap bands based on 1,000 replications.}}
	\label{figure:terra_luna_selected_asset_app}
\end{figure}

\begin{figure}
	\centering
	\includegraphics[width=\textwidth]{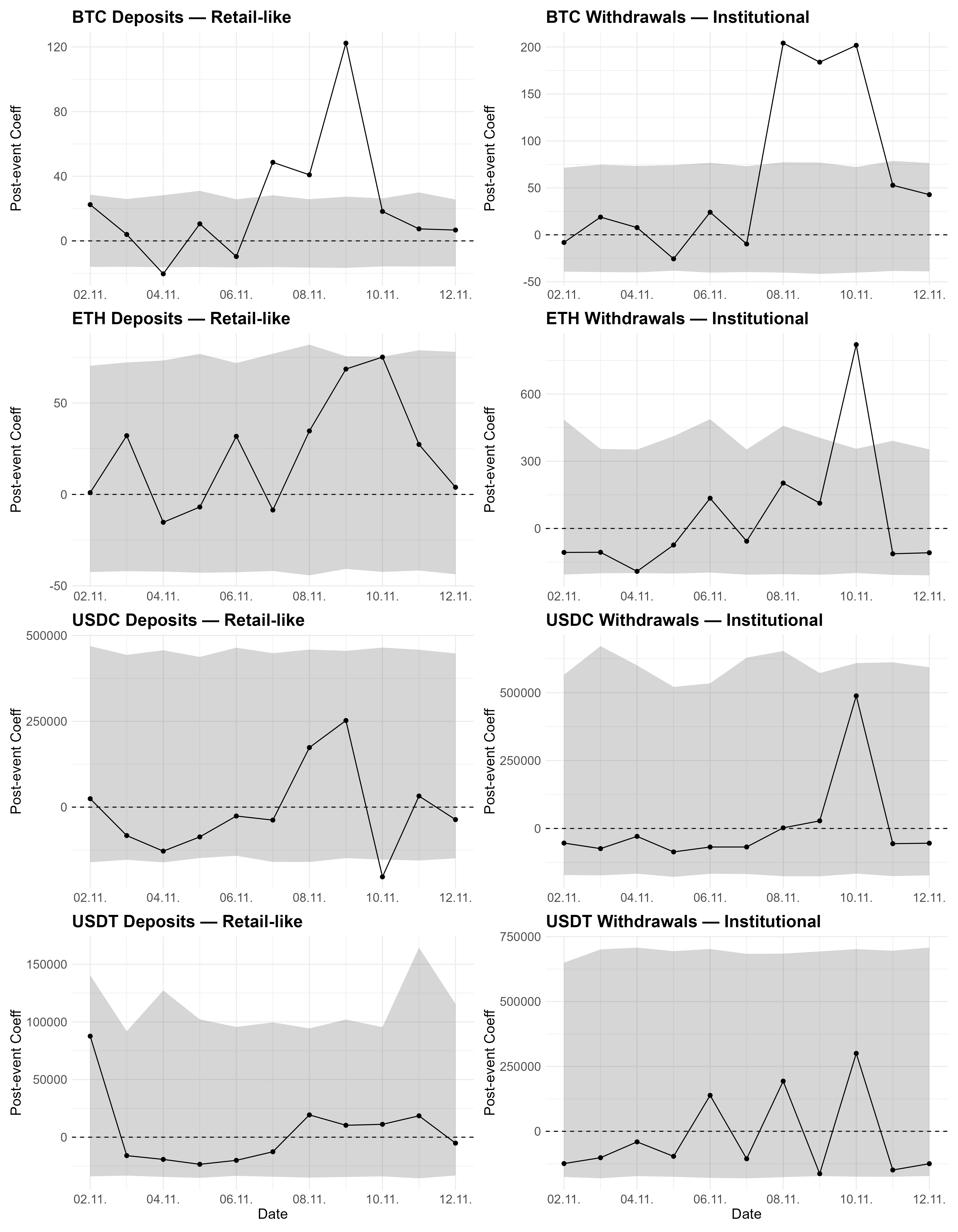}
	\caption{\textbf{FTX collapse: alternative asset-flow responses}. {\small \textit{Notes:} post-event coefficients from the FTX collapse event-study regressions, in asset units. Each panel shows coefficients for days 0 to 10 after the event, using a 45-day pre-event estimation window. The dashed horizontal line marks zero. Shaded areas show bootstrap bands based on 1,000 replications.}}
	\label{figure:ftx_selected_asset_app}
\end{figure}

\begin{figure}
	\centering
	\includegraphics[width=\textwidth]{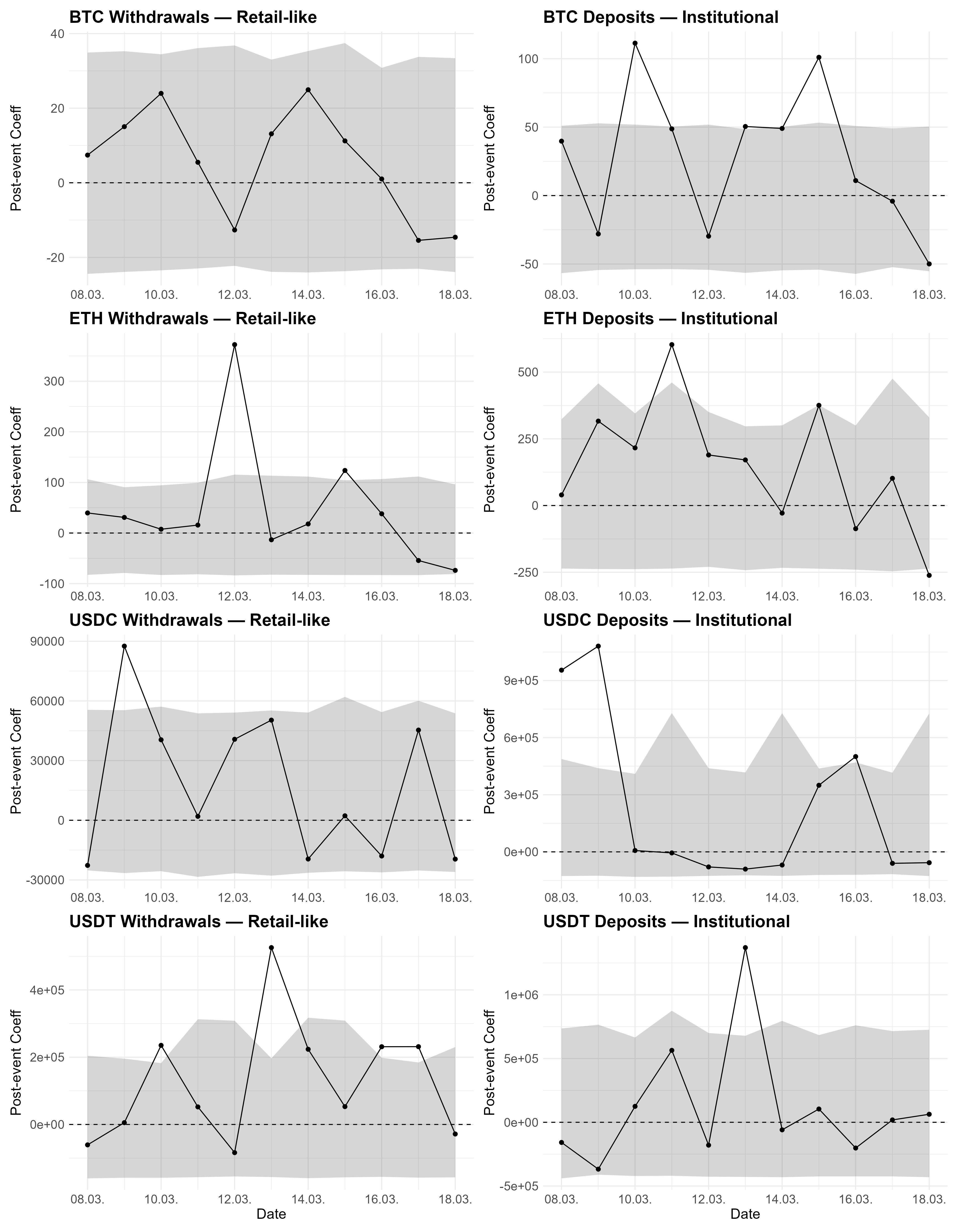}
	\caption{\textbf{SVB failure: alternative asset-flow responses}. {\small \textit{Notes:} post-event coefficients from the SVB failure event-study regressions, in asset units. Each panel shows coefficients for days 0 to 10 after the event, using a 45-day pre-event estimation window. The dashed horizontal line marks zero. Shaded areas show bootstrap bands based on 1,000 replications.}}
	\label{figure:svb_failure_selected_asset_app}
\end{figure}

Next, we report the full set of asset-flow responses for each event, completing the figures shown in the main text (Figures~\ref{figure:EventStudyTerraLunaCrash} to~\ref{figure:EventStudySVBFailure}), so that every combination of asset, group, and flow direction is covered. For the Terra Luna crash, Figure~\ref{figure:terra_luna_selected_asset_app} adds ETH and USDT, again split by group and flow direction; the ETH response echoes BTC, with institutional deposits and withdrawals both rising after May 10, while USDT, like USDC, is volatile and lacks a persistent direction. For the FTX collapse, Figure~\ref{figure:ftx_selected_asset_app} reports the complementary margins (retail-like deposits and institutional withdrawals) across all four assets; these are weaker and less systematic than the ones emphasized in the main text, confirming that the adjustment was concentrated in retail-like withdrawals and institutional deposits. For the SVB failure, Figure~\ref{figure:svb_failure_selected_asset_app} reports retail-like withdrawals and institutional deposits; BTC and ETH remain largely unresponsive on these margins as well, apart from an increase in ETH retail-like withdrawals on March 12, and the two-sided USDC flow pattern is clearest in the retail-deposit, institutional-withdrawal margins. In all cases, findings are consistent with the interpretation given in Section~\ref{sec:event_study}.

\newpage

\begin{figure}[t]
	\centering
	\includegraphics[width=\textwidth]{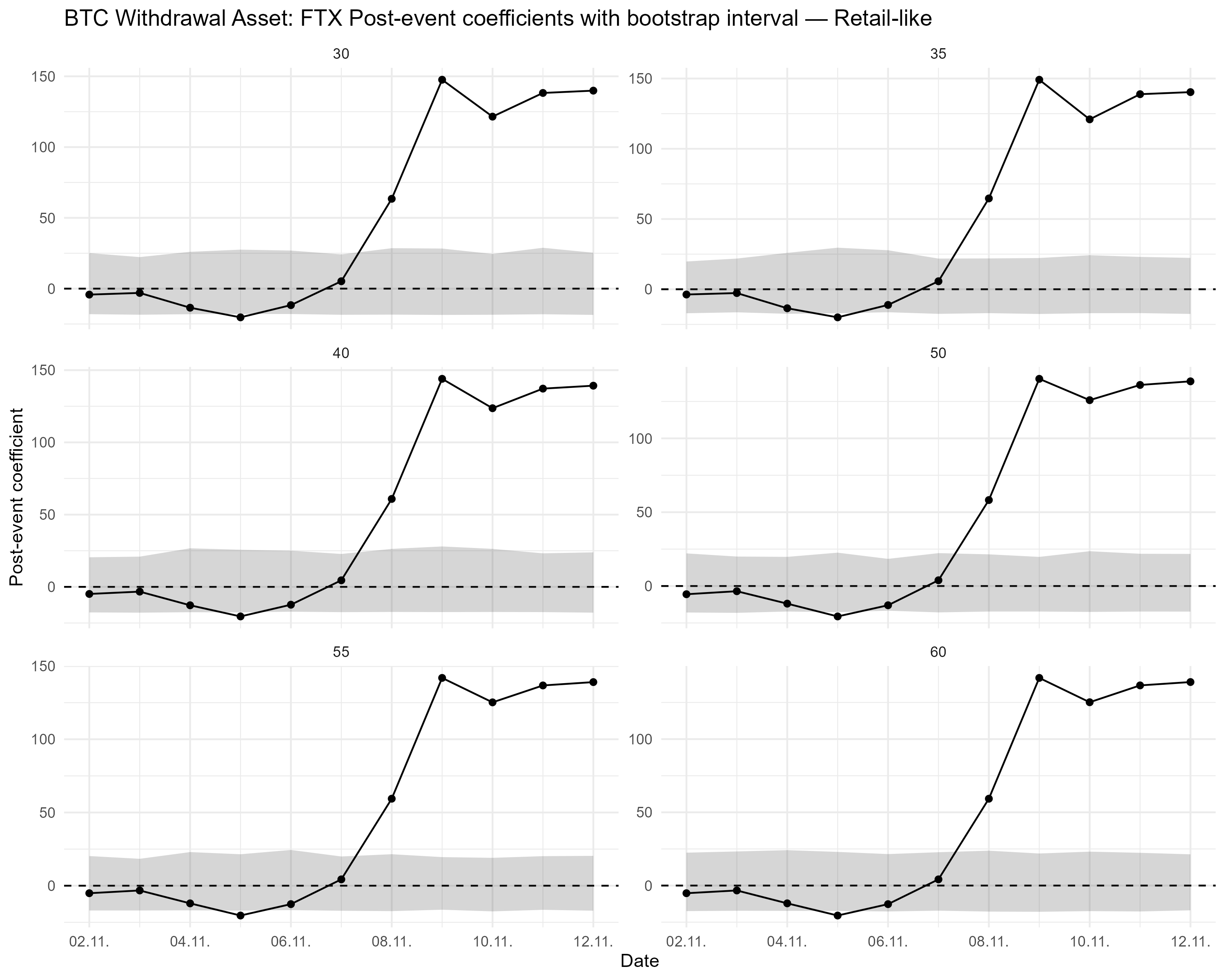}
	\caption{\textbf{FTX collapse: BTC retail-like withdrawals, varying pre-event window}. {\small \textit{Notes:} Post-event coefficients for retail-like BTC withdrawals around the FTX collapse, in asset units. Each panel shows days 0 to 10 using one of six pre-event windows: 30, 35, 40, 50, 55, or 60 days. The dashed horizontal line marks zero. Shaded areas show bootstrap bands based on 1,000 replications. The response for this outcome is similar across the six alternative pre-event windows.}}
	\label{figure:varying_L1}
\end{figure}

We then vary the pre-event estimation window length.
Eq.~(\ref{equ:CASPEventStudy}) estimates normal activity over a fixed 45-day pre-event window. Figure~\ref{figure:varying_L1} re-estimates the response using windows of 30, 35, 40, 50, 55, and 60 days, taking retail-like BTC withdrawals around the FTX collapse as the representative case. The estimated abnormal response is essentially unchanged across all six alternative windows: withdrawals stay within the displayed bootstrap band through November 7 and rise sharply above it from November 8, peaking a few days later. Thus, this result is not an artifact of the baseline window length. We obtain comparable findings in other asset, group, and event specifications.

\begin{figure}
	\centering
	\includegraphics[width=\textwidth]{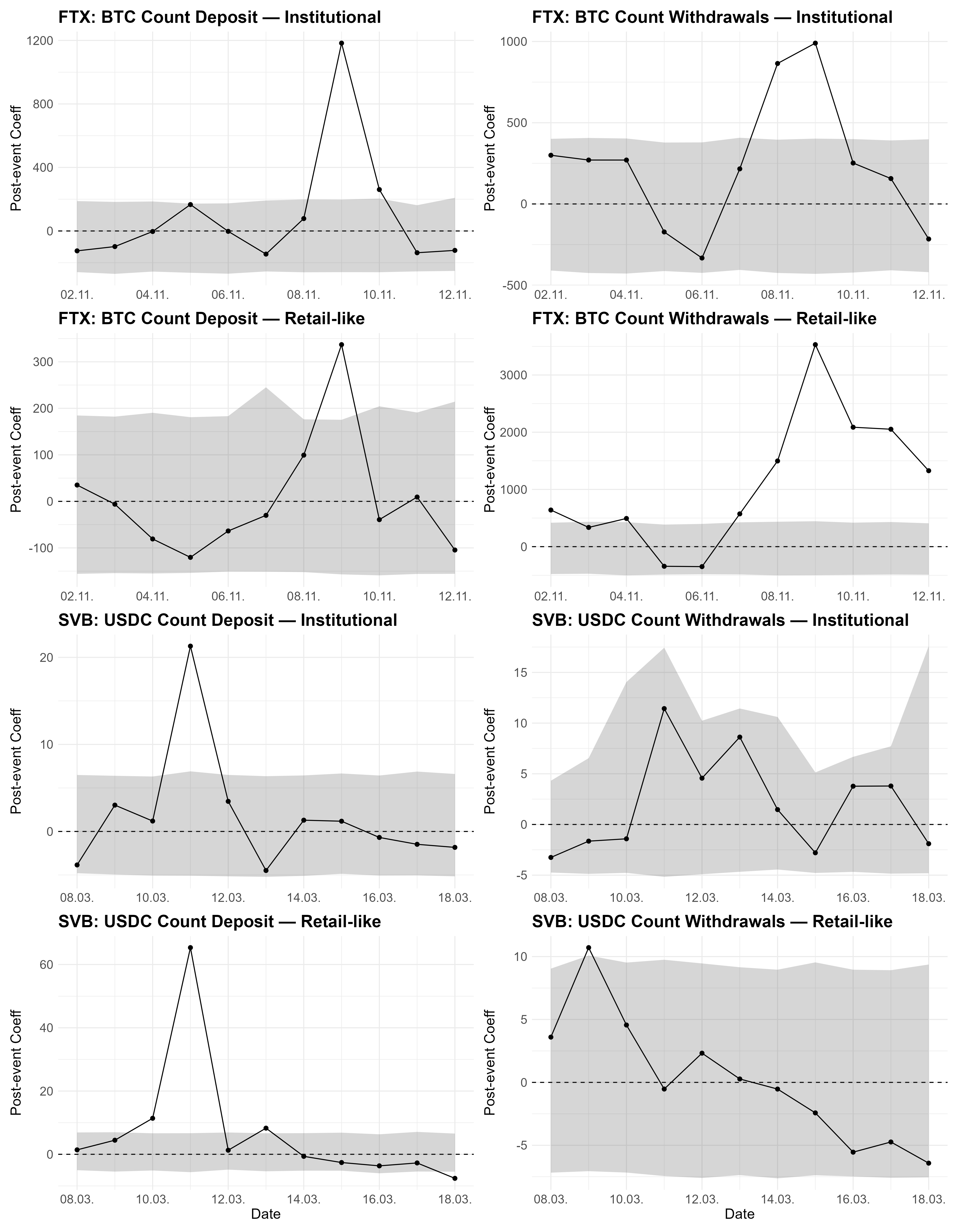}
	\caption{\textbf{FTX and SVB: results with alternative outcomes (transaction count)}. {\small \textit{Notes:} Each panel shows coefficients for days 0 to 10 after the event, using a 45-day pre-event estimation window. The dashed horizontal line marks zero. Shaded areas show bootstrap bands based on 1,000 replications. These results reproduce the patterns obtained with asset-unit outcomes.
    }}
	\label{figure:ftx_svb_selected_asset_app}
\end{figure}

Next, we re-estimate the responses using transaction counts as the outcome variable in place of native asset units (Figure~\ref{figure:ftx_svb_selected_asset_app}). The selected transaction-count panels provide a complementary check on the corresponding results in Section~\ref{sec:event_study}.
The findings reproduce the asset-unit findings: at FTX, retail-like BTC withdrawal counts rise above the band from November 8, while institutional deposit counts spike around November 9–10; at SVB, retail-like USDC deposit counts spike around Circle's disclosure on March 11. Similar patterns in transaction counts indicate that responses to shocks also appear in transfer frequency and are not concentrated in a small number of unusually large transfers.

Finally, we run a placebo test to verify that the abnormal responses documented in Section~\ref{sec:event_study} are specific to the event dates, rather than an artifact of the estimation procedure or of the day-to-day fluctuations that characterize on-chain flows. If the specification mechanically generated abnormal activity, placebo dates without a crisis would display spikes of similar size and frequency.
For each shock, we select a placebo event date exactly one calendar month before the actual event and re-estimate Eq.~(\ref{equ:CASPEventStudy}) with identical estimation and post-event windows. This choice ensures that the placebo window does not overlap with the actual crisis while keeping the comparison sufficiently close in time to limit differences in prevailing market conditions and CASP activity.

\begin{figure}
	\centering
	\includegraphics[width=\textwidth]{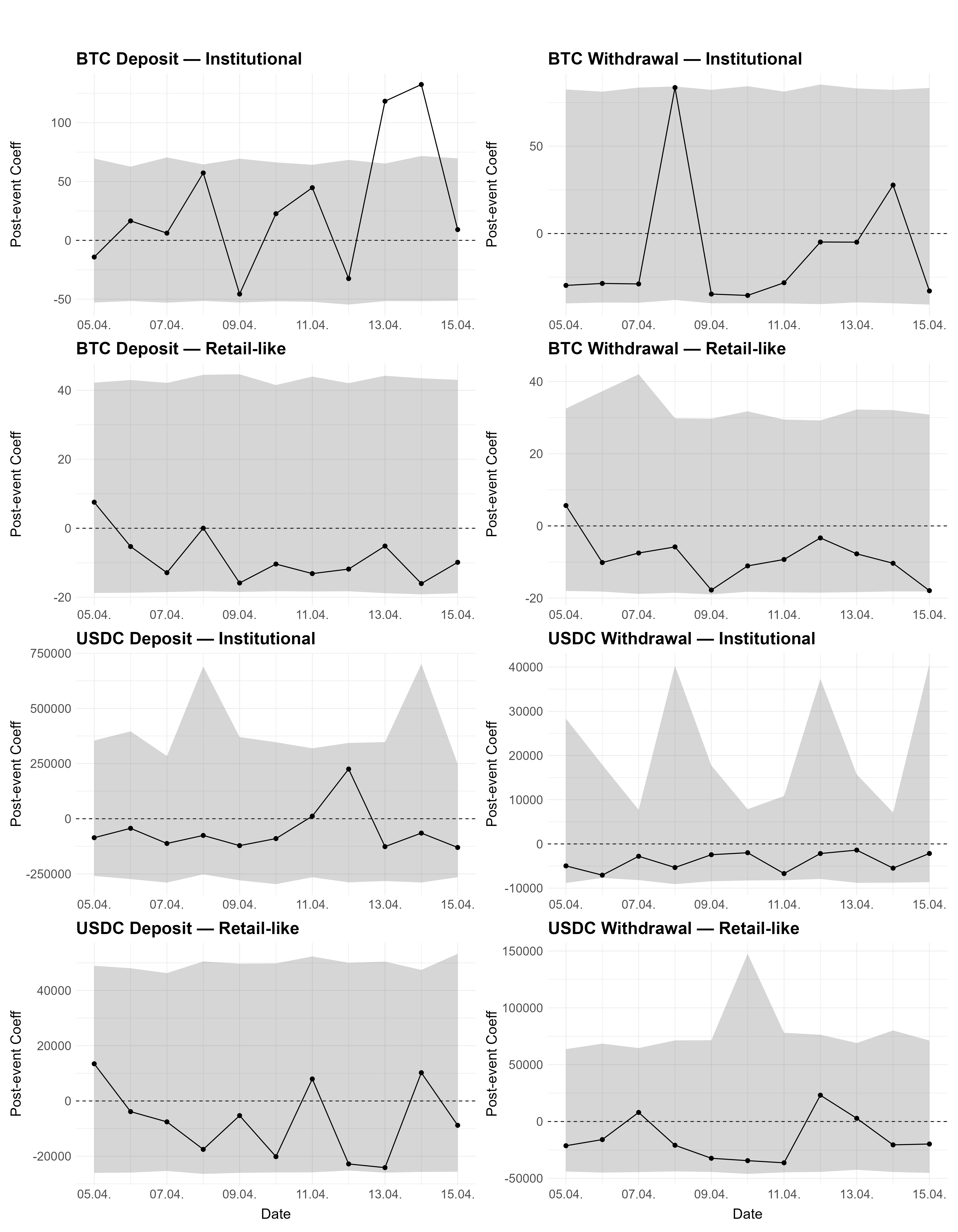}
	\caption{\textbf{Terra Luna crash: placebo test}. {\small \textit{Notes:} Selected post-event coefficients from the Terra Luna crash event-study regressions, in asset units, estimated around a placebo event date of April 5, 2022, one month before the actual event. Each panel shows coefficients for days 0 to 10 after the placebo event, using a 45-day pre-event estimation window. The dashed horizontal line marks zero. Shaded areas show bootstrap bands based on 1,000 replications. Panels correspond to those in Figure~\ref{figure:EventStudyTerraLunaCrash}: rows correspond to BTC and USDC, each split into institutionally mediated and retail-like flows; the left column reports deposits and the right column withdrawals.}}
	\label{figure:placebo_terra_app}
\end{figure}

\begin{figure}
	\centering
	\includegraphics[width=\textwidth]{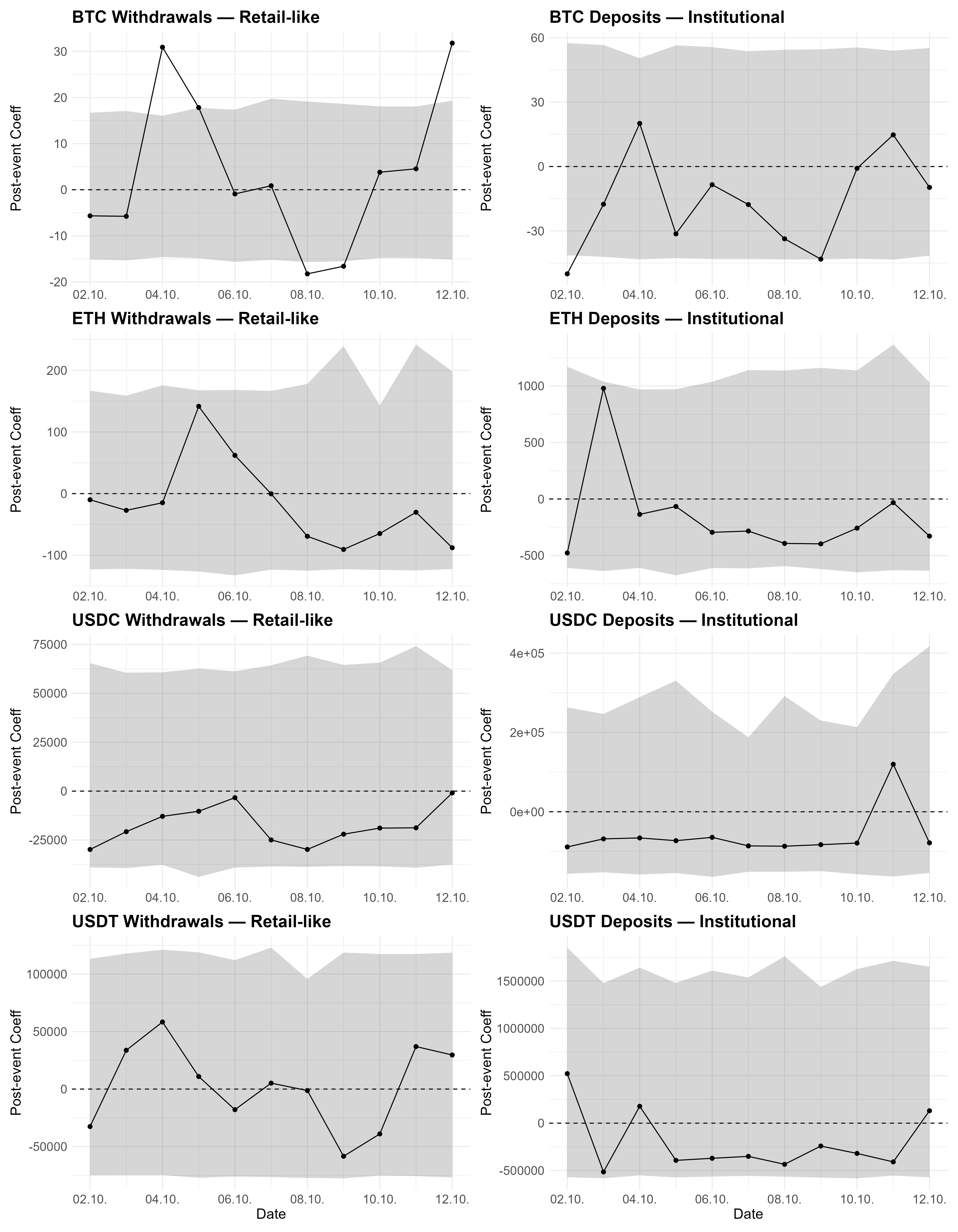}
	\caption{\textbf{FTX collapse: placebo test}. {\small \textit{Notes:} Selected post-event coefficients from the FTX collapse event-study regressions, in asset units, estimated around a placebo event date of October 2, 2022, one month before the actual event. Each panel shows coefficients for days 0 to 10 after the placebo event, using a 45-day pre-event estimation window. The dashed horizontal line marks zero. Shaded areas show bootstrap bands based on 1,000 replications. Panels correspond to those in Figure~\ref{figure:EventStudyFTXCollapse}: the left column reports retail-like withdrawals and the right column institutional deposits. Rows correspond to BTC, ETH, USDC, and USDT, respectively.}}
	\label{figure:placebo_ftx_app}
\end{figure}

\begin{figure}
	\centering
	\includegraphics[width=\textwidth]{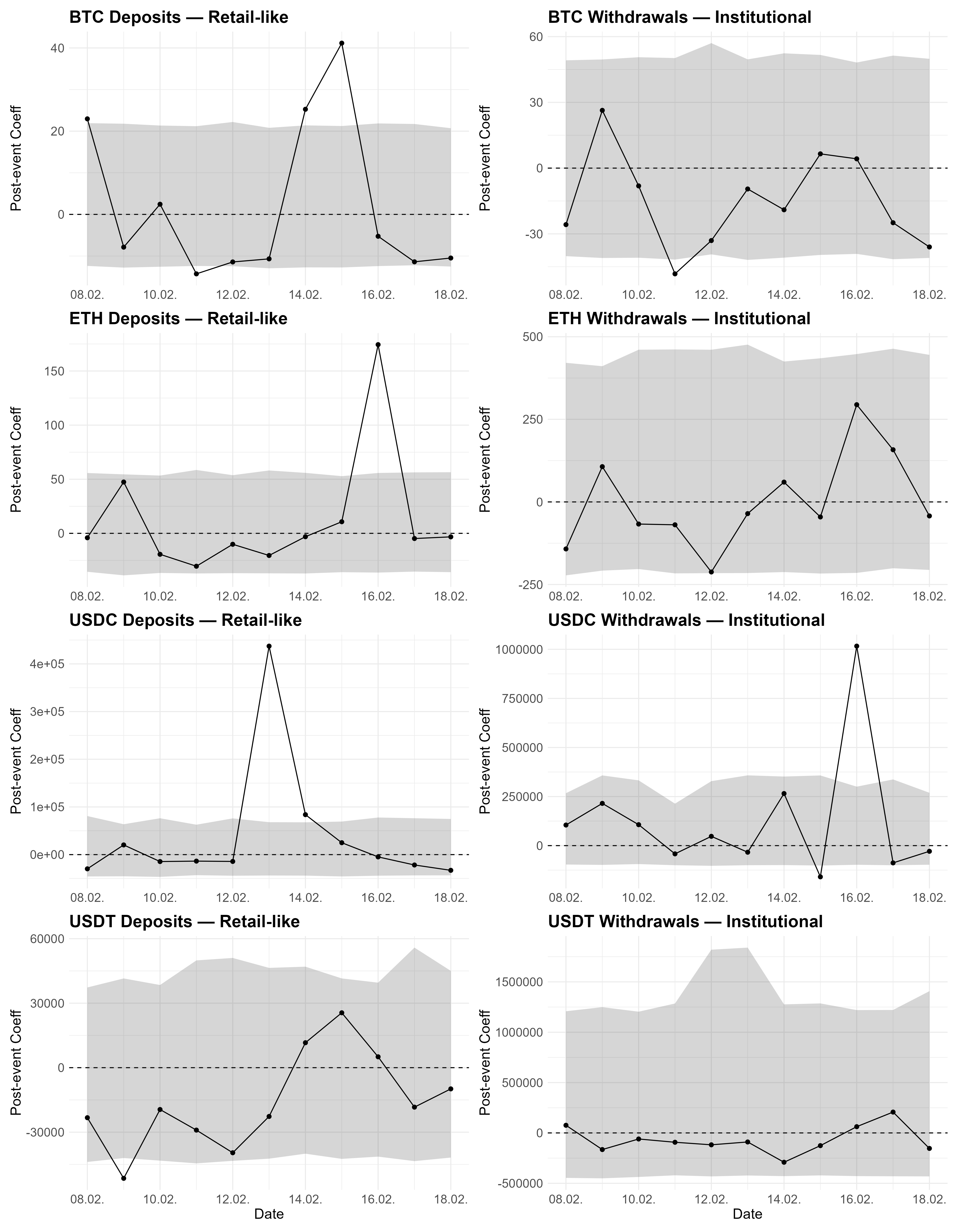}
	\caption{\textbf{SVB failure: placebo test}. {\small \textit{Notes:} Selected post-event coefficients from the SVB failure event-study regressions, in asset units, estimated around a placebo event date of February 8, 2023, one month before the actual event. Each panel shows coefficients for days 0 to 10 after the placebo event, using a 45-day pre-event estimation window. The dashed horizontal line marks zero. Shaded areas show bootstrap bands based on 1,000 replications. Panels correspond to those in Figure~\ref{figure:EventStudySVBFailure}: the left column reports retail-like deposits and the right column institutional withdrawals. Rows correspond to BTC, ETH, USDC, and USDT, respectively.}}
	\label{figure:placebo_svb_app}
\end{figure}

\begin{figure}
	\centering
	\includegraphics[width=\textwidth]{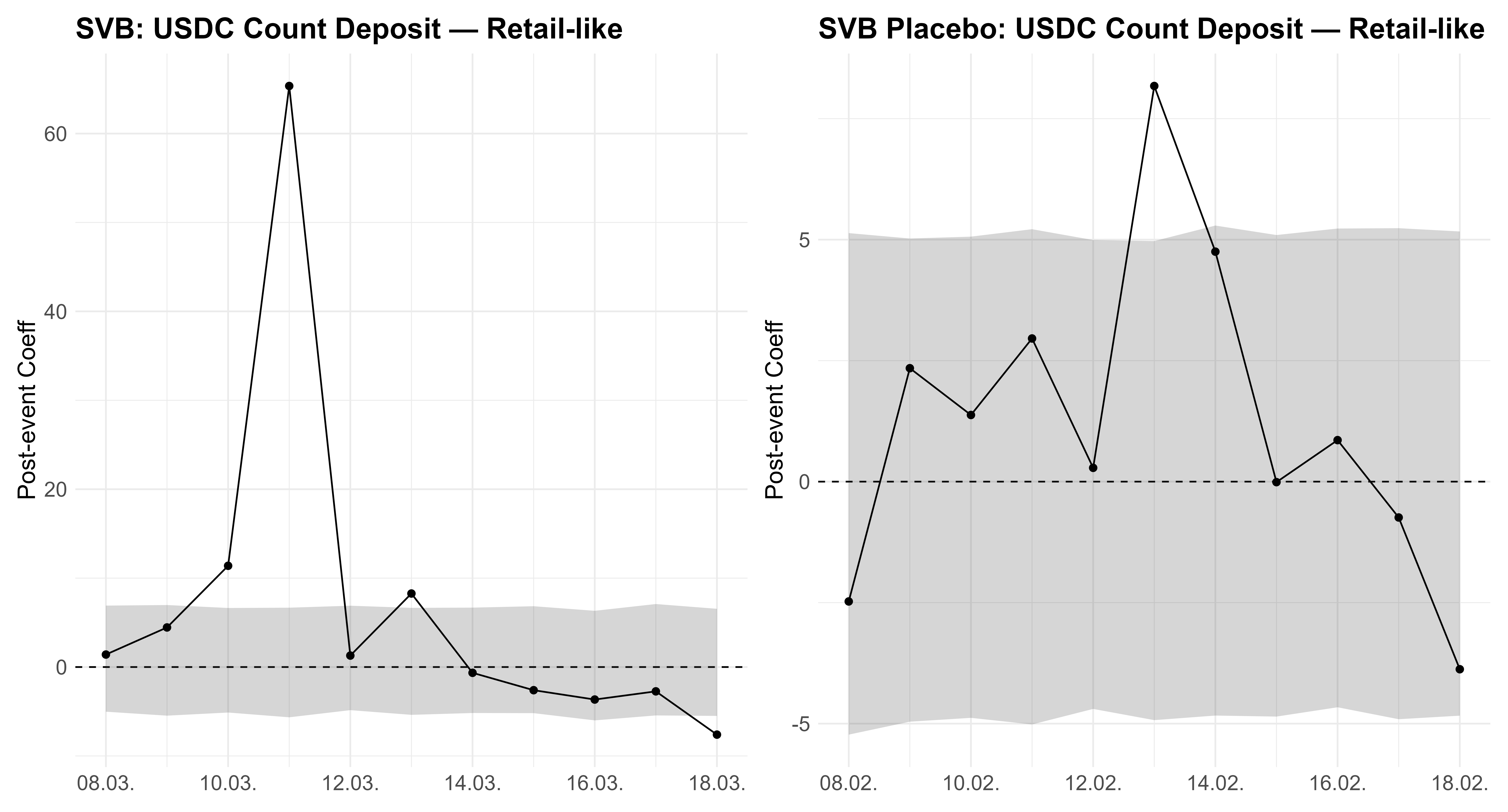}
	\caption{\textbf{SVB failure: retail-like USDC transaction counts, real event (left) and placebo (right)}. {\small \textit{Notes:} Post-event coefficients from the SVB failure event-study regressions, using the number of USDC transactions executed by retail-like counterparties as the outcome variable, estimated around the actual event date (March 8, 2023) and the placebo event date (February 8, 2023). Each panel shows coefficients for days 0 to 10 after the event, using a 45-day pre-event estimation window. The dashed horizontal line marks zero. Shaded areas show bootstrap bands based on 1,000 replications.}}
	\label{figure:placebo_svb_txcount_app}
\end{figure}

We report the results in Figures~\ref{figure:placebo_terra_app}, \ref{figure:placebo_ftx_app}, and~\ref{figure:placebo_svb_app}, which show the same asset, group, and flow-direction panels as Figures~\ref{figure:EventStudyTerraLunaCrash}, \ref{figure:EventStudyFTXCollapse}, and~\ref{figure:EventStudySVBFailure}, respectively.
Most placebo coefficients fall within the bootstrap bands, indicating that flows are largely unaffected around the placebo dates. The few exceedances in the Terra Luna and FTX placebo are confined to one or two days and lack the timing and persistence observed around the actual events.
The SVB placebo shows more single-day exceedances. The largest occurs in retail-like USDC deposits on February 13, with a further single-day spike in institutional USDC withdrawals on February 16. These movements are comparable to the ones observed in Figure~\ref{figure:EventStudySVBFailure} and may partly reflect stablecoin-specific news in the placebo window.
To assess whether the March response is distinct from this pattern, we compare the number of USDC transactions executed by retail-like counterparties around the real and the placebo SVB event dates (Figure~\ref{figure:placebo_svb_txcount_app}). The increase in transaction counts around the real SVB event largely exceeds that around the placebo date. This indicates that the March 2023 response involved an unusually large number of retail-like counterparties, consistent with the run-like pattern discussed in Section~\ref{sec:SVBFailureResults}.
Overall, the placebo results support the interpretation that observed flows are related to the shocks rather than to routine fluctuations in CASP flows.

\end{document}